\documentclass[aps,preprint,nofootinbib]{revtex4}%
\usepackage{amsfonts}
\usepackage{amsmath}
\usepackage{amssymb}
\usepackage{graphicx}%
\providecommand{\U}[1]{\protect\rule{.1in}{.1in}}

\begin{document}
\preprint{ }
\title[Short title for running header]{The Hamiltonian of Einstein affine-metric formulation of General Relativity}
\author{N. Kiriushcheva}
\email{nkiriush@uwo.ca}
\author{S.V. Kuzmin}
\email{skuzmin@uwo.ca}
\affiliation{Faculty of Arts and Social Science, Huron University College, N6G 1H3 and
Department of Applied Mathematics, University of Western Ontario, N6A 5B7,
London, Canada}
\pacs{PACS number}

\begin{abstract}
It is shown that the Hamiltonian of the Einstein affine-metric (first order)
formulation of General Relativity (GR) leads to a constraint structure that
allows the restoration of its unique gauge invariance, four-diffeomorphism,
without the need of any field dependent redefinition of gauge parameters as is
the case for the second order formulation. In the second order formulation of
ADM gravity the need for such a redefinition is the result of the
non-canonical change of variables [arXiv: 0809.0097]. For the first order
formulation, the necessity of such a redefinition \textquotedblleft to
correspond to diffeomorphism invariance\textquotedblright\ (reported by
Ghalati [arXiv: 0901.3344]) is just an artifact of using the
Henneaux-Teitelboim-Zanelli ansatz [Nucl. Phys. B 332 (1990) 169], which is
sensitive to the choice of linear combination of tertiary constraints. This
ansatz cannot be used as an algorithm for finding a gauge invariance, which is
a unique property of a physical system, and it should not be affected by
different choices of linear combinations of non-primary first class
constraints. The algorithm of Castellani [Ann. Phys. 143 (1982) 357] is free
from such a deficiency and it leads directly to four-diffeomorphism invariance
for first, as well as for second order Hamiltonian formulations of GR. The
distinct role of primary first class constraints, the effect of considering
different linear combinations of constraints, the canonical transformations of
phase-space variables, and their interplay are discussed in some detail for
Hamiltonians of the second and first order formulations of metric GR. The
first order formulation of Einstein-Cartan theory, which is the classical
background of Loop Quantum Gravity, is also discussed.

\end{abstract}
\eid{identifier}
\date{\today}
\maketitle


\section{Introduction}

We reconsider the Hamiltonian of General Relativity (GR) by using its
equivalent first order form, the affine-metric formulation of Einstein
\cite{Einstein}. The reason for returning to this old and apparently solved
problem is twofold.

Firstly, in the literature this problem is claimed to have been solved 50
years ago by Arnowitt, Deser and Misner (ADM) \cite{ADM1959}. The comparison
of the ADM Hamiltonian with the Dirac Hamiltonian \cite{Dirac} shows some
similarities (not equivalence) \cite{Myths}; but Dirac based his derivation on
second order, metric, GR. Preliminary results on the GR Hamiltonian for an
equivalent first order formulation, based on slightly different but equivalent
set of variables (a linear combination of affine connections), leads to a
different conclusion; in particular, the necessity to have tertiary
constraints \cite{KK, KKM}, contrary to the ADM treatment of the same problem.
Recently, and for the first time, the Dirac analysis of the first order
formulation was completed by Ghalati and McKeon \cite{G/R} with an explicit
demonstration of the closure of the Dirac procedure \cite{Diracbook} and with
the explicit form of the tertiary constraints given. This differs from the ADM
Hamiltonian formulation and the reason for this discrepancy lies in the
solving of the first class constraints, as indicated and discussed in
\cite{KK, KKM, G/R}. According to the Dirac procedure, only second class
constraints can be solved and the Poisson brackets of the remaining
phase-space variables might be modified (Dirac brackets) \cite{Kurt,
Gitman-Tyutin}.

Secondly, the Dirac analysis for systems with first class constraints cannot
be considered complete without the restoration of gauge transformations that,
in accordance with the Dirac conjecture \cite{Diracbook}, needs all first
class constraints. Some steps of such a restoration and partial transformation
for one phase-space variable, $h^{00}=\sqrt{-g}g^{00}$, was recently reported
by Ghalati \cite{Novel} based on earlier obtained in \cite{G/R} first class
constraints. In \cite{Novel} the transformation of $h^{00}$ was found using
the approach proposed by Henneaux, Teitelboim and Zanelli (HTZ) \cite{HTZ}.
This transformation is different from four-diffeomorphism and the author of
\cite{Novel} concluded that a field dependent redefinition of gauge parameters
is needed for the derived transformation \textquotedblleft to correspond to
diffeomorphism invariance\textquotedblright\ \cite{Novel}. This result is
puzzling because in the Hamiltonian formulation of the second order GR, the
necessity for such a field dependent redefinition is the result of a
non-canonical change of variables \cite{Myths}. Without such changes, the
four-diffeomorphism follows without any redefinition of gauge parameters as
was demonstrated for the Dirac formulation \cite{Dirac}, as well as for the
oldest Hamiltonian formulation of GR due to Pirani, Schild and Skinner (PSS)
\cite{PSS} in \cite{Myths, KKRV}. The equivalence of two formulations was
demonstrated in \cite{FKK}. In formulation of \cite{G/R}, the redefinitions of
phase-space variables that might affect the result obtained in \cite{Novel}
are canonical, so one expects the complete and direct restoration of
four-diffeomorphism without the need for any field dependent redefinition of
gauge parameters, as is found in the second order formulation. The gauge
invariance is a unique characteristic of a theory and equivalent second and
first order formulations should give the same gauge invariance. The only
difference in the second and first order formulations (considered in
\cite{Myths, KKRV} and \cite{Novel}, respectively) lies in the methods of the
restoration of gauge invariance that were used. For the Hamiltonian of the
second order GR, the Castellani algorithm \cite{Castellani} was used for both
Dirac \cite{Myths} and PSS \cite{KKRV} formulations; whereas in \cite{Novel}
the HTZ ansatz was employed.

We want to clarify these two discrepancies simultaneously; and this dictates
our choice of first order formulation. We perform our analysis for the
standard and more familiar affine-metric formulation due to Einstein
\cite{Einstein} because this formulation was the starting point in the ADM
analysis \cite{ADM1959}. The affine-metric formulation, and one used in
\cite{KK, KKM, G/R, Novel}, are both equivalent to the second order
formulation. The equivalence of affine-metric and metric GR was demonstrated
by Einstein \cite{Einstein} and for a different variable which is a linear
combination of affine connections used in \cite{G/R}, the equivalence was
explicitly demonstrated in Appendix A of \cite{KK}. Both first order
formulations lead to similar Hamiltonians, as will become clear in the course
of the calculations   for the affine-metric formulation presented here and
from the comparison of our results with ones obtained in \cite{G/R}. Such a
choice of variables (affine connections or a linear combination of affine
connections) cannot be responsible for appearance of a different gauge
invariance. There are some purely technical advantages in the parts of the
calculations for one formulation over another that we will comment on in the
course of our calculations; but they are not crucial, and neither formulation
gives any overall advantage in calculation efficiency. The only
\textquotedblleft advantages\textquotedblright\ of affine-metric formulation
that we want to mention, is the manifest covariant structure of primary first
class constraints and possibility of performing a direct comparison with known
transformations for affine connections.

We perform this analysis for all dimensions higher than two as a
specialization to the four dimensional case does not have any advantages or
peculiarities. The standard expression \textquotedblleft
four-diffeomorphism\textquotedblright\ that we will use is equally well
applied to all dimensions. The main goal of our article is a thorough analysis
of all the steps of calculation and, in particular, of the canonicity of all
changes of phase-space variables that are used. All of these \textquotedblleft
technicalities\textquotedblright\ are our main concern. Our article is not an
essay on the Hamiltonian formulation, but it is the Hamiltonian formulation
itself, with all the steps of calculations and with sufficient details that
anyone can repeat or check our derivations. So this article is the detailed
proof that the Hamiltonian of the affine-metric GR and its constraints lead
directly to four-diffeomorphism invariance without any redefinition of gauge
parameters. This is exactly as it was shown in \cite{KKRV, Myths} for the
second order formulation of GR, if we use exactly the same algorithm of
restoration of gauge invariance \cite{Castellani}.

The plan of the paper is as follows. In the next Section, by using Dirac
procedure, the Hamiltonian of the affine-metric formulation is obtained by
performing the Hamiltonian reduction, i.e. the elimination of the phase-space
variables associated with second class constraints. In Section 3 closure of
the Dirac procedure is demonstrated for a particular choice of tertiary
constraints and the algebra of Poisson brackets (PBs) among tertiary
constraints is compared with a similar algebra of secondary constraints of the
second order formulation. In Section 4 the effect of different choices of
linear combinations of non-primary\footnote{We call non-primary constraints
all secondary, tertiary constraints, etc. In the literature they are sometimes
all called secondary; but because we consider two particular formulations with
secondary and tertiary (not some general case) this terminology seems to be
preferable to avoid confusion.} first class constraints are considered and
their interplay with canonical transformations of phase-space variables is
discussed. Section 5 provides arguments for a special role of primary first
class constraints in the Hamiltonian formulation of gauge invariant theories.
The examples from the Hamiltonian formulation of the second order metric GR as
well as the Hamiltonian formulation of the first order Einstein-Cartan theory
are discussed and the effect of unjustified manipulations with primary
constraints is illustrated by an example of the Hamiltonian of Loop Quantum
Gravity (LQG). In Section 6 using the Castellani algorithm we restore as in
\cite{Novel} the partial transformations, but for \textit{all} the phase-space
variables of the reduced Hamiltonian and we obtain \textit{all }terms with
temporal derivatives of fields and gauge parameters in the gauge
transformations that coincide with four-diffeomorphism. We show that different
choices of combinations of non-primary constraints do not affect the gauge
transformations if we use the correct method to restore gauge invariance. In
Section 7 we demonstrate the sensitivity of HTZ ansatz to a choice of tertiary
constraints, contrary to the Castellani algorithm. Such an ambiguity and
dependence of gauge invariance on a choice of a linear combination of
non-primary first class constraints is in contradiction to the uniqueness of
gauge invariance, which is an important property of a theory. This explains
the origin of the puzzling result reported in \cite{Novel}. The reason for the
failure of the HTZ ansatz is discussed, which is also related to a special
role of the primary constraints - the true Masters of Hamiltonians for gauge
invariant theories. In the Appendix A the details of the solution for the
secondary second class constraints that were used in Section 2 to find the
reduced Hamiltonian is given.

\section{The Hamiltonian }

We start from the first order, affine-metric, Einstein action \cite{Einstein}%

\begin{equation}
S_{E}\left(  g^{\alpha\beta},\Gamma_{\alpha\sigma}^{\lambda}\right)  =\int
L\left(  g^{\alpha\beta},\Gamma_{\alpha\sigma}^{\lambda}\right)
dx^{D}\label{eqnAM1}%
\end{equation}
with the Lagrange density function%

\begin{equation}
L\left(  g^{\alpha\beta},\Gamma_{\alpha\sigma}^{\lambda}\right)  =-\sqrt
{-g}g^{\alpha\beta}\left(  \Gamma_{\alpha\beta,\lambda}^{\lambda}%
-\Gamma_{\alpha\lambda,\beta}^{\lambda}+\Gamma_{\sigma\lambda}^{\lambda}%
\Gamma_{\alpha\beta}^{\sigma}-\Gamma_{\alpha\sigma}^{\lambda}\Gamma
_{\beta\lambda}^{\sigma}\right)  \label{eqnAM2}%
\end{equation}
where the metric,\ $g^{\mu\nu},$ and affine connection, $\Gamma_{\alpha\beta
}^{\lambda},$ are treated as independent variables, $g=\det\left(  g_{\mu\nu
}\right)  $, and $D$ is the dimension of spacetime. Greek letters are used for
\textquotedblleft spacetime\textquotedblright\ indices ($\mu=0,1,...,D-1$) and
Latin letters for \textquotedblleft space\textquotedblright\ indices
($k=1,...,D-1$).

The first step in passing to the Hamiltonian formulation is the explicit
separation of terms with temporal derivatives (\textquotedblleft
kinetic\textquotedblright\ part of $L$). For (\ref{eqnAM2}) we obtain%

\begin{equation}
L_{kin}=-\sqrt{-g}g^{km}\Gamma_{km,0}^{0}-\sqrt{-g}g^{k0}\left(  \Gamma
_{k0,0}^{0}-\Gamma_{km,0}^{m}\right)  +\sqrt{-g}g^{00}\Gamma_{0k,0}^{k}.
\label{eqnAM5}%
\end{equation}

This suggests the following field redefinition:%

\begin{equation}
\Gamma_{km}^{0}=\Sigma_{km},\text{ \ }\Gamma_{k0}^{0}=2\Sigma_{k0}+\Gamma
_{km}^{m},\text{\ }\Gamma_{0m}^{k}=\Sigma_{0m}^{k}-\frac{1}{D-1}\delta_{m}%
^{k}\Sigma_{00},\label{eqnAM6}%
\end{equation}
where $\Sigma_{0m}^{k}$ is a traceless field, $\Sigma_{0k}^{k}=0$, i.e.
$\Sigma_{00}=-\Gamma_{0k}^{k}$.

This redefinition does not affect the following components:%

\begin{equation}
\Gamma_{00}^{\mu}=\Gamma_{00}^{\mu},\text{ }\Gamma_{kp}^{m}=\Gamma_{kp}^{m}.
\label{eqnAM7}%
\end{equation}

After integration by parts and a change of variables, (\ref{eqnAM6}), the
\textquotedblleft kinetic\textquotedblright\ part of the Lagrangian becomes diagonal%

\begin{equation}
L_{kin}=\left(  \sqrt{-g}g^{km}\right)  _{,0}\Sigma_{km}+2\left(  \sqrt
{-g}g^{k0}\right)  _{,0}\Sigma_{k0}+\left(  \sqrt{-g}g^{00}\right)
_{,0}\Sigma_{00}=\left(  \sqrt{-g}g^{\alpha\beta}\right)  _{,0}\Sigma
_{\alpha\beta}.\label{eqnAM8}%
\end{equation}
and \textquotedblleft potential\textquotedblright\ part of the Lagrangian
(terms without \textquotedblleft velocities\textquotedblright) is%

\begin{equation}
L_{pot}=\left(  \sqrt{-g}g^{00}\right)  _{,k}\Gamma_{00}^{k}+2\left(
\sqrt{-g}g^{p0}\right)  _{,k}\Sigma_{0p}^{k}-\left(  \sqrt{-g}g^{k0}\right)
_{,k}\left(  \Gamma_{00}^{0}-\frac{D-3}{D-1}\Sigma_{00}\right)  \label{eqnAM9}%
\end{equation}

\[
+\left(  \sqrt{-g}g^{pq}\right)  _{,k}\Gamma_{pq}^{k}-2\left(  \sqrt{-g}%
g^{pk}\right)  _{,k}\left(  \Sigma_{0p}+\Gamma_{pm}^{m}\right)
\]

\[
-\sqrt{-g}g^{00}\left(  -\Sigma_{00}\Gamma_{00}^{0}-2\Sigma_{0k}\Gamma
_{00}^{k}-\Sigma_{0k}^{m}\Sigma_{0m}^{k}-\frac{1}{D-1}\Sigma_{00}\Sigma
_{00}\right)
\]

\[
-2\sqrt{-g}g^{k0}\left(  -2\Sigma_{00}\Sigma_{0k}-\Sigma_{00}\Gamma_{km}%
^{m}+\Gamma_{mp}^{p}\Sigma_{0k}^{m}-\Sigma_{km}\Gamma_{00}^{m}-\Gamma_{km}%
^{p}\Sigma_{0p}^{m}\right)
\]

\[
-\sqrt{-g}g^{km}\left[  \Gamma_{00}^{0}\Sigma_{km}+2\Sigma_{0p}\Gamma_{km}%
^{p}+2\Gamma_{pq}^{q}\Gamma_{km}^{p}-4\Sigma_{0k}\Sigma_{0m}\right.
\]

\[
\left.  -4\Sigma_{0k}\Gamma_{mq}^{q}-\Gamma_{kp}^{p}\Gamma_{mq}^{q}%
-2\Sigma_{kp}\Sigma_{0m}^{p}-\frac{D-3}{D-1}\Sigma_{km}\Sigma_{00}-\Gamma
_{kq}^{p}\Gamma_{mp}^{q}\right]  .
\]

Using the above variables and by performing Legendre transformation, we obtain
the total Hamiltonian%

\begin{equation}
H_{T}=\dot{g}^{\alpha\beta}P_{\alpha\beta}+\dot{\Sigma}_{\alpha\beta}%
\Pi^{\alpha\beta}+\dot{\Gamma}_{00}^{\mu}\Pi_{\mu}^{00}+\dot{\Sigma}_{0m}%
^{k}\Pi_{k}^{0m}+\dot{\Gamma}_{kp}^{m}\Pi_{m}^{kp}-\left(  \sqrt{-g}%
g^{\alpha\beta}\right)  _{,0}\Sigma_{\alpha\beta}-L_{pot}. \label{eqnAM12}%
\end{equation}

The new set of independent variables, (\ref{eqnAM6}) and (\ref{eqnAM7}), and
their conjugate momenta obey the fundamental Poisson brackets (PBs)%

\begin{equation}
\left\{  g^{\alpha\beta}\left(  \overrightarrow{x}\right)  ,P_{\mu\nu}\left(
\overrightarrow{y}\right)  \right\}  =\Delta_{\mu\nu}^{\alpha\beta}%
\delta\left(  \overrightarrow{x}-\overrightarrow{y}\right)  ,\text{
\ \ \ }\left\{  \Sigma_{\alpha\beta},\Pi^{\mu\nu}\right\}  =\Delta
_{\alpha\beta}^{\mu\nu}\equiv\frac{1}{2}\left(  \delta_{\alpha}^{\mu}%
\delta_{\beta}^{\nu}+\delta_{\beta}^{\mu}\delta_{\alpha}^{\nu}\right)  ,
\label{eqnAM10}%
\end{equation}

\begin{equation}
\left\{  \Gamma_{00}^{\mu},\Pi_{\nu}^{00}\right\}  =\delta_{\nu}^{\mu},\text{
\ }\left\{  \Sigma_{0m}^{k},\Pi_{q}^{0p}\right\}  =\delta_{q}^{k}\Delta
_{0m}^{0p}-\frac{1}{D-1}\delta_{m}^{k}\Delta_{0q}^{0p},\text{ \ \ \ }\left\{
\Gamma_{kp}^{m},\Pi_{a}^{nq}\right\}  =\delta_{a}^{m}\Delta_{kp}^{nq}.
\label{eqnAM11}%
\end{equation}

Here, only the first PB is written in a complete form. Further we will omit
the delta functions and the dependence on \textquotedblleft
space\textquotedblright\ vectors $\overrightarrow{x}$, $\overrightarrow{y}$ to
shorten our notation, except in the cases where the derivatives of the delta
functions appear; and where it is important to indicate with respect to what
argument the differentiation is performed.

We could equally well start from (\ref{eqnAM2}) to obtain the Hamiltonian and
then perform canonical transformations in phase space from the original pairs
$\left(  g^{\alpha\beta},P_{\alpha\beta}\right)  $ and $\left(  \Gamma
_{\alpha\sigma}^{\lambda},P_{\lambda}^{\alpha\sigma}\right)  $ to a new set of
variables: $\left(  g^{\alpha\beta},P_{\alpha\beta}\right)  ,$ $\left(
\Sigma_{\alpha\beta},\Pi^{\alpha\beta}\right)  ,$ $\left(  \Sigma_{0m}^{k}%
,\Pi_{k}^{0m}\right)  ,$ $\left(  \Gamma_{00}^{\mu},\Pi_{\mu}^{00}\right)  $
and $\left(  \Gamma_{kp}^{m},\Pi_{m}^{kp}\right)  $. This is a canonical
transformation that is automatically guaranteed for a linear and invertible
redefinition of fields.

As in any first order formulation, the Hamiltonian analysis leads to primary
constraints equal in number to the number of independent fields%

\begin{equation}
P_{\alpha\beta}-\sqrt{-g}E_{\alpha\beta}^{\mu\nu}\Sigma_{\mu\nu}%
\approx0,\text{ \ }\Pi^{\alpha\beta}\approx0,\text{ \ }\Pi_{\mu}^{00}%
\approx0,\text{ \ \ }\Pi_{k}^{0m}\approx0,\text{ \ }\Pi_{m}^{kp}%
\approx0\label{eqmAM14}%
\end{equation}
where we used%

\begin{equation}
\left(  \sqrt{-g}g^{\alpha\beta}\right)  _{,0}=\sqrt{-g}E_{\mu\nu}%
^{\alpha\beta}g_{,0}^{\mu\nu}\label{eqnAM15}%
\end{equation}
with%

\begin{equation}
E_{\mu\nu}^{\alpha\beta}\equiv\Delta_{\mu\nu}^{\alpha\beta}-\frac{1}{2}%
g_{\mu\nu}g^{\alpha\beta}. \label{eqnAM15a}%
\end{equation}

Among the primary constraints (\ref{eqmAM14}) we have one pair of second class
constraints with the following PB%

\begin{equation}
\left\{  P_{\alpha\beta}-\sqrt{-g}E_{\alpha\beta}^{\mu\nu}\Sigma_{\mu\nu}%
,\Pi^{\rho\sigma}\right\}  =-\sqrt{-g}E_{\alpha\beta}^{\rho\sigma
}~,\label{eqnAM16}%
\end{equation}
which can be easily eliminated (they are of a special form; and therefore, the
Dirac brackets among the remaining variables are the same as the corresponding
PBs \cite{Gitman-Tyutin}). The solution for this pair is%

\begin{equation}
\Pi^{\rho\sigma}=0,\text{ \ \ \ }\Sigma_{\mu\nu}=\frac{1}{\sqrt{-g}}I_{\mu\nu
}^{\alpha\beta}P_{\alpha\beta}\label{eqnAM17}%
\end{equation}
where%

\begin{equation}
I_{\mu\nu}^{\alpha\beta}\equiv\Delta_{\mu\nu}^{\alpha\beta}-\frac{1}%
{D-2}g_{\mu\nu}g^{\alpha\beta},\text{ \ \ }I_{\mu\nu}^{\alpha\beta}%
E_{\alpha\beta}^{\gamma\sigma}=\Delta_{\mu\nu}^{\gamma\sigma}. \label{eqnAM18}%
\end{equation}

Substitution of solution (\ref{eqnAM17}) into the Hamiltonian (\ref{eqnAM12})
leads to (the first Hamiltonian reduction)%

\begin{equation}
H_{T}=\dot{\Gamma}_{00}^{\mu}\Pi_{\mu}^{00}+\dot{\Sigma}_{0m}^{k}\Pi_{k}%
^{0m}+\dot{\Gamma}_{kp}^{m}\Pi_{m}^{kp}-L_{pot}\left(  \Sigma_{\mu\nu}%
=\frac{1}{\sqrt{-g}}I_{\mu\nu}^{\alpha\beta}P_{\alpha\beta}\right)  .
\label{eqnAM18a}%
\end{equation}

The appearance of the combinations $\sqrt{-g}g^{\alpha\beta}$ and $\frac
{1}{\sqrt{-g}}I_{\mu\nu}^{\alpha\beta}$ in the Hamiltonian and the
invertability of $I_{\mu\nu}^{\alpha\beta}$ suggest the following canonical
change of variables:%

\begin{equation}
h^{\alpha\beta}=\sqrt{-g}g^{\alpha\beta},\text{ \ \ }\pi_{\mu\nu}=\frac
{1}{\sqrt{-g}}I_{\mu\nu}^{\alpha\beta}P_{\alpha\beta}.\label{eqnAM19}%
\end{equation}
Such a change obviously preserves the relation (using (\ref{eqnAM15}) and
(\ref{eqnAM18}))%

\begin{equation}
\dot{h}^{\alpha\beta}\pi_{\alpha\beta}=\dot{g}^{\alpha\beta}P_{\alpha\beta
}.\label{eqnAM20}%
\end{equation}
So, this transformation is canonical \cite{Lanczos} and it can be also checked
explicitly that%

\begin{equation}
\left\{  h^{\alpha\beta},\pi_{\mu\nu}\right\}  =\left\{  \sqrt{-g}%
g^{\alpha\beta},\frac{1}{\sqrt{-g}}I_{\mu\nu}^{\rho\sigma}P_{\rho\sigma
}\right\}  _{g^{\alpha\beta},P_{\rho\sigma}}=\Delta_{\mu\nu}^{\alpha\beta}.
\label{eqnAM21}%
\end{equation}

Equation (\ref{eqnAM20}) is a necessary and sufficient condition for
canonicity \cite{Lanczos}. With such a canonical change, the Hamiltonian
becomes much simpler; moreover, because the Lagrangian (\ref{eqnAM2}) is
linear in $\sqrt{-g}g^{\alpha\beta}$, the Hamiltonian written in terms of new
fields is polynomial. (So, the often stated polynomiality of the tetrad
Hamiltonian constraints in Ashtekar variables \cite{Polynom} as one of the
advantages of the Hamiltonian formulation of the first order, tetrad-spin
connection, Einstein-Cartan theory, is actually not something special since
the first order affine-metric formulation is also polynomial in fields after
the canonical transformation (\ref{eqnAM19}) is performed).

Substitution of (\ref{eqnAM19}) into (\ref{eqnAM12}) constitutes the first
reduction and gives the total Hamiltonian with fewer variables%

\begin{equation}
H_{T}=\dot{\Gamma}_{00}^{\mu}\Pi_{\mu}^{00}+\dot{\Sigma}_{0m}^{k}\Pi_{k}%
^{0m}+\dot{\Gamma}_{kp}^{m}\Pi_{m}^{kp}+H_{c}~,\label{eqnAM22}%
\end{equation}
where the canonical Hamiltonian, $H_{c},$ is%

\begin{equation}
H_{c}=-\Gamma_{00}^{k}\left(  h_{,k}^{00}+2h^{00}\pi_{k0}+2h^{m0}\pi
_{km}\right)  -\Gamma_{00}^{0}\left(  -h_{,k}^{0k}+h^{00}\pi_{00}-h^{km}%
\pi_{km}\right)  \text{ } \label{eqnAM23}%
\end{equation}

\[
-h^{00}\frac{1}{D-1}\pi_{00}\pi_{00}-\frac{D-3}{D-1}\left(  h_{,k}^{k0}%
+h^{km}\pi_{km}\right)  \pi_{00}-\left(  4h^{k0}\pi_{00}+4h^{km}\pi
_{0m}-2h_{,p}^{pk}\right)  \pi_{0k}%
\]

\[
-h^{00}\Sigma_{0k}^{m}\Sigma_{0m}^{k}+2h^{k0}\left(  \Gamma_{mp}^{p}%
\Sigma_{0k}^{m}-\Gamma_{km}^{p}\Sigma_{0p}^{m}\right)  -2h^{km}\pi_{kp}%
\Sigma_{0m}^{p}-2h_{,k}^{p0}\Sigma_{0p}^{k}%
\]

\[
+h^{km}\left(  2\Gamma_{pq}^{q}\Gamma_{km}^{p}-\Gamma_{kp}^{p}\Gamma_{mq}%
^{q}-\Gamma_{kq}^{p}\Gamma_{mp}^{q}\right)  -2h^{k0}\pi_{00}\Gamma_{km}%
^{m}+2h^{km}\left(  \pi_{0p}\Gamma_{km}^{p}-2\pi_{0k}\Gamma_{mq}^{q}\right)
-h_{,k}^{pq}\Gamma_{pq}^{k}+2h_{,k}^{pk}\Gamma_{pm}^{m}.
\]

After the elimination of the primary second class constraints (\ref{eqnAM17})
(the first Hamiltonian reduction) and after performing the canonical
transformation (\ref{eqnAM19}), we must continue the Dirac procedure and
consider the time development of the remaining primary constraints. Two of
them give the following secondary constraints:%

\begin{equation}
\dot{\Pi}_{0}^{00}=\left\{  \Pi_{0}^{00},H_{c}\right\}  =-h_{,k}^{0k}%
+h^{00}\pi_{00}-h^{km}\pi_{km}\equiv\chi_{0}^{00}, \label{eqnAM24}%
\end{equation}

\begin{equation}
\dot{\Pi}_{k}^{00}=\left\{  \Pi_{k}^{00},H_{c}\right\}  =h_{,k}^{00}%
+2h^{00}\pi_{0k}+2h^{m0}\pi_{km}\equiv\chi_{k}^{00};\label{eqnAM25}%
\end{equation}
and the only non-zero PB among them is%

\begin{equation}
\left\{  \chi_{0}^{00},\chi_{k}^{00}\right\}  =-\chi_{k}^{00}. \label{eqnAM26}%
\end{equation}

The secondary constraints $\chi_{\mu}^{00}$ obviously (just from their field
content) have zero PBs with all of the primary constraints:%

\begin{equation}
\left\{  \chi_{\mu}^{00},\Pi_{\nu}^{00}\right\}  =0,\text{ \ \ \ }\left\{
\chi_{\mu}^{00},\Pi_{k}^{0m}\right\}  =0,\text{ \ \ \ }\left\{  \chi_{\mu
}^{00},\Pi_{m}^{kp}\right\}  =0,\label{eqnAM27}%
\end{equation}
so they are first class, at least, at this stage of the Dirac procedure.

Two remaining primary constraints,\ $\Pi_{k}^{0m}\approx0$ and\ $\Pi_{m}%
^{kp}\approx0$, lead to the secondary constraints%

\begin{equation}
\dot{\Pi}_{k}^{0m}=\left\{  \Pi_{k}^{0m},H_{c}\right\}  =-\frac{\delta H_{c}%
}{\delta\Sigma_{0m}^{k}}=\chi_{k}^{0m}, \label{eqnAM28}%
\end{equation}

\begin{equation}
\dot{\Pi}_{m}^{kp}=\left\{  \Pi_{m}^{kp},H_{c}\right\}  =-\frac{\delta H_{c}%
}{\delta\Gamma_{kp}^{m}}=\chi_{m}^{kp}.\label{eqnAM29}%
\end{equation}
And, because $H_{c}$ has quadratic contributions in the corresponding
coordinates $\left(  \Sigma_{0m}^{k},\Gamma_{kp}^{m}\right)  $, we have
secondary constraints, which are second class, and two additional pairs of
constraints, $\left(  \Pi_{k}^{0m},\chi_{k}^{0m}\right)  $ and $\left(
\Pi_{kp}^{m},\chi_{m}^{kp}\right)  ,$ which can be solved, and the
corresponding pairs of canonical variables, $\left(  \Pi_{k}^{0m},\Sigma
_{0m}^{k}\right)  $ and $\left(  \Pi_{kp}^{m},\Gamma_{kp}^{m}\right)  ,$ can
be eliminated without affecting the PBs of remaining fields, as pairs $\left(
\Pi_{k}^{0m},\chi_{k}^{0m}\right)  $ and $\left(  \Pi_{kp}^{m},\chi_{m}%
^{kp}\right)  $ are also of a special form. Solution of these constraints is
given in Appendix A. Substitution of the solutions into the Hamiltonian (the
second Hamiltonian reduction) gives us the following total Hamiltonian%

\begin{equation}
H_{T}\left(  \Gamma_{00}^{\mu},\Pi_{\mu}^{00},h^{\alpha\beta},\pi_{\alpha
\beta}\right)  =\dot{\Gamma}_{00}^{\mu}\Pi_{\mu}^{00}+H_{c}\label{eqnAM31}%
\end{equation}
with the canonical part $H_{c}$%

\begin{equation}
H_{c}=-\Gamma_{00}^{\mu}\chi_{\mu}^{00}+H_{c}^{\prime},\label{eqnAM32}%
\end{equation}
where $H_{c}^{\prime}$ is the part of the canonical Hamiltonian after
extracting the term $-\Gamma_{00}^{\mu}\chi_{\mu}^{00}$ (secondary
constraints)$:$%

\[
H_{c}^{\prime}=-h^{00}\frac{1}{D-1}\pi_{00}\pi_{00}-\frac{D-3}{D-1}D_{k}%
^{0k}\pi_{00}-\left(  4h^{k0}\pi_{00}+4h^{km}\pi_{0m}-2h_{,p}^{pk}\right)
\pi_{0k}%
\]

\begin{equation}
+\frac{1}{h^{00}}D_{k}^{0n}D_{n}^{0k}-\frac{1}{D-1}\frac{1}{h^{00}}D_{n}%
^{0n}D_{k}^{0k} \label{eqnAM32a}%
\end{equation}

\[
+\frac{1}{2}D_{y}^{xb}h_{bz}D_{x}^{yz}-\frac{1}{4}h_{ay}h_{bz}e^{nx}D_{n}%
^{ab}D_{x}^{yz}+\frac{1}{4}\frac{1}{D-2}e^{nx}h_{ab}D_{n}^{ab}h_{yz}D_{x}^{yz}%
\]

Here we use the notation introduced by Dirac in the second order formulation
of GR \cite{Dirac} as it naturally arises in the course of calculation (see
Appendix A):%

\begin{equation}
e^{nq}\equiv h^{nq}-\frac{h^{0n}h^{0q}}{h^{00}},\text{ \ }e^{nq}h_{qp}%
=\delta_{p}^{n}.\text{\ } \label{eqnAM33}%
\end{equation}

Note that $e^{nq}$ is a short-hand notation, not a new variable. We introduce
the combinations $D_{a}^{0b}$ and $D_{n}^{ab}$, and their explicit forms are
calculated in Appendix and given by:%

\begin{equation}
D_{m}^{0k}=h_{,m}^{k0}+h^{bk}\pi_{bm}, \label{eqnAM34}%
\end{equation}

\begin{equation}
D_{m}^{kq}=-2h^{kq}\pi_{0m}+h_{,m}^{kq}-\frac{h^{q0}}{h^{00}}D_{m}^{0k}%
-\frac{h^{k0}}{h^{00}}D_{m}^{0q}+\frac{1}{D-1}\left(  \frac{1}{h^{00}}%
D_{c}^{0c}-\pi_{00}\right)  \left(  h^{0q}\delta_{m}^{k}+h^{k0}\delta_{m}%
^{q}\right)  . \label{eqnAM35}%
\end{equation}

The next step of the Dirac procedure is the time development of the secondary
first class constraints $\dot{\chi}_{\mu}^{00}=\left\{  \chi_{\mu}^{00}%
,H_{T}\right\}  $, to check whether they produce tertiary constraints. At this
step of the analysis, we can proceed in two different ways. The direct way is
to obtain $\left\{  \chi_{\mu}^{00},H_{T}\right\}  ,$ and single out already
known secondary constraints and form tertiary constraints from what is left.
Another way is to isolate terms with secondary constraints, in addition to
$\Gamma_{00}^{\mu}\chi_{\mu}^{00}$, in the total Hamiltonian before
calculating $\left\{  \chi_{\mu}^{00},H_{T}\right\}  $. The terms in $H_{T}$
proportional to the secondary constraints $A_{00}^{\nu}\chi_{\nu}^{00}$ will
give PB $\left\{  \chi_{\mu}^{00},A_{00}^{\nu}\chi_{\nu}^{00}\right\}  $ with
the result which is proportional to the secondary constraints because of
(\ref{eqnAM26}). In this case, in order to find the tertiary constraints (if
any), we have to consider the PBs of secondary constraints with the rest of
the Hamiltonian, what is left after all terms with secondary constraints were
isolated. Note that considering linear combinations of constraints is
perfectly consistent with the Dirac procedure (as a linear combination of
constraints is also a constraint).

Performing the direct calculation, we use the following simple PBs of the
secondary constraint $\chi_{0}^{00}$ with the combinations presented in
(\ref{eqnAM32})-(\ref{eqnAM35}):%

\begin{equation}
\left\{  \chi_{0}^{00},\frac{1}{h^{00}}\right\}  =\frac{1}{h^{00}}\text{,
}\left\{  \chi_{0}^{00},h^{00}\right\}  =-h^{00}\text{, }\left\{  \chi
_{0}^{00},\pi_{00}\right\}  =\pi_{00},\text{ } \label{eqnAM37}%
\end{equation}

\begin{equation}
\left\{  \chi_{0}^{00},h^{0k}\right\}  =0,\text{ }\left\{  \chi_{0}^{00}%
,D_{m}^{0k}\right\}  =0, \label{eqnAM38}%
\end{equation}

\begin{equation}
\left\{  \chi_{0}^{00},4h^{k0}\pi_{00}+4h^{km}\pi_{0m}-2h_{,p}^{pk}\right\}
=4h^{k0}\pi_{00}+4h^{km}\pi_{0m}-2h_{,p}^{pk},\label{eqnAM39}%
\end{equation}
as well as%

\begin{equation}
\left\{  \chi_{0}^{00},h_{,m}^{kq}-2h^{kq}\pi_{0m}\right\}  =h_{,m}%
^{kq}-2h^{kq}\pi_{0m}\label{eqnAM40}%
\end{equation}
that, in combination with (\ref{eqnAM37}) and (\ref{eqnAM38}), gives%

\begin{equation}
\left\{  \chi_{0}^{00},D_{m}^{pk}\right\}  =D_{m}^{pk}.\label{eqnAM41}%
\end{equation}
Using definition (\ref{eqnAM33}) we also find:%

\begin{equation}
\left\{  \chi_{0}^{00},e^{km}\right\}  =e^{km},\text{ \ }\left\{  \chi
_{0}^{00},h_{km}\right\}  =-h_{km}.\label{eqnAM42}%
\end{equation}
The first PB in (\ref{eqnAM42}) requires calculations; but the second is just
the result of $e^{km}h_{kn}=\delta_{n}^{m}$. Similar PBs for $\chi_{k}^{00}$ are:%

\begin{equation}
\left\{  \chi_{b}^{00},e^{km}\right\}  =0\text{, \ }\left\{  \chi_{b}%
^{00},h_{km}\right\}  =0,\label{eqnAM43}%
\end{equation}

\begin{equation}
\left\{  \chi_{b}^{00},h^{00}\right\}  =0.\label{eqnAM43a}%
\end{equation}

We would like to mention here that we originally analyzed the first order of
the EH action in different variables \cite{KK, KKM} where we used another
combination\footnote{This is generalization of change of variables that we
found considering the Hamiltonian of 2D in \cite{2D}. Later we learned that
such a combination was known before (see Horava \cite{Horava} for any
dimension and Kijowski for four-dimesional case \cite{Kijowski}).}%

\begin{equation}
\xi_{\alpha\beta}^{\lambda}=\Gamma_{\alpha\beta}^{\lambda}-\frac{1}{2}\left(
\delta_{\alpha}^{\lambda}\Gamma_{\beta\sigma}^{\sigma}+\delta_{\beta}%
^{\lambda}\Gamma_{\alpha\sigma}^{\sigma}\right)  \label{eqnAM7a}%
\end{equation}
and discussed the unavoidable appearance of tertiary constraints. These
variables, (\ref{eqnAM7a}), simplify the calculations because the
\textquotedblleft kinetic\textquotedblright\ part of the Lagrangian becomes
diagonal automatically: $L_{kin}=-\sqrt{-g}g^{\alpha\beta}\xi_{\alpha\beta
,0}^{0}$. But this simplification appears only in the first steps of the
Hamiltonian analysis. These variables were used by Ghalati and McKeon to find
the Hamiltonian for the first order formulation of GR and, for the first time,
they explicitly demonstrated the closure of the Dirac procedure \cite{G/R}.
Firstly, our choice to start from the affine connections that lead to
combinations (\ref{eqnAM6}) and (\ref{eqnAM7}) is dictated by our goal to
compare our results with \cite{ADM1959}, where the authors claimed that they
used Palatini formulation,\footnote{This formulation was originally introduced
by Einstein \cite{Einstein}, but continues to be mistakenly attributed to
Palatini \cite{Frang} (see also Palatini's original paper \cite{Palatini} and
its English translation \cite{Palatini-English}).} which is the affine-metric
formulation due to Einstein. Secondly, the variables (\ref{eqnAM6}%
)-(\ref{eqnAM7}) also diagonalize the \textquotedblleft
kinetic\textquotedblright\ part of the Lagrangian (\ref{eqnAM8}); and, what is
more important, the primary first class constraints $\Pi_{\mu}^{00}$ appear in
the covariant form. In addition, if we used $\xi_{\alpha\beta}^{\lambda}$ from
(\ref{eqnAM7a}) instead of $\Gamma_{\alpha\beta}^{\lambda}$ then some of the
brackets (\ref{eqnAM37})-(\ref{eqnAM41}) would be more complicated. For
example, (\ref{eqnAM41}) became non-local (proportional to derivatives of
delta functions). In this case, it would be impossible to use simple
associative properties of PBs for the terms of $H_{c}^{\prime}$ in the second
and third lines of (\ref{eqnAM32a}) when calculating $\left\{  \chi_{\mu}%
^{00},H_{c}^{\prime}\right\}  $. Whereas, using the Hamiltonian in terms of
$\Gamma_{\alpha\beta}^{\lambda}$ and properties of the above PBs the
calculation of $\left\{  \chi_{0}^{00},H_{c}^{\prime}\right\}  $ is greatly
simplified, for example, from (\ref{eqnAM41}) and (\ref{eqnAM42}) it
immediately follows that $\left\{  \chi_{0}^{00},D_{y}^{xb}h_{bz}D_{x}%
^{yz}\right\}  =D_{y}^{xb}h_{bz}D_{x}^{yz}$, etc. One more advantage in using
the original variables $\Gamma_{\alpha\beta}^{\lambda}$, is the simplification
of restoration of gauge invariance; we will not need to restore the gauge
transformations of $\Gamma_{\alpha\beta}^{\lambda}$ from that of $\xi
_{\alpha\beta}^{\lambda}$. The above arguments are mainly technical and the
real advantage is just a manifestly covariant form of the primary first class
constraints. As we worked with both $\xi_{\alpha\beta}^{\lambda}$ and
$\Gamma_{\alpha\beta}^{\lambda}$ , we have to admit that there is no overall
advantage in a particular choice of variables and the total amount of
calculation (difficulties) is conserved.

For the time development of $\chi_{0}^{00}$ we have%

\begin{equation}
\dot{\chi}_{0}^{00}=\left\{  \chi_{0}^{00},H_{T}\right\}  =\left\{  \chi
_{0}^{00},-\Gamma_{00}^{\mu}\chi_{\mu}^{00}\right\}  +\left\{  \chi_{0}%
^{00},H_{c}^{\prime}\right\}  =\Gamma_{00}^{k}\chi_{k}^{00}+\left\{  \chi
_{0}^{00},H_{c}^{\prime}\right\}  \label{eqnAM44}%
\end{equation}
where using properties (\ref{eqnAM37})-(\ref{eqnAM42}), the last PB can be
just read off%

\begin{equation}
\left\{  \chi_{0}^{00},H_{c}^{\prime}\right\}  =H_{c}^{\prime}+\left(
2h^{k0}\pi_{00}+2h^{km}\pi_{0m}-h_{,p}^{pk}\right)  _{,k}~.\label{eqnAM45}%
\end{equation}
Note that the right-hand side of (\ref{eqnAM45}) is not the Hamiltonian
density and we cannot neglect this spatial derivative.

Because all PBs (\ref{eqnAM37})-(\ref{eqnAM42}) are local (no derivatives of
delta functions), using them and associative properties of PB makes the result
(\ref{eqnAM45}) almost obvious. The first term in (\ref{eqnAM44}),
$\Gamma_{00}^{k}\chi_{k}^{00}$, is proportional to the secondary constraints.
The rest is given by (\ref{eqnAM45}) and to find out whether we have closure
of Dirac procedure at this stage or if the next generation of constraints
appears, we first have to find in (\ref{eqnAM45}) the combinations
proportional to the secondary constraints. If there are contributions, which
are not proportional to secondary constraints in (\ref{eqnAM45}), then we have
tertiary constraints.

We proceed as follows. Using secondary constraints (\ref{eqnAM24}) and
(\ref{eqnAM25}) we express some fields in terms of constraints and remaining
variables, substitute them into the Hamiltonian, isolate the contributions,
which are proportional to secondary first class constraints, and then work
with what is left. Note that according to the Dirac procedure we cannot solve
first class constraints, as it was done in \cite{ADM1959} and which is shown
especially clearly by Faddeev in \cite{Faddeev}; and the consequence of
solving first class constraints was discussed in \cite{KK, KKM, G/R}. These
re-expressions, to single out contributions proportional to secondary first
class constraints, after calculation of PB (\ref{eqnAM45}) or before, as it
was done in \cite{G/R}, is consistent with the possibility of using linear
combinations of non-primary first class constraints (the role primary first
class constraints will be discussed in Section 5).

The only two phase-space variables that can be unambiguously re-expressed (by
algebraic operations) using secondary constraints (\ref{eqnAM24}),
(\ref{eqnAM25}) are two momenta:%

\begin{equation}
\pi_{00}=\frac{1}{h^{00}}D_{k}^{0k}+\frac{1}{h^{00}}\chi_{0}^{00},
\label{eqnAM50}%
\end{equation}

\begin{equation}
\pi_{0k}=-\frac{1}{2}\frac{1}{h^{00}}h_{,k}^{00}-\frac{1}{h^{00}}h^{m0}%
\pi_{km}+\frac{1}{2}\frac{1}{h^{00}}\chi_{k}^{00}. \label{eqnAM51}%
\end{equation}

After substitution of (\ref{eqnAM50}) and (\ref{eqnAM51}) into the Hamiltonian
(\ref{eqnAM32a}), we separate terms proportional to $\chi_{\mu}^{00}$ and
write the rest of the Hamiltonian as a sum of three contributions of different
order in $\pi_{km}$ (e.g. $H_{c}^{\prime\prime}\left(  2\right)  $ is of
the\ second order in $\pi_{km}$, etc.). Performing the same operations with
(\ref{eqnAM45}) we obtain the following form of the Hamiltonian $H_{c}%
^{\prime}$%

\begin{equation}
H_{c}^{\prime}=H_{c}^{\prime\prime}\left(  2\right)  +H_{c}^{\prime\prime
}\left(  1\right)  +H_{c}^{\prime\prime}\left(  0\right)  +A^{\mu}\chi_{\mu
}^{00}+S_{,m}^{m}\label{eqnAM54a}%
\end{equation}
and equation (\ref{eqnAM45}) can be rewritten as%

\begin{equation}
\left\{  \chi_{0}^{00},H_{c}^{\prime}\right\}  =H_{c}^{\prime\prime}\left(
2\right)  +H_{c}^{\prime\prime}\left(  1\right)  +H_{c}^{\prime\prime}\left(
0\right)  +\left\{  \chi_{0}^{00},A^{\mu}\chi_{\mu}^{00}\right\}
,\label{eqnAM54}%
\end{equation}
where%

\begin{equation}
H_{c}^{\prime\prime}\left(  2\right)  =\frac{1}{h^{00}}\left[  e^{kp}%
e^{mq}\left(  \pi_{mp}\pi_{kq}-\pi_{kp}\pi_{mq}\right)  \right]  ,
\label{eqnAM55}%
\end{equation}

\begin{equation}
H_{c}^{\prime\prime}\left(  1\right)  =\frac{h^{0k}}{h^{00}}\left[  2\left(
e^{mq}\pi_{mq}\right)  _{,k}-2\left(  e^{mq}\pi_{kq}\right)  _{,m}-\frac
{1}{h^{00}}\left(  e^{np}h^{00}\right)  _{,k}\pi_{np}\right]  ,
\label{eqnAM56}%
\end{equation}

\begin{equation}
H_{c}^{\prime\prime}\left(  0\right)  =\frac{1}{h^{00}}\left[  -\left(
e^{km}h^{00}\right)  _{,mk}+\frac{1}{2}\left(  e^{xb}h^{00}\right)
_{,y}h_{bz}\frac{1}{h^{00}}\left(  e^{yz}h^{00}\right)  _{,x}\right.
\label{eqnAM57}%
\end{equation}

\[
\left.  -\frac{1}{4}h_{ay}h_{bz}e^{nx}\left(  e^{ab}h^{00}\right)  _{,n}%
\frac{1}{h^{00}}\left(  e^{yz}h^{00}\right)  _{,x}+\frac{1}{4}\frac{1}%
{D-2}e^{nx}h_{ab}\left(  e^{ab}h^{00}\right)  _{,n}h_{yz}\frac{1}{h^{00}%
}\left(  e^{yz}h^{00}\right)  _{,x}\right]  ,\text{ }%
\]

\begin{equation}
A^{0}=-\frac{1}{D-1}\pi_{00}-\frac{D-2}{D-1}\frac{D_{k}^{0k}}{h^{00}}%
+2\frac{h^{k0}}{h^{00}}\frac{h_{,k}^{00}}{h^{00}}+4\frac{h^{k0}}{h^{00}}%
\frac{h^{m0}}{h^{00}}\pi_{km} \label{eqnAM58}%
\end{equation}

\[
-\frac{1}{D-1}\frac{1}{h^{00}}\left(  h^{0y}\delta_{x}^{z}+h^{z0}\delta
_{x}^{y}\right)  \left[  \frac{1}{2}D_{y}^{xb}h_{bz}-\frac{1}{4}h_{ay}%
h_{bz}e^{nx}D_{n}^{ab}+\frac{1}{4}\frac{1}{D-2}e^{nx}h_{ab}D_{n}^{ab}%
h_{yz}\right]
\]

\[
-\frac{1}{D-1}\frac{1}{h^{00}}\left(  h^{0a}\delta_{n}^{b}+h^{b0}\delta
_{n}^{a}\right)  \left[  \frac{1}{2}h_{bz}\overset{\approx}{D}_{a}^{nz}%
-\frac{1}{4}h_{ay}h_{bz}e^{nx}\overset{\approx}{D}_{x}^{yz}+\frac{1}{4}%
\frac{1}{D-2}e^{nx}h_{ab}h_{yz}\overset{\approx}{D}_{x}^{yz}\right]  ,
\]

\begin{equation}
A^{k}=-2\frac{h^{k0}}{h^{00}}\pi_{00}-2\frac{h^{km}}{h^{00}}\pi_{0m}%
+\frac{h_{,p}^{pk}}{h^{00}}+\frac{h^{km}}{h^{00}}\frac{h_{,m}^{00}}{h^{00}%
}+2\frac{h^{km}}{h^{00}}\frac{h^{p0}}{h^{00}}\pi_{mp} \label{eqnAM59}%
\end{equation}

\[
-\frac{h^{yz}}{h^{00}}\left[  \frac{1}{2}D_{y}^{kb}h_{bz}-\frac{1}{4}%
h_{ay}h_{bz}e^{nk}D_{n}^{ab}+\frac{1}{4}\frac{1}{D-2}e^{nk}h_{ab}D_{n}%
^{ab}h_{yz}\right]
\]

\[
-\frac{h^{ab}}{h^{00}}\left[  \frac{1}{2}h_{bz}\overset{\approx}{D}_{a}%
^{kz}-\frac{1}{4}h_{ay}h_{bz}e^{kx}\overset{\approx}{D}_{x}^{yz}+\frac{1}%
{4}\frac{1}{D-2}e^{kx}h_{ab}h_{yz}\overset{\approx}{D}_{x}^{yz}\right]
\]
and%

\[
\overset{\approx}{D}_{m}^{kq}=h^{kq}\frac{1}{h^{00}}h_{,m}^{00}+2h^{kq}%
\frac{1}{h^{00}}h^{p0}\pi_{mp}+h_{,m}^{kq}-\frac{h^{q0}}{h^{00}}D_{m}%
^{0k}-\frac{h^{k0}}{h^{00}}D_{m}^{0q}.
\]
$A^{0}$ and $A^{k}$ are the functions which depend on the fields and
derivatives and their explicit form is not needed for further calculations in
this article; they are given only for completeness. Our main interest is in
the contributions, which are not proportional to secondary constraints,
$H_{c}^{\prime\prime}\left(  i\right)  $. The explicit form of $S^{m}$ is
found by comparison of the parts not proportional to the secondary constraints
in (\ref{eqnAM45}) with $H_{c}^{\prime}$%

\begin{equation}
S^{m}=-2\frac{h^{0m}}{h^{00}}e^{kq}\pi_{kq}+2\frac{h^{0k}}{h^{00}}e^{mq}%
\pi_{kq}-h_{,p}^{0p}\frac{h^{0m}}{h^{00}}+h_{,p}^{0m}\frac{h^{0p}}{h^{00}%
}+\frac{1}{h^{00}}\left(  e^{km}h^{00}\right)  _{,k} \label{eqnAM60}%
\end{equation}

And, as in the previous steps of the Hamiltonian reduction, writing $h^{km}$
in terms of $e^{km}$ makes the expressions more transparent. Note that
(\ref{eqnAM54a}), contrary to (\ref{eqnAM45}), is the Hamiltonian density and
a surface term can be neglected in subsequent calculations.

Using (\ref{eqnAM54a}) and (\ref{eqnAM55})-(\ref{eqnAM57}) the calculation of
the time development of $\chi_{k}^{00}$ is straightforward%

\begin{equation}
\dot{\chi}_{k}^{00}=\left\{  \chi_{k}^{00},H_{c}\right\}  =-\Gamma_{00}%
^{0}\chi_{k}^{00}+\left\{  \chi_{k}^{00},H_{c}^{\prime}\right\}
\label{eqnAM65}%
\end{equation}
where%

\begin{equation}
\left\{  \chi_{k}^{00},H_{c}^{\prime}\right\}  =2\left(  e^{mq}\pi
_{mq}\right)  _{,k}-2\left(  e^{mq}\pi_{kq}\right)  _{,m}-\frac{1}{h^{00}%
}\left(  e^{np}h^{00}\right)  _{,k}\pi_{np}+\left\{  \chi_{p}^{00},A^{\mu}%
\chi_{\mu}^{00}\right\}  .\label{eqnAM66}%
\end{equation}
Note that $\left\{  \chi_{k}^{00},H_{c}^{\prime}\left(  2\right)  \right\}
=\left\{  \chi_{k}^{00},H_{c}^{\prime}\left(  0\right)  \right\}  =0$ which is
based on simple PBs of $\chi_{k}^{00}$ with combinations of fields presented
in (\ref{eqnAM43})-(\ref{eqnAM43a}).

The last term in (\ref{eqnAM66}) gives contributions proportional to
$\chi_{\mu}^{00}$, whereas the first three terms in (\ref{eqnAM66}) cannot be
expressed as a linear combination of $\chi_{\mu}^{00}$; moreover they coincide
with the expression in square brackets of $H_{c}^{\prime\prime}\left(
1\right)  $ in (\ref{eqnAM56}). This part of (\ref{eqnAM66}), which is not
proportional to the secondary constraints, can be chosen to be called a
tertiary constraint%

\begin{equation}
\tau_{k}^{00}\equiv2\left(  e^{mq}\pi_{mq}\right)  _{,k}-2\left(  e^{mq}%
\pi_{kq}\right)  _{,m}-\frac{1}{h^{00}}\left(  e^{np}h^{00}\right)  _{,k}%
\pi_{np}. \label{eqnAM70}%
\end{equation}

Taking into account (\ref{eqnAM70}), the bracket $\left\{  \chi_{0}^{00}%
,H_{c}^{\prime}\right\}  $ can be rewritten as%

\[
\left\{  \chi_{0}^{00},H_{c}^{\prime}\right\}  =H_{c}^{\prime\prime}\left(
2\right)  +\frac{h^{0k}}{h^{00}}\tau_{k}^{00}+H_{c}^{\prime\prime}\left(
0\right)  +\left\{  \chi_{0}^{00},A^{\mu}\chi_{\mu}^{00}\right\}  ,
\]
so the terms which are not proportional to already known constraints should be
called a new, tertiary constraint. Using the exact expressions of
$H_{c}^{\prime\prime}\left(  2\right)  $ and $H_{c}^{\prime\prime}\left(
0\right)  $ from (\ref{eqnAM55}) and (\ref{eqnAM57}) we can name the following
combination as a tertiary constraint%

\begin{equation}
\tau_{0}^{00}\equiv e^{kp}e^{mq}\left(  \pi_{mp}\pi_{kq}-\pi_{kp}\pi
_{mq}\right)  -\left(  e^{km}h^{00}\right)  _{,mk}+\frac{1}{2}\left(
e^{xb}h^{00}\right)  _{,y}h_{bz}\frac{1}{h^{00}}\left(  e^{yz}h^{00}\right)
_{,x} \label{eqnAM72}%
\end{equation}

\[
-\frac{1}{4}h_{ay}h_{bz}e^{nx}\left(  e^{ab}h^{00}\right)  _{,n}\frac
{1}{h^{00}}\left(  e^{yz}h^{00}\right)  _{,x}+\frac{1}{4}\frac{1}{D-2}%
e^{nx}h_{ab}\left(  e^{ab}h^{00}\right)  _{,n}h_{yz}\frac{1}{h^{00}}\left(
e^{yz}h^{00}\right)  _{,x}.
\]

The canonical Hamiltonian written in terms of the constraints $\chi_{\mu}%
^{00}$ and $\tau_{\mu}^{00}$ is%

\begin{equation}
H_{c}=-\Gamma_{00}^{\mu}\chi_{\mu}^{00}+\frac{1}{h^{00}}\tau_{0}^{00}%
+\frac{h^{0k}}{h^{00}}\tau_{k}^{00}+A^{\mu}\chi_{\mu}^{00}. \label{eqnAM74}%
\end{equation}

Note that the choice of tertiary constraints is not unique. For example, if we
start from $\dot{\chi}_{0}^{00}$ we would name the whole combination
$H_{c}^{\prime\prime}\left(  2\right)  +H_{c}^{\prime\prime}\left(  1\right)
+H_{c}^{\prime\prime}\left(  0\right)  $ in (\ref{eqnAM54}) a tertiary
constraint $\tau_{0}^{00}$, because in this case $\tau_{k}^{00}$ has not yet
been found. Such arbitrariness in the choice of constraints looks ambiguous.
Firstly, it does not contradict the Dirac procedure as any linear combination
of constraints is also a constraint. Secondly, and we will show this below, if
the correct method of the restoration of gauge invariance is used then the
final result does not depend on a choice of tertiary constraints.

We have chosen such a form of tertiary constraints, (\ref{eqnAM70}) and
(\ref{eqnAM72}), because calculations of PBs among secondary and these
tertiary constraints is almost manifest due to the simple properties of the
PBs of the secondary constraints with their combinations presented here (e.g.
(\ref{eqnAM37})-(\ref{eqnAM42}) and (\ref{eqnAM43})-(\ref{eqnAM43a})). It is
easy to show that%

\begin{equation}
\left\{  \chi_{\mu}^{00},\tau_{\nu}^{00}\right\}  =0. \label{eqnAM76}%
\end{equation}

The brackets $\left\{  \tau_{\nu}^{00},A^{\mu}\chi_{\mu}^{00}\right\}  $ are
proportional to the secondary constraints because of (\ref{eqnAM76}). In
addition, because of the relatively simple form of $\tau_{\mu}^{00}$
((\ref{eqnAM70}) and (\ref{eqnAM72})), the calculation of PBs among them is
not inordinately tedious. We must find these PBs to prove the closure of the
Dirac procedure; and if it closes for one choice of tertiary constraints, then
it closes for any combination of them. For these constraints, $\tau_{\mu}%
^{00},$ the only possibility to have a closure is to demonstrate that%

\begin{equation}
\left\{  \tau_{\mu}^{00},\tau_{\nu}^{00}\right\}  =0\text{ \ \ \ or }\sim
\tau_{\sigma}^{00},\label{eqnAM78}%
\end{equation}
as PBs of combinations (\ref{eqnAM70}) and (\ref{eqnAM72}) cannot form
secondary constraints just because of their field content. In the next
Section, we consider the calculations of PBs (\ref{eqnAM78}) and the closure
of the Dirac procedure.

\section{Calculation of $\left\{  \tau_{\mu}^{00},\tau_{\nu}^{00}\right\}  $,
closure of the Dirac procedure and algebra of tertiary constraints}

Even considering relatively simple combinations $\tau_{\mu}^{00}$
((\ref{eqnAM70}) and (\ref{eqnAM72})) as a choice of tertiary constraints, the
calculation of the PBs among them is a laborious procedure. This fact and a
variety of other choices of tertiary constraints were the reason why the first
attempts to prove a closure of the Dirac procedure for first order formulation
of GR were not finished \cite{KK, KKM} where the variables (\ref{eqnAM7a})
were used; only later the proof was completed in \cite{G/R0, G/R} and the
expectations outlined in \cite{KK, KKM} are thus realized. The presence of
derivatives of fields imposes some additional complications and the best way
of dealing with such calculations is to use test functions \cite{Kurt} that
were demonstrated in some detail for constraints of Yang-Mills theory
\cite{KMAOP}. More details of calculation using test functions were given and
applied to the first order formulation of GR in \cite{G/R, Novel}. \ The PBs
among constraints can be written in the form with the explicit presence of
test functions (e.g. see work of Faddeev \cite{Faddeev}, Section 3), which
might be useful for further calculations (e.g. restoration of gauge
invariance, analysis of different choices of tertiary constraints, etc.)
compared with the standard form which contains derivatives of delta functions.

Introducing test functions, $f\left(  \overrightarrow{x}\right)  $ and
$g\left(  \overrightarrow{y}\right)  $, that have a zero PB with phase space
variables, we calculate%

\[
\int\int d\overrightarrow{x}d\overrightarrow{y}\left\{  f\left(
\overrightarrow{x}\right)  \tau_{\mu}^{00}\left(  \overrightarrow{x}\right)
,g\left(  \overrightarrow{y}\right)  \tau_{\nu}^{00}\left(  \overrightarrow
{y}\right)  \right\}  =
\]

\begin{equation}
\int\int d\overrightarrow{x}d\overrightarrow{y}\left(  ...\right)
_{\overrightarrow{x},\overrightarrow{y}}\delta\left(  \overrightarrow
{x}-\overrightarrow{y}\right)  =\int d\overrightarrow{x}\left(  ...\right)
_{\overrightarrow{x}}\label{eqnAM80}%
\end{equation}
where, to shorten the notation, we omit the integrals, i.e.%

\begin{equation}
\left\{  f\left(  \overrightarrow{x}\right)  \tau_{\mu}^{00}\left(
\overrightarrow{x}\right)  ,g\left(  \overrightarrow{y}\right)  \tau_{\nu
}^{00}\left(  \overrightarrow{y}\right)  \right\}  =\left(  ...\right)
_{\overrightarrow{x}}~. \label{eqnAM85}%
\end{equation}

After long but straightforward calculation, (\ref{eqnAM80}) can be presented
in the following form for $\mu,\nu=0,i$:%

\begin{equation}
\left\{  f\left(  \overrightarrow{x}\right)  \tau_{i}^{00}\left(
\overrightarrow{x}\right)  ,g\left(  \overrightarrow{y}\right)  \tau_{j}%
^{00}\left(  \overrightarrow{y}\right)  \right\}  =f_{,j}g\tau_{i}^{00}%
-f\tau_{j}^{00}g_{,i}~,\label{eqnAM81}%
\end{equation}

\begin{equation}
\left\{  f\left(  \overrightarrow{x}\right)  \tau_{0}^{00}\left(
\overrightarrow{x}\right)  ,g\left(  \overrightarrow{y}\right)  \tau_{0}%
^{00}\left(  \overrightarrow{y}\right)  \right\}  =fh^{00}e^{kp}g_{,k}\tau
_{p}^{00}-f_{,k}h^{00}e^{kp}g\tau_{p}^{00}~,\label{eqnAM82}%
\end{equation}

\begin{equation}
\left\{  f\left(  \overrightarrow{x}\right)  \tau_{0}^{00}\left(
\overrightarrow{x}\right)  ,g\left(  \overrightarrow{y}\right)  \tau_{i}%
^{00}\left(  \overrightarrow{y}\right)  \right\}  =f_{,i}g\tau_{0}%
^{00}-fg_{,i}\tau_{0}^{00}~,\label{eqnAM83}%
\end{equation}

\begin{equation}
\left\{  f\left(  \overrightarrow{x}\right)  \tau_{i}^{00}\left(
\overrightarrow{x}\right)  ,g\left(  \overrightarrow{y}\right)  \tau_{0}%
^{00}\left(  \overrightarrow{y}\right)  \right\}  =f_{,i}g\tau_{0}%
^{00}-fg_{,i}\tau_{0}^{00}~.\label{eqnAM83a}%
\end{equation}

From these expressions we can also obtain the standard form of PBs with delta
functions if we rearrange (\ref{eqnAM80}) in the form without derivatives of
the test functions (note that to do this for (\ref{eqnAM81})-(\ref{eqnAM83a})
derivatives of delta functions are unavoidable)%

\begin{equation}
\int\int d\overrightarrow{x}d\overrightarrow{y}\left\{  f\left(
\overrightarrow{x}\right)  \tau_{\mu}^{00}\left(  \overrightarrow{x}\right)
,g\left(  \overrightarrow{y}\right)  \tau_{\nu}^{00}\left(  \overrightarrow
{y}\right)  \right\}  =\int\int d\overrightarrow{x}d\overrightarrow{y}f\left(
\overrightarrow{x}\right)  \left(  ...\right)  _{\overrightarrow
{x},\overrightarrow{y}}g\left(  \overrightarrow{y}\right)  .\label{eqnAM84}%
\end{equation}
For (\ref{eqnAM81})-(\ref{eqnAM83}) we obtain:%

\begin{equation}
\left\{  \tau_{i}^{00}\left(  \overrightarrow{x}\right)  ,\tau_{j}^{00}\left(
\overrightarrow{y}\right)  \right\}  =\tau_{j}^{00}\left(  x\right)
\partial_{i}^{y}\delta\left(  x-y\right)  -\tau_{i}^{00}\left(  y\right)
\partial_{j}^{x}\delta\left(  x-y\right)  ,\label{eqnAM86}%
\end{equation}

\begin{equation}
\left\{  \tau_{0}^{00}\left(  \overrightarrow{x}\right)  ,\tau_{0}^{00}\left(
\overrightarrow{y}\right)  \right\}  =-h^{00}\left(  x\right)  e^{kp}\left(
x\right)  \tau_{p}^{00}\left(  x\right)  \partial_{k}^{y}\delta\left(
x-y\right)  +h^{00}\left(  y\right)  e^{kp}\left(  y\right)  \tau_{p}%
^{00}\left(  y\right)  \partial_{k}^{x}\delta\left(  x-y\right)
,\label{eqnAM87}%
\end{equation}

\begin{equation}
\left\{  \tau_{0}^{00}\left(  \overrightarrow{x}\right)  ,\tau_{i}^{00}\left(
\overrightarrow{y}\right)  \right\}  =\tau_{0}^{00}\left(  x\right)
\partial_{i}^{y}\delta\left(  x-y\right)  -\tau_{0}^{00}\left(  y\right)
\partial_{i}^{x}\delta\left(  x-y\right)  ,\label{eqnAM88}%
\end{equation}

\begin{equation}
\left\{  \tau_{i}^{00}\left(  \overrightarrow{x}\right)  ,\tau_{0}^{00}\left(
\overrightarrow{y}\right)  \right\}  =\tau_{0}^{00}\left(  x\right)
\partial_{i}^{y}\delta\left(  x-y\right)  -\tau_{0}^{00}\left(  y\right)
\partial_{i}^{x}\delta\left(  x-y\right)  \label{eqnAM88a}%
\end{equation}
(here we use the notation: $\partial_{k}^{x}=\frac{\partial}{\partial x^{k}}$
or $\partial_{i}^{y}=\frac{\partial}{\partial y^{i}}$). As a consistency check
we can integrate the above expressions with $\int d\overrightarrow{y}f\left(
\overrightarrow{x}\right)  g\left(  \overrightarrow{y}\right)  \left(
...\right)  _{\overrightarrow{x},\overrightarrow{y}}$ which leads us back to
the previous form (\ref{eqnAM81})-(\ref{eqnAM83}). So, these, (\ref{eqnAM81}%
)-(\ref{eqnAM83a}) and (\ref{eqnAM86})-(\ref{eqnAM88a}) are two different but
equivalent forms of the constraint algebra.

This algebra of constraints (in one form or another) is equivalent with the
algebra found in \cite{G/R}, which becomes clear after canonical
transformations are performed (see next Section) and it is also the same as
the algebra of constraints (but secondary) given by Faddeev \cite{Faddeev} and
Teitelboim \cite{Teitelboim} (despite different expressions for constraints themselves).

The conventional form of the algebra of secondary constraints for the
conventional Hamiltonian formulation of the second order EH action, known also
as \textquotedblleft Dirac's algebra\textquotedblright\ or \textquotedblleft
hypersurface deformation algebra\textquotedblright,\ is:%

\begin{equation}
\left\{  \mathcal{H}_{L}\left(  x\right)  ,\mathcal{H}_{L}\left(  x^{\prime
}\right)  \right\}  =e^{rs}\left(  x\right)  \mathcal{H}_{s}\left(  x\right)
\delta_{,r\left(  x\right)  }\left(  x-x^{\prime}\right)  -e^{rs}\left(
x^{^{\prime}}\right)  \mathcal{H}_{s}\left(  x^{\prime}\right)  \delta
_{,r\left(  x^{\prime}\right)  }\left(  x-x^{\prime}\right)  ,
\label{eqnAM88b}%
\end{equation}

\begin{equation}
\left\{  \mathcal{H}_{s}\left(  x\right)  ,\mathcal{H}_{L}\left(  x^{\prime
}\right)  \right\}  =\mathcal{H}_{L}\left(  x\right)  \delta_{,s\left(
x\right)  }\left(  x-x^{\prime}\right)  ,\text{ } \label{eqnAM88c}%
\end{equation}

\begin{equation}
\left\{  \mathcal{H}_{r}\left(  x\right)  ,\mathcal{H}_{s}\left(  x^{\prime
}\right)  \right\}  =\mathcal{H}_{s}\left(  x\right)  \delta_{,r\left(
x\right)  }\left(  x-x^{\prime}\right)  -\mathcal{H}_{r}\left(  x^{\prime
}\right)  \delta_{,s\left(  x^{\prime}\right)  }\left(  x-x^{\prime}\right)
;\label{eqnAM88d}%
\end{equation}
where $\mathcal{H}_{L}$ and $\mathcal{H}_{s}$ are the \textquotedblleft
Hamiltonian\textquotedblright\ and \textquotedblleft
diffeomorphism\textquotedblright\ constraints, respectively. The algebra
(\ref{eqnAM88b})-(\ref{eqnAM88d}) can be found in slightly different forms,
for example, in \cite{Diracbook, DeWitt, Kuchar, Teitelboim-preprint,
Teitelboim-AOP, Relativity}.

Comparing (\ref{eqnAM88b})-(\ref{eqnAM88d}) with (\ref{eqnAM86}%
)-(\ref{eqnAM88a}) one can notice a difference. It is not related to the
variety of notations used in the literature, but to the obviously less
symmetric form of (\ref{eqnAM88c}) in the conventional algebra, contrary to
(\ref{eqnAM88a}) and (\ref{eqnAM83a}). This discrepancy must be clarified. Let
us trace out the origin of the algebra (\ref{eqnAM88b})-(\ref{eqnAM88d})
starting from its first name \textquotedblleft Dirac's
algebra\textquotedblright. It appeared for the first time in the Dirac book
\cite{Diracbook}, where he referred to his paper \cite{Dirac-CJM}. But\ in
\cite{Dirac-CJM} he derived this algebra for the motion of space-like
surfaces, as PBs among tangential and normal to a surface variables, not as
PBs among the secondary constrains of GR. Kuchar in \cite{Kuchar}, by
geometrical reasoning, showed how $\mathcal{H}_{L}$ and $\mathcal{H}_{s}$
(\textquotedblleft super-Hamiltonian\textquotedblright\ and \textquotedblleft
super-momenta\textquotedblright, in his terminology) \textquotedblleft
represent the set of deformations of space-like
hypersurfaces\textquotedblright. Probably, after that the name
\textquotedblleft hypersurface deformation algebra\textquotedblright%
\ appeared. In \cite{Teitelboim-AOP} Teitelboim reconstructed this algebra by
\textquotedblleft a simple geometrical argument based exclusively on the path
independence of the dynamical evolution\textquotedblright, i.e. on
\textquotedblleft the `motion'\ of a three-dimensional cut in a
four-dimensional manifold of hyperbolic signature\textquotedblright. In his
later work \cite{Teitelboim} this algebra was altered by another one where the
PB (\ref{eqnAM88c}) was replaced by (\ref{eqnAM87}). This transition was not
explained and left unnoticed, which is strange, especially because when
presenting the new algebra he referred to his old paper \cite{Teitelboim-AOP}
where the algebra is different. This algebra of PBs among secondary
constraints was derived by Faddeev in \cite{Faddeev} where he considered the
first order formulation of GR; in his paper, the algebra among constraints is
written in the form of (\ref{eqnAM81})-(\ref{eqnAM83a}). The same algebra was
presented by Ghalati and McKeon \cite{G/R}. Our calculation also results in
(\ref{eqnAM81})-(\ref{eqnAM83a}), or equivalently in (\ref{eqnAM86}%
)-(\ref{eqnAM88a}).

In \cite{Katanaev} analyzing the Dirac derivation of \cite{Dirac-CJM}, the
author made a conclusion that \textquotedblleft Dirac's derivation of the
constraint algebra cannot be considered satisfactory\textquotedblright. This
statement must be clarified. Dirac did not derive the algebra of constraints
as they were not even known and appeared only a few years after \cite{Dirac};
and in this article there is no statement that constraints satisfy
\textquotedblleft Dirac algebra\textquotedblright\ of \cite{Dirac-CJM}. This
conclusion was made by other authors and without calculation.

To answer the question why the discrepancy, the difference between
(\ref{eqnAM88a}) and (\ref{eqnAM88c}), appears, we are planning to revisit our
analysis of the Dirac formulation of the second order of EH action given in
\cite{Myths}, where a different choice of secondary constraints was used. The
results will be reported elsewhere.

If we are interested only in a demonstration of closure of the Dirac procedure
and in restoration of four-diffeomorphism, we need to consider the time
development of all first class constraints started from primary, i.e. to
calculate PB with Hamiltonian. In this case, the algebra of particularly
chosen tertiary constraints is not important. In addition, one can completely
avoid the non-locality of these PBs (i.e. derivatives of test functions in
(\ref{eqnAM81})-(\ref{eqnAM83a}) or derivatives of delta functions in
(\ref{eqnAM86})-(\ref{eqnAM88a})) as the PB of a constraint with the
Hamiltonian is always local (one integration has to be performed as
Hamiltonian in field theories is integral of Hamiltonian density). For
example, PBs between tertiary constraints and the Hamiltonian is defined as%

\[
\overset{\cdot}{\tau}_{\mu}^{00}=\left\{  \tau_{\mu}^{00},H_{T}\right\}
=\left\{  f\left(  \overrightarrow{x}\right)  \tau_{\mu}^{00}\left(
\overrightarrow{x}\right)  ,\int d\overrightarrow{y}g\left(  \overrightarrow
{y}\right)  H_{T}\left(  \overrightarrow{y}\right)  \right\}  .
\]

As the Hamiltonian for the EH action is a linear combination of the
constraints, what we actually need to calculate are the following PBs%

\begin{equation}
\left\{  f\left(  \overrightarrow{x}\right)  \tau_{\mu}^{00}\left(
\overrightarrow{x}\right)  ,\int d\overrightarrow{y}g^{\nu}\left(
\overrightarrow{y}\right)  \tau_{\nu}^{00}\left(  \overrightarrow{y}\right)
\right\}  \label{eqnAM90}%
\end{equation}
where, according to our choice of constraints (\ref{eqnAM74}), we have to put
$g^{0}\left(  \overrightarrow{y}\right)  =\frac{1}{h^{00}\left(
\overrightarrow{y}\right)  }$ and $g^{k}\left(  \overrightarrow{y}\right)
=\frac{h^{0k}\left(  \overrightarrow{y}\right)  }{h^{00}\left(
\overrightarrow{y}\right)  }.$ Using (\ref{eqnAM86})-(\ref{eqnAM88a}) we can
easily find:%

\begin{equation}
\left\{  \tau_{i}^{00}\left(  \overrightarrow{x}\right)  ,\int
d\overrightarrow{y}\tau_{j}^{00}\left(  \overrightarrow{y}\right)
\frac{h^{0j}\left(  y\right)  }{h^{00}\left(  y\right)  }\right\}  =-\left(
\tau_{i}^{00}\frac{h^{0j}}{h^{00}}\right)  _{,j}-\tau_{j}^{00}\left(
\frac{h^{0j}}{h^{00}}\right)  _{,i}~,\label{eqnAM91}%
\end{equation}

\begin{equation}
\left\{  \tau_{0}^{00}\left(  \overrightarrow{x}\right)  ,\int
d\overrightarrow{y}\tau_{0}^{00}\left(  \overrightarrow{y}\right)  \frac
{1}{h^{00}\left(  y\right)  }\right\}  =\left(  e^{kp}\tau_{p}^{00}\right)
_{,k}+h^{00}e^{kp}\tau_{p}^{00}\left(  \frac{1}{h^{00}}\right)  _{,k}%
~,\label{eqnAM92}%
\end{equation}

\begin{equation}
\left\{  \tau_{0}^{00}\left(  \overrightarrow{x}\right)  ,\int
d\overrightarrow{y}\tau_{i}^{00}\left(  \overrightarrow{y}\right)
\frac{h^{0i}\left(  y\right)  }{h^{00}\left(  y\right)  }\right\}  =-\tau
_{0}^{00}\left(  \frac{h^{0i}}{h^{00}}\right)  _{,i}-\left(  \tau_{0}%
^{00}\frac{h^{0i}}{h^{00}}\right)  _{,i}~,\label{eqnAM93}%
\end{equation}

\begin{equation}
\left\{  \tau_{i}^{00}\left(  \overrightarrow{x}\right)  ,\int
d\overrightarrow{y}\tau_{0}^{00}\left(  \overrightarrow{y}\right)  \frac
{1}{h^{00}\left(  y\right)  }\right\}  =-\tau_{0}^{00}\left(  \frac{1}{h^{00}%
}\right)  _{,i}-\left(  \tau_{0}^{00}\frac{1}{h^{00}}\right)  _{,i}%
~.\label{eqnAM94}%
\end{equation}

Equally well we can use the PBs, (\ref{eqnAM81})-(\ref{eqnAM83a}), where by
integration by parts we have to move the derivative from a test function $f$,
and put $f=1$ at the end of the calculations (see, e.g. \cite{KMAOP}). This
will result in the same expressions (\ref{eqnAM91})-(\ref{eqnAM94}). We also
want to emphasize that these calculations can be performed directly, without
any reference to a particular form of the algebra of constraints, which
depends on our choice of constraints (see next Section).

The above PBs, (\ref{eqnAM91})-(\ref{eqnAM94}), complete the proof of closure
of the Dirac procedure that the PBs of the tertiary constraints with the
Hamiltonian are proportional to already known constraints%

\begin{equation}
\left\{  \tau_{\mu}^{00},H_{T}\right\}  =\left\{  \tau_{\mu}^{00},\frac
{1}{h^{00}}\tau_{0}^{00}\right\}  +\left\{  \tau_{\mu}^{00},\frac{h^{0i}%
}{h^{00}}\tau_{i}^{00}\right\}  +\left\{  \tau_{\mu}^{00},A^{\nu}\right\}
\chi_{\nu}~. \label{eqmAM98}%
\end{equation}

In next two Sections we will discuss the role of different choices of
constraints and canonical transformations, which is important in general, but
also will be needed for the discussion of methods of restoration \ of gauge
symmetry (Sections 6 and 7) that, in accordance with the Dirac conjecture
\cite{Diracbook}, is generated by the full set of first class constraints.

Both equivalent forms of the PB algebra among tertiary constraints, which we
chose out of many possible combinations, (\ref{eqnAM81})-(\ref{eqnAM83a}) and
(\ref{eqnAM86})-(\ref{eqnAM88a}), might be useful in calculations that we need
to perform, especially for the general analysis of a role of different linear
combinations of non-primary first class constraints. A particular choice of
tertiary constraints can lead to considerable simplification in some parts of
the analysis; but the algebra with derivatives of delta functions or test
functions can be completely avoided \cite{KKRV, Myths}, and are not needed for
proof of closure nor for the restoration of gauge invariance using the
Castellani algorithm \cite{Castellani}.

\section{Linear combinations of tertiary constraints, canonical
transformations and their interplay}

Considering the time development of secondary constraints in Section 2 we
demonstrated that tertiary constraints can be defined in different ways and
all such combinations do not contradict the Dirac procedure. All the different
choices are linear combinations of each other, i.e. the Hamiltonian
formulation of affine-metric GR provides an example of the theory with the
non-artificial appearance of different linear combinations of constraints. We
would like to discuss this apparent ambiguity of the Hamiltonian procedure. We
restrict our discussion to different combinations of tertiary constraints that
we initially defined in (\ref{eqnAM70}), (\ref{eqnAM72}) and the corresponding
part of the canonical Hamiltonian%

\begin{equation}
H_{c}^{\prime\prime}=\frac{1}{h^{00}}\tau_{0}^{00}+\frac{h^{0k}}{h^{00}%
}\text{\ }\tau_{k}^{00}.\label{eqnAM99}%
\end{equation}

Of course, more choices exist if the secondary constraints are also used in
such redefinitions; but it will just make the calculations more involved. Some
conclusions can be made based on simple examples. Note that the role of first
class primary constraints is quite special and will be discussed in next Section.

One possible choice of tertiary constraints is%

\begin{equation}
\bar{\tau}_{0}^{00}=\frac{1}{h^{00}}\tau_{0}^{00},\text{ \ }\bar{\tau}%
_{k}^{00}=\tau_{k}^{00}.\label{eqnAM100}%
\end{equation}
In terms of these constraints, the part, $H_{c}^{\prime\prime}$, of the
canonical Hamiltonian (\ref{eqnAM99}) is%

\begin{equation}
\bar{H}_{c}^{\prime\prime}=\bar{\tau}_{0}^{00}+\frac{h^{0k}}{h^{00}}%
\text{\ }\bar{\tau}_{k}^{00}.\label{eqnAM101}%
\end{equation}

The PBs among constraints $\bar{\tau}_{0}$ and $\bar{\tau}_{k}$ can be easily
found by using their relations with the original choice (\ref{eqnAM100}) and
the corresponding algebra of constraints (\ref{eqnAM81})-(\ref{eqnAM83a}), e.g.%

\begin{equation}
\left\{  f\left(  \overrightarrow{x}\right)  \bar{\tau}_{0}\left(
\overrightarrow{x}\right)  ,g\left(  \overrightarrow{y}\right)  \bar{\tau}%
_{i}^{00}\left(  \overrightarrow{y}\right)  \right\}  =-f\frac{1}{h^{00}%
}h_{,i}^{00}g\bar{\tau}_{0}^{00}+f_{,i}g\bar{\tau}_{0}^{00}-f\bar{\tau}%
_{0}^{00}g_{,i}~.\label{eqnAM102}%
\end{equation}

The whole algebra of $\bar{\tau}_{0}^{00}$ and $\bar{\tau}_{k}^{00}$ can be
calculated and the closure of the Dirac procedure can be demonstrated using
these combinations.

Another choice is:%

\begin{equation}
\tilde{\tau}_{0}^{00}=\frac{1}{h^{00}}\tau_{0}^{00}+\frac{h^{0k}}{h^{00}}%
\tau_{k}^{00},\text{ \ }\tilde{\tau}_{k}^{00}=\tau_{k}^{00}\label{eqnAM110}%
\end{equation}
that leads to a very simple expression for the corresponding part of the
canonical Hamiltonian%

\begin{equation}
\tilde{H}_{c}^{\prime\prime}=\tilde{\tau}_{0}^{00},\label{eqnAM111}%
\end{equation}
and again the PB algebra of constraints $\tilde{\tau}_{0}^{00}$ and
$\tilde{\tau}_{k}^{00}$ can be found using (\ref{eqnAM110}) and (\ref{eqnAM81}%
)-(\ref{eqnAM83a}), e.g.%

\[
\left\{  f\left(  \overrightarrow{x}\right)  \tilde{\tau}_{0}^{00}\left(
\overrightarrow{x}\right)  ,g\left(  \overrightarrow{y}\right)  \tilde{\tau
}_{i}^{00}\left(  \overrightarrow{y}\right)  \right\}  =
\]

\begin{equation}
f_{,i}g\tilde{\tau}_{0}^{00}-f\tilde{\tau}_{0}^{00}g_{,i}+f\frac{h^{0k}%
}{h^{00}}\tilde{\tau}_{k}^{00}g_{,i}-f\frac{h^{0k}}{h^{00}}\tilde{\tau}%
_{i}^{00}g_{,k}+fg\frac{h_{,i}^{0k}}{h^{00}}\tilde{\tau}_{k}^{00}%
-fg\frac{h_{,i}^{00}}{h^{00}}\tilde{\tau}_{0}^{00}~.\label{eqnAM112}%
\end{equation}

Is there any physical significance in a particular choice of constraints? Of
course, there are some possible technical (computational) advantages; but
considering different linear combinations of tertiary constraints should not
affect the physical results. The simplest argument is to convert our reduced
total Hamiltonian into the corresponding Lagrangian\ by inverse Legendre transformation%

\begin{equation}
L=\dot{\Gamma}_{00}^{\mu}\Pi_{\mu}^{00}+\dot{h}^{\alpha\beta}\pi_{\alpha\beta
}-H_{T}=\dot{h}^{\alpha\beta}\pi_{\alpha\beta}-\Gamma_{00}^{\alpha}%
\chi_{\alpha}^{00}-H^{\prime\prime}-A_{00}^{\mu}\chi_{\mu}^{00}%
,\label{eqnAM120}%
\end{equation}
where $H^{\prime\prime}$ is a functional, $H^{\prime\prime}\left(
h^{00},h^{0k},h^{km},\pi_{km}\right)  $. Whatever combination we consider,
(\ref{eqnAM99}), (\ref{eqnAM101}) or (\ref{eqnAM111}), we have the same
Lagrangian ($H^{\prime\prime}=\bar{H}^{\prime\prime}=\tilde{H}^{\prime\prime}%
$) and by calling some parts of these Lagrangians by `tertiary constraints',
which is nothing more than a short notation at the Lagrangian level, we cannot
affect the physics; and in particular, the gauge invariance should not change.
So, if a choice of tertiary constraints cannot influence gauge invariance,
then an algebra of PBs for this particular choice should not bear any physical
significance. For the Hamiltonian formulation of the second order metric GR,
we can make the same conclusion, but, of course, for possible choices of
secondary first class constraints and their algebra, contrary to a broadly
accepted view that a particular choice of constraints/algebra has some
physical significance that is even reflected in special names given to one
particular choice, \textquotedblleft Hamiltonian\textquotedblright\ and
\textquotedblleft spatial diffeomorphism\textquotedblright\ constraints. These
particular combinations become special only after a non-canonical change of
variables is made (see discussion in next Section). Related to the
combinations of constraints idea of the Master Constraint Programme is of
limited interest as any physical results cannot depend on a particular choice
of non-primary first class constraints; in other words, in the
\textquotedblleft society\textquotedblright\ of non-primary constraints there
is no place for a Master. One additional conclusion that is connected to the
freedom to choose combinations of tertiary constraints is more technical and
related to the methods of the restoration of gauge symmetry, based on a full
set of first class constraints (the Dirac conjecture). Gauge invariance should
be independent of a choice of non-primary first class constraints. So any
method, which is sensitive to redefinition of constraints (gives different
transformations) is not correct (see, Sections 6, 7).

For any Hamiltonian formulation, if change of phase-space variables is
performed then it must be canonical. Also such a change should not affect the
physical properties of a system, in particular, it should not change its gauge
invariance. In the second order formulation of GR we considered the connection
of two Hamiltonians, Dirac's and PSS, and demonstrated that they are related
by a canonical transformation and both lead to the same gauge transformation
which is four diffeomorphism invariance \cite{KKRV, Myths, FKK}. In the
passage from PSS to the Dirac formulation, we followed Dirac's idea: to
simplify primary constraints. He achieved this by modifying the original
Lagrangian. We worked in phase space and performed the canonical
transformation \cite{FKK}. In first order formulation of GR that we consider
here, the primary first class constraints already have the simplest possible
form: pure momenta conjugate to $\Gamma_{00}^{\mu}$, and any further
simplification is impossible. We should have different reasons to look for
canonical transformations. One such a reason is to simplify the expressions
for the secondary constraints, for example:%

\begin{equation}
\chi_{0}^{00}=-h_{,k}^{0k}+h^{00}\pi_{00}-h^{km}\pi_{km}=-h_{,k}^{0k}%
+h^{00}\tilde{\pi}_{00}~=\tilde{\chi}_{0}^{00}, \label{eqnAM130}%
\end{equation}

\begin{equation}
\chi_{k}^{00}=h_{,k}^{00}+2h^{00}\pi_{0k}+2h^{m0}\pi_{km}=h_{,k}^{00}%
+2h^{00}\tilde{\pi}_{0k}=\tilde{\chi}_{k}^{00},\label{eqnAM131}%
\end{equation}
i.e. to introduce new momenta%

\begin{equation}
\tilde{\pi}_{00}=\pi_{00}-\frac{h^{km}}{h^{00}}\pi_{km}~, \label{eqnAM132}%
\end{equation}

\begin{equation}
\tilde{\pi}_{0k}=\pi_{0k}+\frac{h^{m0}}{h^{00}}\pi_{km}~. \label{eqnAM133}%
\end{equation}

These two redefinitions, (\ref{eqnAM132}) and (\ref{eqnAM133}), are algebraic
and invertible; but this is not enough to preserve the canonicity of
phase-space variables, and it must be accompanied by a change of the remaining
phase-space variables, which are involved in redefinitions (\ref{eqnAM132}%
)-(\ref{eqnAM133}). Such a necessary and sufficient condition for the
transformation to be canonical is \cite{Lanczos}%

\begin{equation}
h_{,0}^{\alpha\beta}\pi_{\alpha\beta}=\tilde{h}_{,0}^{\alpha\beta}\tilde{\pi
}_{\alpha\beta}~. \label{eqnAM135}%
\end{equation}

Let us restrict our search for canonical transformations by assuming that%

\begin{equation}
h^{00}=\tilde{h}^{00}\text{, \ }h^{0k}=\tilde{h}^{0k}.\label{eqnAM136}%
\end{equation}
(This restriction has to be relaxed if it is not possible to satisfy
(\ref{eqnAM135}).) After substitution of new variables in terms of old into
(\ref{eqnAM135}), and some simple rearrangements we obtain%

\begin{equation}
\tilde{h}_{,0}^{km}\tilde{\pi}_{km}=\left(  h^{00}h^{km}-h^{0k}h^{0m}\right)
_{,0}\frac{1}{h^{00}}\pi_{km}~.\label{eqmAM137}%
\end{equation}
To fulfill this condition we have to define:%

\begin{equation}
\tilde{\pi}_{km}=\frac{1}{h^{00}}\pi_{km}\text{, \ }\tilde{h}^{km}%
=h^{00}h^{km}-h^{0k}h^{0m}=h^{00}e^{km}. \label{eqnAM138}%
\end{equation}

Of course, that the transformation which involves (\ref{eqnAM132}),
(\ref{eqnAM133}), (\ref{eqnAM136}) and (\ref{eqnAM138}) is canonical can be
checked by direct calculation of the PBs among all new variables; and this
should lead to:%

\begin{equation}
\left\{  \tilde{h}^{\alpha\beta},\tilde{\pi}_{\nu\mu}\right\}  =\left\{
\tilde{h}^{\alpha\beta}\left(  h^{\alpha\beta},\pi_{\nu\mu}\right)
,\tilde{\pi}_{\nu\mu}\left(  h^{\alpha\beta},\pi_{\nu\mu}\right)  \right\}
_{h^{\alpha\beta},\pi_{\nu\mu}}=\left\{  h^{\alpha\beta},\pi_{\nu\mu}\right\}
, \label{eqnAM139}%
\end{equation}

\[
\left\{  \tilde{h}^{\alpha\beta},\tilde{h}^{\nu\mu}\right\}  =\left\{
\tilde{h}^{\alpha\beta}\left(  h^{\alpha\beta},\pi_{\nu\mu}\right)  ,\tilde
{h}^{\nu\mu}\left(  h^{\alpha\beta},\pi_{\nu\mu}\right)  \right\}
_{h^{\alpha\beta},\pi_{\nu\mu}}=0,
\]

\[
\left\{  \tilde{\pi}_{\alpha\beta},\tilde{\pi}_{\nu\mu}\right\}  =\left\{
\tilde{\pi}_{\alpha\beta}\left(  h^{\alpha\beta},\pi_{\nu\mu}\right)
,\tilde{\pi}_{\nu\mu}\left(  h^{\alpha\beta},\pi_{\nu\mu}\right)  \right\}
_{h^{\alpha\beta},\pi_{\nu\mu}}=0.
\]

Note, that the second equation in (\ref{eqnAM138}) was used by Faddeev
\cite{Faddeev}, but without any discussion of canonicity, i.e. necessary
changes for the rest of the variables. In Faddeev's approach all of the
problems related to such a change of variables are hidden because, in
addition, some first class constraints were solved. This is against the Dirac
procedure and it eliminates the possibility to restore gauge invariance as all
first class constraints are needed (a simple example can be found in \cite{KK}).

Equally well, instead of a simplification of secondary constraints, we can try
to use a combination $e^{km}$ which naturally appeared when the secondary
second class constraints were solved (the same combination, $e^{km}$, was used
in the second order GR by Dirac \cite{Dirac}). We used $e^{km}$ as a
short-hand notation; but it is possible to find such a canonical
transformation that converts $e^{km}$ into a new variable. Let us introduce a variable%

\begin{equation}
\bar{h}^{km}\equiv e^{km}=h^{km}-\frac{h^{0k}h^{0m}}{h^{00}}.\label{eqnAM140}%
\end{equation}
As in the previous case, we will restrict our search by imposing%

\begin{equation}
h^{00}=\bar{h}^{00}\text{, \ }h^{0k}=\bar{h}^{0k}\label{eqnAM141}%
\end{equation}
and use the same condition as (\ref{eqnAM135})%

\begin{equation}
h_{,0}^{\alpha\beta}\pi_{\alpha\beta}=\bar{h}_{,0}^{\alpha\beta}\bar{\pi
}_{\alpha\beta}~.\label{eqnAM142}%
\end{equation}
Substitution of $h^{km}$ from (\ref{eqnAM140}) into (\ref{eqnAM142}) gives%

\begin{equation}
\bar{h}_{,0}^{km}\pi_{km}+2h_{,0}^{0k}\left(  \pi_{0k}+\frac{h^{0m}}{h^{00}%
}\pi_{km}\right)  +h_{,0}^{00}\left(  \bar{\pi}_{00}-\frac{h^{0k}h^{0m}%
}{h^{00}h^{00}}\pi_{km}\right)  =\bar{h}_{,0}^{km}\bar{\pi}_{\alpha\beta
}+2\bar{h}_{,0}^{0k}\bar{\pi}_{0k}+\bar{h}_{,0}^{00}\bar{\pi}_{00}%
\label{eqnAM143}%
\end{equation}
and the redefinition of momenta follows:%

\begin{equation}
\bar{\pi}_{km}=\pi_{km}~, \label{eqnAM144}%
\end{equation}

\begin{equation}
\bar{\pi}_{0k}=\pi_{0k}+\frac{h^{0m}}{h^{00}}\pi_{km}~, \label{eqnAM145}%
\end{equation}

\begin{equation}
\bar{\pi}_{00}=\pi_{00}-\frac{h^{0k}h^{0m}}{h^{00}h^{00}}\pi_{km}%
~.\label{eqnAM146}%
\end{equation}
It is easy to check that (\ref{eqnAM144})-(\ref{eqnAM146}), together with
(\ref{eqnAM140}) and (\ref{eqnAM141}), give the same PBs as (\ref{eqnAM139})
and so these transformations are canonical.

Another possible argument to find canonical transformations is to look at
expressions for tertiary constraints (or rather our first choice of tertiary
constraints) (\ref{eqnAM70}), (\ref{eqnAM72}) and try to simplify them. There
is one obvious combination, $h^{00}e^{km}$. And it can be used to build
canonical transformations; but it will lead to the same transformation that we
have already considered in the first example (\ref{eqnAM138}). If we use it
together with (\ref{eqnAM132})-(\ref{eqnAM133}), we will get the canonical
transformations, which simplify secondary and tertiary constraints simultaneously.

Substitution of (\ref{eqnAM138}) into constraints (\ref{eqnAM70}) and
(\ref{eqnAM72}) gives:%

\begin{equation}
\tilde{\tau}_{k}^{00}=2\left(  \tilde{h}^{mq}\tilde{\pi}_{mq}\right)
_{,k}-2\left(  \tilde{h}^{mq}\tilde{\pi}_{kq}\right)  _{,m}-\tilde{h}%
_{,k}^{np}\tilde{\pi}_{np}~, \label{eqnAM150}%
\end{equation}

\[
\tilde{\tau}_{0}^{00}=\tilde{h}^{kp}\tilde{h}^{mq}\left(  \pi_{mp}\pi_{kq}%
-\pi_{kp}\pi_{mq}\right)
\]

\begin{equation}
-\tilde{h}_{,km}^{km}+\frac{1}{2}\tilde{h}_{,y}^{xb}\tilde{h}_{bz}\tilde
{h}_{,x}^{yz}-\frac{1}{4}\tilde{h}_{ay}\tilde{h}_{bz}\tilde{h}^{nx}\tilde
{h}_{,n}^{ab}\tilde{h}_{,x}^{yz}+\frac{1}{4}\frac{1}{D-2}\tilde{h}^{nx}%
\tilde{h}_{ab}\tilde{h}_{,n}^{ab}\tilde{h}_{yz}\tilde{h}_{,x}^{yz}%
\label{eqnAM151}%
\end{equation}
where $\tilde{h}_{km}$ is defined as $\tilde{h}^{qm}\tilde{h}_{mp}=\delta
_{p}^{q}$ and related to the original variables by $\tilde{h}_{km}%
=\frac{h_{km}}{h^{00}}$.

These constraints, (\ref{eqnAM150}) and (\ref{eqnAM151}), are very similar to
the constraints obtained in \cite{Novel} for the first order formulation based
on variables (\ref{eqnAM7a}). This is not a surprise, as both first order
formulations are equivalent to the EH action. Up to a simple rearrangement and
with a different notation the expressions corresponding to (\ref{eqnAM150}%
)-(\ref{eqnAM151}) in \cite{Novel} are the following: (\ref{eqnAM150}) is
equivalent with Eq. (56) of \cite{Novel} and in (\ref{eqnAM151}), only the
first term differs in sign from the corresponding term in Eq. (58) of
\cite{Novel}.

The effect of canonical transformations on the algebra of constraints is
simple and it preserves its form (form-invariance). In general, a canonical
change of variables leads to changes in constraints \ %

\begin{equation}
\left(  q,p\right)  \rightarrow\left(  Q,P\right)  :\psi_{\mu}\left(
q,p\right)  \rightarrow\Psi_{\mu}\left(  Q,P\right)  \label{eqnAM160}%
\end{equation}
and if the algebra of constraints in old variables is%

\begin{equation}
\left\{  \psi_{\mu},\psi_{\nu}\right\}  =c_{\mu\nu}^{\gamma}\psi_{\gamma
}~,\label{eqnAM161}%
\end{equation}
where $c_{\mu\nu}^{\gamma}$ are structure functions, then in new variables its
form should be preserved and given by%

\begin{equation}
\left\{  \Psi_{\mu},\Psi_{\nu}\right\}  =C_{\mu\nu}^{\gamma}\Psi_{\gamma
}\label{eqnAM162}%
\end{equation}
with the simple condition on the structure functions%

\begin{equation}
C_{\mu\nu}^{\gamma}\left(  Q,P\right)  =c_{\mu\nu}^{\gamma}\left(  q,p\right)
_{q=q\left(  Q,P\right)  ,p=p\left(  Q,P\right)  }~. \label{eqnAM163}%
\end{equation}

For the first time such properties were demonstrated for the canonical
transformation in linearized gravity \cite{GKK} and later for a complete
formulation \cite{FKK}. In the Hamiltonian formulation of the first order EH
action, the form-invariance of the algebra of constraints is also preserved
after canonical transformations. For example, our secondary $\chi_{\mu}^{00}$
(\ref{eqnAM24})-(\ref{eqnAM25}) and tertiary $\tau_{\mu}^{00}$ (\ref{eqnAM70}%
)-(\ref{eqnAM72}) constraints have the algebra of PBs given in (\ref{eqnAM26}%
), (\ref{eqnAM76}), (\ref{eqnAM86})-(\ref{eqnAM88a}). Performing the canonical
transformations (\ref{eqnAM132}), (\ref{eqnAM133}), (\ref{eqnAM136}),
(\ref{eqnAM138}), we obtain new secondary $\tilde{\chi}_{\mu}^{00}$
(\ref{eqnAM130})-(\ref{eqnAM131}) and tertiary $\tilde{\tau}_{\mu}^{00}$
(\ref{eqnAM150})-(\ref{eqnAM151}) constraints and calculate new PBs among
them. The algebra of new PBs is related to the old one exactly as is described
by (\ref{eqnAM160})-(\ref{eqnAM163}), i.e. it is form-invariant.

In the first order formulation of the EH action, we consider the canonical
change of variables that does not affect the  primary variables - fields for
which their corresponding momenta are primary first class constraints. So the
corresponding algebra of all first class primary constraints is form-invariant
(in new and old variables primary first class constraints have zero PBs with
the rest of constraints). In the second order, PSS/Dirac, formulation,
canonical transformations also affect the primary constraints; but the
form-invariance of the whole algebra is preserved (see more details in
\cite{KKRV, Myths, FKK} and discussion in next Section).

In this Section we considered two operations: using different linear
combinations of tertiary constraints and canonical transformations that
involve only non-primary variables. What is the relationship between these two
operations? We showed that they are independent in the following sense.
Considering different choices of tertiary constraints for the same canonical
transformation produces different algebra among constraints, but preserves its
form-invariance. For example, compare $\bar{\tau}_{\mu}^{00}$ (\ref{eqnAM100})
with $\tau_{\mu}^{00}$ for the canonical transformation given by
(\ref{eqnAM132}), (\ref{eqnAM133}), (\ref{eqnAM136}), (\ref{eqnAM138}). The
PBs for them are different (compare, for example, (\ref{eqnAM102}) and
(\ref{eqnAM112})); but the form-invariance is preserved in accordance with
(\ref{eqnAM160})-(\ref{eqnAM163}). Applying different canonical
transformations to the same choice of tertiary constraints modifies the
constraints and structure functions, but it also preserves the form-invariance
of the algebra of constraints. We briefly discussed examples of such
operations with tertiary first class constraints and canonical transformations
that did not involve primary variables (and primary first class constraints).
We will consider operations with them in the next Section.

\section{The role of primary first class constraints}

Let us discuss the special properties and distinct role of primary first class
constraints in the Hamiltonian formulation of gauge invariant theories. In the
previous Section we discussed and demonstrated, by examples, the independence
of two operations: the choice of combinations of non-primary first class
constraints and canonical transformations of phase-space variables. We will
show that these two operations are not independent any more for primary first
class constraints, which are either pure canonical momenta as, for example, in
the Hamiltonians of the  affine-metric formulation and the metric formulation
due to Dirac \cite{Dirac}, or pure momenta plus some extra contributions, as
in the metric formulation due to Pirani, Schild and Skinner (PSS) \cite{PSS}
(the oldest one). In the PSS formulation, primary constraints are originated
from terms in the GR Lagrangian linear in the temporal derivatives
(\textquotedblleft velocities\textquotedblright) of the $g_{0\mu}$ components
of the metric tensor, and in the Dirac formulation or any first order
formulation of gauge invariant theories (e.g. affine-metric or tetrad-spin
connection), from variables without temporal derivatives in the corresponding
Lagrangians. Momenta conjugate to such variables are primary constraints and
at the same time they are part of a phase space of a considered system. This
part of a phase space is often and mistakenly neglected in Hamiltonian
formulations of GR (e.g. see discussion on p. 47 of \cite{Myths} and
references therein).

In the Dirac approach to the Hamiltonian formulation of constrained systems,
all variables are treated on an equal footing and each variable has the
corresponding momentum. Moreover, variables that are often neglected in the
Hamiltonian formulation of GR even have a special name given by Bergmann:
\textquotedblleft\textit{primary}\textquotedblright,\ that reflects their
importance. In monographs on constrained dynamics, e.g. \cite{Kurt,
Gitman-Tyutin}, and in non-GR Hamiltonians (Maxwell, Yang-Mills), primary
constraints are always present and are part of the \textit{total Hamiltonian}
(name given by Dirac \cite{Diracbook}).

For the Hamiltonian of first order metric-affine GR any change of the primary
first class constraints would be artificial as they are already in the
simplest possible form. So, to discuss canonical transformations that involve
primary first class constraints we refer to two Hamiltonian formulations of
the same theory, PSS and Dirac. Constraints and structure functions of their
algebra are quite different \cite{KKRV, Myths}; but in both cases the complete
sets of first class constraints lead to the same gauge invariance, as it
should be. And this invariance, derivable from the constraints, is the
four-dimensional diffeomorphism that follows directly with no need for a field
dependent redefinition of the gauge parameters. For the ADM formulation the
gauge transformations differ from four-diffeomorphism and the only so-called
\textquotedblleft correspondence\textquotedblright\ with diffeomorphism
\cite{Saha}, or \textquotedblleft diffeomorphism-induced\textquotedblright%
\ \cite{PonsSS}, or \textquotedblleft specific metric-dependent
diffeomorphism\textquotedblright\ \cite{Pons}, etc. can be accomplished.

The two total Hamiltonians of PSS and Dirac, $H_{T}^{PSS}$ and $H_{T}^{Dirac}$
respectively, are given by \cite{KKRV} and \cite{Myths}:%

\begin{equation}
H_{T}^{PSS}=\dot{g}_{0\rho}\left(  \pi_{PSS}^{0\rho}-\phi^{0\rho}\left(
\pi,g\right)  \right)  +g_{0\rho}\mathcal{H}_{PSS}^{0\rho}\left(  \pi
^{km},g_{\mu\nu}\right)  , \label{eqnAM200}%
\end{equation}

\begin{equation}
H_{T}^{Dirac}=\dot{g}_{0\rho}\pi_{Dirac}^{0\rho}+g_{0\rho}\mathcal{H}%
_{Dirac}^{0\rho}\left(  \pi^{km},g_{\mu\nu}\right)  . \label{eqnAM201}%
\end{equation}

In \cite{FKK} it was explicitly shown that the phase-space variables of the
two formulations are related by a canonical transformation that can be
performed if one wants to simplify the primary first class constraints of the
PSS formulation. Note that Dirac \cite{Dirac} found a suitable change in the
EH Lagrangian by adding to it two total derivatives, which does not affect the
equations of motion, but the primary constraints can be brought into simple
form. In \cite{FKK} it was shown that the same simplification can be
accomplished at the Hamiltonian level using canonical transformations in its
phase space and, of course, with the same result. What is important is that
the comparison of two formulations \cite{FKK} with different constraints and
structure functions in the algebras of constraints gives exactly relations
(\ref{eqnAM160})-(\ref{eqnAM163}). Note that primary first class constraints
of the Dirac and PSS formulations have non-zero PBs with a particular choice
of secondary constraints that were used in \cite{KKRV} and \cite{Myths}; but
this part of the algebra and structure functions also preserves
form-invariance under a canonical transformation. So, conditions
(\ref{eqnAM160})-(\ref{eqnAM163}) are satisfied by all first class
constraints; and this is in complete correspondence with the Dirac conjecture
\cite{Diracbook} that all first class constraints generate gauge symmetry. It
is to be expected that all relations amongst them must be preserved under
canonical transformations to keep invariance in tact.

As in a first order formulation, different combinations of secondary
constraints for the Dirac Hamiltonian (\ref{eqnAM201}) can also be considered;
and one particular combination was discussed in \cite{Myths} %

\begin{equation}
H_{T}=\dot{g}_{0\rho}\pi^{0\rho}+\left(  -g^{00}\right)  ^{-1/2}%
\mathcal{\tilde{H}}-\frac{g^{0i}}{g^{00}}\mathcal{\tilde{H}}_{i}~.
\label{eqnAM202}%
\end{equation}

Note, there are some similarities with our first choice of tertiary
constraints for part $H_{c}^{\prime\prime}$ (\ref{eqnAM99}). One particular
choice, out of many possible linear combinations of constraints, as in the
case of first order formulation (where a few choices of tertiary constraints
were considered in previous Section), is not special and cannot affect the
gauge invariance of this Hamiltonian; so it has to lead to a complete
restoration of four-dimensional diffeomorphism with field independent gauge
parameters as was stated in \cite{Myths}. Of course, the algebra of these
combinations is different compared with algebra of constraints for
(\ref{eqnAM201}); but all the different choices should not affect the physical
results. Similar combinations can also be constructed for (\ref{eqnAM200});
and under canonical transformations, the form-invariance is also preserved,
which is the same interplay of linear combinations of non-primary first class
constraints and canonical transformations that we illustrated for tertiary
constraints in previous Section.

Now we briefly discuss a connection of (\ref{eqnAM202}) with the conventional
formulation due to Arnowitt, Deser and Misner (ADM) \cite{ADM1959} in which
the total Hamiltonian is (see e.g. \cite{Castellani})%

\begin{equation}
H_{T}=\dot{N}P+\dot{N}^{i}P_{i}+N\mathcal{\tilde{H}}+N^{i}\mathcal{\tilde{H}%
}_{i}\label{eqnAM203}%
\end{equation}
where the secondary constraints are exactly the same as Dirac's and are known
as the \textquotedblleft Hamiltonian\textquotedblright\ $\mathcal{\tilde{H}}$
and the \textquotedblleft spatial diffeomorphism constraint\textquotedblright%
\ $\mathcal{\tilde{H}}_{i}$. According to Pullin \cite{Pulin2008}
\textquotedblleft It [ADM paper \cite{ADM}] bases the formulation on the
Palatini action principle\textquotedblright,\ which is actually the
affine-metric formulation of Einstein \cite{Einstein} (see footnote 3). But it
is clear that it differs from our results for the affine-metric formulation
(\ref{eqnAM31}), (\ref{eqnAM74}) and rather obviously have, at least, some
similarities with Dirac's second order formulation (\ref{eqnAM202}) where the
coefficients in front of constraints for one out of many possible combinations
were redefined and called new variables:%

\begin{equation}
N=\left(  -g^{00}\right)  ^{-1/2}, \label{eqnAM204}%
\end{equation}

\begin{equation}
N^{i}=-\frac{g^{0i}}{g^{00}}\label{eqnAM205}%
\end{equation}
which are known as \textquotedblleft lapse\textquotedblright\ and
\textquotedblleft shift\textquotedblright\ functions\footnote{Note that even
the names used in this formulation manifest their non-covariant nature and
shows the distinction of these variables, or different roles that they play in
the ADM formulation. Their names appeared soon after the original works of ADM
were published. To the best of our knowledge, it was Wheeler who coined the
names of these variables, \textquotedblleft lapse\textquotedblright\ and
\textquotedblleft shift\textquotedblright\ in \cite{Relativity}. DeWitt in
\cite{DeWitt} reserved the name \textquotedblleft Hamiltonian
constraint\textquotedblright\ only for $\mathcal{\tilde{H}}$ as it is a
\textquotedblleft particularly important constraint\textquotedblright,
probably, to reflect its distinction from\ $\mathcal{\tilde{H}}_{i}$, although
$\mathcal{\tilde{H}}$ and $\mathcal{\tilde{H}}_{i}$ are both the part of the
Hamiltonian. }.

As in the examples from the previous Section, because we are working in a
phase space we cannot just use some invertible transformations like
(\ref{eqnAM204})-(\ref{eqnAM205}) and simply write the new total Hamiltonian
as (\ref{eqnAM202}). To preserve canonicity, the change of variables
(\ref{eqnAM204})-(\ref{eqnAM205}), which is obviously invertible, has to be
accompanied by a change of the rest of phase-space variables or, at least,
some of them (as in the examples considered in previous Section). But, as in
ADM approach, the space-space components of the metric tensor, $g_{km}$, are
not changed and are exactly the same as in the Dirac formulation%

\begin{equation}
\left(  g_{km},\pi^{km}\right)  _{Dirac}=\left(  g_{km},\pi^{km}\right)
_{ADM}~;\label{eqnAM206}%
\end{equation}
and the search for new momenta has to be restricted by%

\begin{equation}
\dot{g}_{0\rho}\pi^{0\rho}=\dot{N}P+\dot{N}^{i}P_{i}~.\label{eqnAM207}%
\end{equation}
But it is impossible to find such a transformation that preserves
(\ref{eqnAM207}) with the additional condition (\ref{eqnAM206}). This
unavoidably leads to a conclusion that passing from (\ref{eqnAM202}) to
(\ref{eqnAM203}) is not a canonical transformation (an interested reader can
find more detail in Section 4 of \cite{Myths}). The calculation of one simple
PB is enough to prove non-canonicity (see Eq. (152) of \cite{Myths})%

\begin{equation}
\left\{  N,\pi^{km}\right\}  =\left\{  \left(  -g^{00}\right)  ^{-1/2}%
,\pi^{km}\right\}  \neq0. \label{eqnAM208}%
\end{equation}

So, because the ADM Hamiltonian and Dirac Hamiltonian of GR are not related
canonically, any connection between them is lost; but the disappearance of
four-dimensional diffeomorphism cannot be explained just by this fact. Even if
the ADM formulation is considered as a model, not related to the Hamiltonian
formulation of metric GR, one can argue that it still might have
diffeomorphism invariance and base his arguments on the fact that the number
of the primary first class constraints are still the same, and this number
defines the number of gauge parameters. Note, if the ADM Hamiltonian
(\ref{eqnAM203}) is treated as a model, then the lapse and shift functions are
canonical variables of this formulation. But if one claims that
(\ref{eqnAM203}) is a canonical formulation of GR, then it is not the case
because ADM variables are not related canonically to the metric tensor and its
momentum. To accommodate these two possible understandings we will use
quotation marks for \textquotedblleft canonical\textquotedblright\ ADM
Hamiltonian\footnote{The common statements as in \cite{Pullin}
\textquotedblleft Unfortunately, the canonical treatment breaks the symmetry
between space and time in general relativity and the resulting algebra of
constraints is not the algebra of four diffeomorphism\textquotedblright\ has a
double meaning. If \textquotedblleft canonical\textquotedblright\ is
understood as the formulation of ADM this is a true statement; but for GR, for
which ADM formulation is not canonical, Pullin's statement is wrong. We would
like to note that in the Hamiltonian formulation of GR the covariance is not
manifest, however, it is not broken as the gauge symmetry of GR,
four-diffeomorphism, is recovered in manifestly covariant form (see
\cite{KKRV, Myths}). So, the Hamiltonian formulation of GR does not break the
main property of the Einstein GR: general covariance. }.

It is not possible to find a canonical transformation, part of which
constitutes (\ref{eqnAM204})-(\ref{eqnAM205}) with the corresponding momenta
(\ref{eqnAM207}) and which simultaneously preserves the condition that the
space-space components of $g_{\mu\nu}$ remains the same as in the Dirac
formulation (\ref{eqnAM206}). But by relaxing this too restrictive condition
(\ref{eqnAM206}), the canonical transformation can be found and its form was
given in \cite{Myths} (see Eqs. (156)-(158)) and the problems that arise with
such transformations were discussed \cite{Myths}. Note that such a
transformation, of course, converts the Dirac Hamiltonian into a form which is
different from ADM anyway. So, for systems with constraints, even canonical
transformations can lead to some problems; and for such systems the canonicity
of the transformations is the only \textit{necessary} condition to have
equivalent formulations. Note that our conclusion based on a particular model,
is in contradiction with general discussion of canonical transformations for
constraints systems of \cite{GomisLR} where the authors stated that the
condition of canonicity is \textquotedblleft too strong\textquotedblright\ for
constraint systems; but in our opinion, based on a particular theory, it is
\textit{too} \textit{weak}, at least for covariant theories with first class constraints.

One obvious, at least for covariant theories, problem with the ADM variables
is in the original transformation (\ref{eqnAM204})-(\ref{eqnAM205}) and it is
related to simple dimensional analysis and to a special role of the primary
constraints. If we can find canonical transformations, they have to preserve
(\ref{eqnAM207}) so, in particular, we will have the following part in $H_{T}$%

\begin{equation}
H_{T}=\dot{N}P+\dot{N}^{i}P_{i}+...\label{eqnAM210}%
\end{equation}
with simple primary first class constraints $P$ and $P_{i}$. However, if
fields have a physical dimension, then the components of the metric tensor
should have the same dimension. It is obvious that the lapse and shift
functions defined by (\ref{eqnAM204})-(\ref{eqnAM205}) have different
dimensions in terms of the dimension of the metric tensor ($\dim N^{i}=0$ and
$\dim N=1/\sqrt{\dim g^{00}}$); so the corresponding momenta in
(\ref{eqnAM210}) (which are primary constraints in such formulation) should
have different dimensions as well. And such a dimensional mismatch of primary
constraints for a covariant theory guarantees the failure of this formulation
to preserve covariance and so such a transformation, even being canonical, has
to be rejected in any Hamiltonian formulation of a covariant system. This
conclusion is based on the following arguments related to the special role of
primary constraints in the derivation of gauge invariance. In the Castellani
procedure \cite{Castellani}, which we used to restore four-diffeomorphism
invariance for the PSS and Dirac formulations \cite{KKRV} and \cite{Myths},
the gauge generator is started from primary first class constraints%

\begin{equation}
G=\partial_{0}^{\left(  n\right)  }\varepsilon^{\mu}P_{\mu}%
+...\label{eqnAM212}%
\end{equation}
where $\partial_{0}^{\left(  n\right)  }$ is the temporal derivative of $n$-th
order of the gauge parameters $\varepsilon^{\mu}$, $n$ depends on the number
of generation of the constraints (if secondary constraints are present then
$n=1$, if tertiary: $n=2$, etc.) For a covariant theory $\varepsilon^{\mu}$
should be a true four-vector. Again based on dimensional analysis, if the
primary constraints have a different dimension, then the components of the
gauge parameters also have different dimensions. It is the well-known fact
that four quantities combined together do not necessary form a true
four-vector as their components must transform in the same way under general
coordinate transformations \cite{DiracGRbook}. As we can see from
(\ref{eqnAM204}) and (\ref{eqnAM205}), $N$ and $N^{i}$ transform differently,
because they are defined in terms of particular components of the metric
tensor in non-covariant way (see again (\ref{eqnAM204}) and (\ref{eqnAM205})).
This property is also transferred to the corresponding momenta and gauge
parameters. If, at least one gauge parameter has a different dimension or
transforming property from the remaining parameters (which is exactly the case
here), it is just impossible to combine them into the four-vector gauge
parameter, which is needed for four-diffeomorphism. To conclude: even in the
case of canonical change of variables, but with a mismatch of their dimensions
(at least for variables which correspond to primary constraints) the
covariance is lost. So, introduction of lapse and shifts functions by itself,
whether they are a part of a canonical transformation or not, unavoidably
destroys covariance and in turn the equivalence with the original covariant
theory. Please note that there are different approaches to the restoration of
gauge invariance, where a generator is built on other principles and is
actually started from the end of the constraint chains (e.g. see \cite{HTZ}),
i.e. from secondary constraints (in second order formulation) multiplied by
the gauge parameters%

\begin{equation}
G=...+\varepsilon\mathcal{\tilde{H}}+\varepsilon^{i}\mathcal{\tilde{H}}_{i}~.
\label{eqnAM214}%
\end{equation}

In this case the conclusion is the same, because if lapse and shift functions
have different dimensions, so do the \textquotedblleft
Hamiltonian\textquotedblright\ and \textquotedblleft spatial
diffeomorphism\textquotedblright\ constraints and the corresponding gauge
parameters in (\ref{eqnAM214}). We will discuss in detail the methods of
restoration of gauge invariance in Sections 6 and 7 with application to a
first order affine-metric formulation of GR. Here we would like just to add,
that in both methods, the initial assumption of either the Castellani
algorithm or HTZ ansatz is \textit{the independence of gauge parameters of
fields}; and this assumption is used in the iterative procedure to a find
generator. So any field dependent redefinition of gauge parameters, as
advocated in many articles (i.e. \cite{Saha, Pons, PonsSS}), that is performed
\textit{after} completion of the procedure, is in complete contradiction with
the initial assumptions for both approaches.

Now we will consider one additional example, the Hamiltonian formulation of
the Einstein-Cartan (EC) theory, which is a little bit aside of the main topic
of this article. But it provides an illustration of even further (possibly
general) restriction on the manipulations with primary constraints (canonical
variables), which cannot be illustrated using metric or affine-metric formulations.

The Hamiltonian formulation of the Einstein-Cartan theory in its first order
form, the so-called tetrad-spin connection formulation was discussed in many
works (i.e. \cite{CNP},\cite{Peldan}). As in the Hamiltonian formulation of
affine-metric GR, it leads to second class constraints that should be
eliminated, and after the Hamiltonian reduction leads to the following total
Hamiltonian (up to a total spatial derivative) \cite{3D, Report}%

\begin{equation}
H_{T}=\dot{e}_{0\left(  \rho\right)  }\pi^{0\left(  \rho\right)  }+\dot
{\omega}_{0\left(  \alpha\beta\right)  }\pi^{0\left(  \alpha\beta\right)
}+e_{0\left(  \rho\right)  }\chi^{0\left(  \rho\right)  }+\omega_{0\left(
\alpha\beta\right)  }\chi^{0\left(  \alpha\beta\right)  }. \label{eqnAM220}%
\end{equation}

In the $3D$ case, to obtain (\ref{eqnAM220}) is a simple task because there
are no secondary second class constraints \cite{3D}. In the $4D$ case, the
different methods (specific to this dimension) of solving secondary
constraints were used. One particularly transparent method is the introduction
of Darboux coordinates (specifically constructed \textit{only} for $4D$) due
to Ba\~{n}ados and Contreras \cite{Banados}. Of course, the Dirac procedure
can be used in any dimension higher than two,\footnote{As for metric EH
action, the second and the first order, affine-metric, formulations are
equivalent only in dimensions higher than two, the same thing happens for EC
action: tetrad and tetrad-spin connection formulations are equivalent also
when $D>2$ (see \cite{Report}, Section II).} which gives the same Hamiltonian
(\ref{eqnAM220}), and that was shown in \cite{Report}. Direct calculations are
involved and a considerable simplification occurs when the Darboux coordinates
(common to all dimensions) are used \cite{Darboux}.

The notation can vary from paper to paper, but it is usually explained in
detail. The equation (\ref{eqnAM220}) is written in the notation used in
\cite{Report}; and it is very close to\ what can be found in the first
\textquotedblleft gauge-free\textquotedblright%
\ formulation\footnote{\textquotedblleft Gauge-free\textquotedblright\ means
without fixing a gauge at the beginning of analysis. Such a fixing is in
contradiction with the Dirac procedure as a gauge cannot be fixed before a
gauge symmetry is found.} of the Einstein-Cartan Hamiltonian, due to
Castellani, van Nieuwenhuizen and Pilati \cite{CNP}, where the canonical
variables (after elimination of secondary second class constraints) are
\begin{equation}
e_{\mu\left(  \rho\right)  },\pi^{\mu\left(  \rho\right)  },\omega_{0\left(
\alpha\beta\right)  },\pi^{0\left(  \alpha\beta\right)  }.\label{eqnAM220a}%
\end{equation}

The form of (\ref{eqnAM220}), which is a nice covariant expression for the
total Hamiltonian of the first order EC action, was known for a long time. But
it is difficult to find it in this form, with a few rare exceptions (e.g.
\cite{Nicolic} where the first two terms of (\ref{eqnAM220}) are given in Eq.
(2.4) and the last two in Eq. (3.3)). Because the \textquotedblleft
canonical\textquotedblright\ \ formulation, in accordance with the
conventional wisdom, is equivalent with the presence of lapse and shift
functions, the change of variables is always performed to introduce
them\footnote{Such a transition is justified by either \textquotedblleft it is
more convenient\textquotedblright\ \cite{Nicolic} or \textquotedblleft it is
useful\textquotedblright\ \cite{CNP}.} instead of completion of the Dirac
analysis (proof of closure) and restoration of gauge invariance for
(\ref{eqnAM220}). (Some steps in the analysis of (\ref{eqnAM220}) can be found
in \cite{3D, Report}.)

The Hamiltonian in terms of lapse and shift functions for tetrads becomes
(this form is much easier to find in literature, contrary to (\ref{eqnAM220}),
e.g. \cite{DiSR})%

\begin{equation}
H_{T}=\dot{N}P+\dot{N}^{i}P_{i}+\dot{\omega}_{0\left(  \alpha\beta\right)
}\pi^{0\left(  \alpha\beta\right)  }+N\mathcal{\tilde{H}}+N^{i}\mathcal{\tilde
{H}}_{i}+\omega_{0\left(  \alpha\beta\right)  }\chi^{0\left(  \alpha
\beta\right)  }. \label{eqnAM221}%
\end{equation}

Note that introduction of lapse and shift functions guarantees the
disappearance of covariance in the formulation. We have already discussed the
effect of the ADM variables, which have different dimensions; and any hope to
have a covariant formulation is lost. They also, as in metric formulations,
are not canonical for a transformation from (\ref{eqnAM220}) to
(\ref{eqnAM221}) if they are introduced with restriction on the rest of tetrad
components and corresponding momenta (which is always the case).

The canonicity of transformation in the phase space, i.e. from the complete
set of canonical variables of (\ref{eqnAM220}) ($e_{\mu\left(  \rho\right)
},\pi^{\mu\left(  \rho\right)  },\omega_{0\left(  \alpha\beta\right)  }%
,\pi^{0\left(  \alpha\beta\right)  }$) \cite{CNP, Nicolic, Report} to
variables of (\ref{eqnAM221}) ($N,N^{i},P,P_{i},e_{k\left(  \rho\right)  }%
,\pi^{k\left(  \rho\right)  },\omega_{0\left(  \alpha\beta\right)  }%
,\pi^{0\left(  \alpha\beta\right)  }$), has never been discussed and it has
the same deficiency as for passing from the Dirac to ADM formulations in the
second order metric EH action. Moreover, exactly as in the metric case, only
the part of variables is involved in such a change%

\begin{equation}
e_{0\left(  \rho\right)  },\pi^{0\left(  \rho\right)  }\rightarrow
N,N^{i},P,P_{i}~, \label{eqnAM222}%
\end{equation}

\begin{equation}
(e_{k\left(  \rho\right)  },\pi^{k\left(  \rho\right)  },\omega_{0\left(
\alpha\beta\right)  },\pi^{0\left(  \alpha\beta\right)  })_{EC}=(e_{k\left(
\rho\right)  },\pi^{k\left(  \rho\right)  },\omega_{0\left(  \alpha
\beta\right)  },\pi^{0\left(  \alpha\beta\right)  })_{ADM}\label{eqnAM222a}%
\end{equation}
where the lapse (\ref{eqnAM204}) and shift (\ref{eqnAM205}) functions have to
be expressed in terms of the original phase-space variables (\ref{eqnAM220a})
of the reduced Hamiltonian (\ref{eqnAM220}) using $g^{\mu\nu}=e_{\left(
\rho\right)  }^{\mu}e^{\nu\left(  \rho\right)  }$%

\begin{equation}
N=\left(  -e_{\left(  \rho\right)  }^{0}e^{0\left(  \rho\right)  }\right)
^{-1/2},\label{eqnAM223}%
\end{equation}

\begin{equation}
N^{i}=-\frac{e_{\left(  \gamma\right)  }^{0}e^{i\left(  \gamma\right)  }%
}{e_{\left(  \rho\right)  }^{0}e^{0\left(  \rho\right)  }}.\label{eqnAM224}%
\end{equation}

As in the metric case, there are no canonical transformations for the subset
of phase-space variables (\ref{eqnAM222}) if the rest of variables of
(\ref{eqnAM220a}) is not involved. As in the transition from the Dirac to ADM
variables it was enough to find one PB (\ref{eqnAM208}) which proves
non-canonicity of the ADM variables, for the tetrad formulation at least one
PB is also non-zero%

\begin{equation}
\left\{  N^{i},\pi^{k\left(  \lambda\right)  }\right\}  =\frac{\delta N^{i}%
}{\delta e_{k\left(  \lambda\right)  }}\neq0.\label{eqnAM225}%
\end{equation}
So, introduction of lapse and shift functions in the tetrad formulation is
also a non-canonical transformation. Actually, based on previous analysis, it
is obvious that such a formulation (even with adjustments for canonicity that
will change constraints) will create a dimensional mismatch of variables and
gauge parameters, and so destroy the covariance in exactly the same way as
lapse and shift functions destroy it for the metric GR. The combination
\textquotedblleft Dirac-ADM\textquotedblright\ is not correct, as well as
\textquotedblleft ADM-Einstein-Hilbert\textquotedblright, so the ADM and
Einstein-Cartan formulations are not compatible. All problems related to the
\textquotedblleft spatial diffeomorphism\textquotedblright\ constraint safely
propagate into the tetrad formulation of GR and such the Hamiltonian
(\ref{eqnAM221}) is not the Hamiltonian of the original theory. In addition,
this new theory is not covariant by construction. However, the use of ADM
variables for first order tetrad-spin connection formulation is much more
interesting example compared to its metric counterpart; and this is the main
reason to include the Einstein-Cartan Hamiltonian in our discussion about the
role of primary first class constraints.

Here one can make an additional and simple observation related to the role of
primary first class constraints. Constructing the generator for formulation
(\ref{eqnAM220}), one obtains%

\begin{equation}
G=\dot{\varepsilon}_{\left(  \rho\right)  }\pi^{0\left(  \rho\right)  }%
+\dot{\varepsilon}_{\left(  \alpha\beta\right)  }\pi^{0\left(  \alpha
\beta\right)  }+...\label{eqnAM230}%
\end{equation}
with two gauge parameters, \textquotedblleft rotational\textquotedblright%
\ ($\varepsilon_{\left(  \alpha\beta\right)  }$) and \textquotedblleft
translational\textquotedblright\ ($\varepsilon_{\left(  \rho\right)  }$), both
with \textquotedblleft internal\textquotedblright\ indices (which correspond
to motion in the tangent space), and which lead to translational and
rotational invariance in the \textit{internal space} in $3D$ \cite{Witten,
3D}, as well as in all higher dimensions \cite{Report, Trans}. It must be
emphasized that among the many papers on the Hamiltonian formulation of EC
action, the work of \cite{CNP} is an exception because this is the only one
where lapse and shift functions are just a short-hand notation, so
non-canonical transformations (\ref{eqnAM223}), (\ref{eqnAM224}) were not
performed. Should one apply the Castellani procedure (see next Section)
starting from the primary constraints of \cite{CNP} the gauge invariance,
translation and rotation in the internal space, would follow (not a
diffeomorphism, which is a symmetry, but not the \textit{gauge} symmetry of
the EC action).

In contrast, if the generator is built for (\ref{eqnAM221}) \cite{DiSR} (let
us forget for a moment about non-canonicity) then it becomes%

\begin{equation}
G=\dot{\varepsilon}P+\dot{\varepsilon}^{i}P_{i}+\dot{\varepsilon}_{\left(
\alpha\beta\right)  }\pi^{0\left(  \alpha\beta\right)  }+...\label{eqnAM231}%
\end{equation}
which also has two gauge parameters. One, as before, corresponds to rotation
in the \textit{internal space} ($\varepsilon_{\left(  \alpha\beta\right)  }$);
but the second (actually, there are two of them, $\varepsilon$ and
$\varepsilon^{i}$) corresponds to the translation in the \textit{external
space}: $\varepsilon$ for time lapse and $\varepsilon^{i}$\textit{ }for shift
in a space-like surface. Because $P$ and $P_{i}$ are momenta conjugate to
lapse and shift functions, which have different dimensions and rules of
transformation, then $P$ and $P_{i}$ also do not form a true four-vector, and
so are the parameters $\varepsilon$ and $\varepsilon^{i}$ (it follows from
(\ref{eqnAM231})). Even if by relaxing the conditions of (\ref{eqnAM222a}) and
finding some canonical transformations, and even with a match of dimensions,
such a change of gauge symmetry, from translation in an internal space to
lapse and shift in an external space would be strange.

Is it possible to have equivalence in such a case? This would mean that the
Hamiltonian formulation does not give a unique \textit{gauge} invariance; but
this conclusion is hard to accept. Why does one, having the simplest possible
constraints presented in (\ref{eqnAM220}), have to perform some manipulations,
besides the desire to have a \textquotedblleft canonical\textquotedblright%
\ formulation, i.e. to have lapse and shift functions that destroy covariance?
We think that for field theories, changes of variables (even canonical, if we
can find such) should be restricted by the requirement to also preserve a
tensorial character of primary variables. Maybe, such a condition is
equivalent to the preservation of form-invariance of the algebra of
constraints (\ref{eqnAM160})-(\ref{eqnAM163}). There are still many questions
that need to be clarified. Dirac introduced his procedure as an outline of a
general approach, together with his famous conjecture about the connection of
first class constraints and gauge invariance. Only after the methods of
restoration of gauge symmetry were developed, it became clear that the Dirac
conjecture is correct and it can be used as a procedure. But there are still
many other questions, especially related to field theories, that remain to be answered.

Based on the examples considered in this Section, we can make a conclusion
that for constrained field theories the canonicity of change of phase-space
variables is a \textit{necessary}, but not a sufficient condition. To keep an
equivalence of two Hamiltonian formulations, the canonical transformations
that lead to the mismatch of the dimensions for primary variables (which is
permissible for mechanical systems or field theories without first class
constraints); and also for canonical transformations that change tensorial
character of primary variables, must be disregarded. Primary constraints are
the true Masters of the Hamiltonian formulations of gauge theories. They do
not need a master constraint programme as they are \textit{primary}
constraints; and manipulations with them are restrictive, especially for
covariant theories. In the formulations considered here where primary
constraints are in the simplest possible form (e.g. $\left(  \Gamma_{00}^{\mu
},\Pi_{\mu}^{00}\right)  $), it is better not to change them at all, to keep
in tact the \textit{primary} properties of the system (or there should be a
very good reason to do the changes).

The reasons behind converting covariant expressions for the total Hamiltonian
of the second order metric GR (\ref{eqnAM200})-(\ref{eqnAM202}) and the first
order tetrad-spin connection formulation (\ref{eqnAM220}) into non-covariant
expressions (\ref{eqnAM203}), and (\ref{eqnAM221}), which destroy any hope of
having covariant results (e.g. four-diffeomorphism for metric GR), remains a
mystery to us. (We prefer to avoid speculation and leave it for the History of
Science to figure out what happened 50 years ago, and why Einstein's general
covariance became unimportant.) But what is even more mysterious is the
unquestionable acceptance of the ADM formulation by the majority of
practitioners despite its inconsistencies and despite (not often) the strong
voices that express concern about its deficiency, contradiction with GR, and
even give \textquotedblleft hints\textquotedblright\ about the source of the
problems. We provide just a few such statements. Hawking 30 years ago
concluded \cite{Hawking}: \textquotedblleft The split into three spatial
dimensions and one time dimension seems to be contrary to the whole spirit of
relativity.\textquotedblright\ There appear more recently the statements of
Pons \cite{Pons}: \textquotedblleft Being non-intrinsic, the 3+1 decomposition
is somewhat at odds with a generally covariant formalism, and difficulties
arise for this reason\textquotedblright\ and Rovelli \cite{Rovelli}:
\textquotedblleft The very foundation of general covariant physics is the idea
that the notion of a simultaneity surface over the universe is devoid of
physical meaning\textquotedblright. The warning about the sources of problems
in further constructions based on the ADM formulation was given by Landsman
\cite{Landsman}: \textquotedblleft the lack of covariance of the ADM approach,
which is especially dangerous in connection with quantum field
theory\textquotedblright. 

Finally, we repeat the \textquotedblleft hint\textquotedblright\ given by
Isham and Kuchar \cite{Isham} \textquotedblleft Thus the full group of
spacetime diffeomorphism has somehow got lost in making the transition from
the Hilbert action to the Dirac-ADM action\textquotedblright.\ This statement
(without long calculation or a long chain of logical constructions)
immediately leads to the conclusion that if a transition from one action
(Einstein-Hilbert) to \textit{another} (ADM) was performed by a change of
variables and something got lost then \textquotedblleft
somehow\textquotedblright\ is exactly the change of variables that were
performed in such a transition. We strongly object to the use of combination
\textquotedblleft Dirac-ADM\textquotedblright, as nothing \textquotedblleft
got lost\textquotedblright\ in the Dirac Hamiltonian formulation based on the
Einstein-Hilbert action \cite{Dirac}. 

Last year, 50 years after publication of the Dirac Hamiltonian of GR
\cite{Dirac}\footnote{We are thankful to Shestakova \cite{Shestakova} for this
observation and in this article we are also trying do keep track of important
anniversaries.}, \textit{ }it was explicitly demonstrated \cite{Myths} that
his Hamiltonian of the second order metric GR leads directly to
four-diffeomorphism without any, even numerical, redefinition of gauge
parameters. This clearly demonstrates that four-diffeomorphism has not
\textquotedblleft got lost\textquotedblright\ if one abandons the idea of
making a transition from the Einstein-Hilbert action to ADM action (i.e. if
the ADM variables \textquotedblleft got lost\textquotedblright\ instead). In
\cite{Myths} we argued (see Eq. (163) of \cite{Myths}) that, based on
equivalence of Hamiltonian and Lagrangian methods, it should be possible to
demonstrate non-equivalence of the ADM and EH actions also at the pure
Lagrangian level (without going to a phase space); but we did not formulate
the criteria and did not show a proof. Recently we demonstrated that the
Einstein-Cartan action is invariant under translation in a tangent space
\cite{Trans} using the pure Lagrangian method (to verify the strong indication
of the presence of this gauge symmetry in the Hamiltonian formulation
\cite{Report}). This also allows us to formulate the condition for equivalence
of two actions at the pure Lagrangian level for singular systems: if
transition (field redefinition) from one singular action (e.g.
Einstein-Hilbert) to another singular action (e.g. ADM-inspired tetrad) is
performed by an invertible change of variables (the necessary condition for
equivalence) we can find differential identities for both of them and obtain
the corresponding transformations for their sets of variables. These
transformations, for equivalent formulations, must be derivable one from
another by using the same (or inverse) redefinition of fields as in transition
from one action to another\footnote{The \textquotedblleft
loss\textquotedblright\ of four-diffeomorphism in the ADM formulation was
demonstrated by Banerjee at al \cite{Banerjee} (see Section 6 of
\cite{Banerjee}) where symmetries of the ADM Lagrangian and the
Einstein-Hilbert Lagrangian were considered. It is not a surprise (taking into
account equivalence of Lagrangian and Hamiltonian methods) that the same gauge
invariance as in the case of the ADM Hamiltonian was found also for the ADM
Lagrangian. The transformations found using the ADM Lagrangian are not
convertible (using the original change of variables) into four-diffeomorphism
invariance of the EH Lagrangian obtained by the same method \cite{Samanta}.
Only after the same manipulations with a field dependent redefinition of gauge
parameters \cite{Saha} (which should also be field independent in construction
of the generator from differential identities) invariance of the ADM
Lagrangian can lead to \textquotedblleft correspondence\textquotedblright%
\ with diffeomorphism. The authors of \cite{Banerjee} call this
\textquotedblleft the equivalence between the gauge and diff parameters by
devising of the one to one mapping\textquotedblright.}. 

This year, one year after 50 anniversary of Dirac's paper \cite{Dirac}, it was
a celebration dedicated to another paper on the same subject \cite{ADM1959}
that took place in Texas A\&M University on November 7 and 8, 2009
\textquotedblleft ADM-50: A celebration of current GR Innovation". This event
marked 50 years of loss of four-diffeomorphism and covariance in the
\textquotedblleft canonical\textquotedblright\ formulation of covariant GR. In
the above mentioned article \cite{ADM1959} the authors claimed to consider the
Hamiltonian formulation of GR in Palatini form, i.e. the formulation of
Einstein \cite{Einstein}. This is also the main subject of our paper, but with
different results and without loss of four-diffeomorphism (see next Section). 

The transition from the EH action to the ADM action, which is responsible for
a loss of four-diffeomorphism, continues to propagate into new fields. A
particular example is Loop Quantum Gravity (LQG) - one of the major players in
the quest for quantization of gravity. According to Thiemann \cite{Inside}
\textquotedblleft One of the reasons why LQG is gaining in its degree of
popularity as compared to string theory is that LQG has `put its cards on the
table'\textquotedblright. Let us look at the LQG \textquotedblleft
cards\textquotedblright\ that were put together in the recent article
\cite{Han} (see first paragraph of Introduction) that describes the meaning of
LQG: \textquotedblleft...LQG is a mathematically rigorous\footnote{In our
taste, \textquotedblleft rigorous\textquotedblright\ in combination with
mathematics is a tautology, but because the rival of LQG employs elegant
mathematics, the use of just mathematics, probably, sounds a little bit weak.}
quantization of general relativity (GR)... It is inspired by the formulation
of GR as a canonical dynamical theory... The total Hamiltonian of GR is a
linear combination of the Gauss constraint, the spatial diffeomorphism
constraints and the Hamiltonian constraint. Thus the dynamics of GR are
essentially the gauge transformations generated by
constraints\textquotedblright. Is it canonical formulation of the covariant GR
if the \textquotedblleft spatial diffeomorphism\textquotedblright\ constraint
is present? It is (\ref{eqnAM221}) that cannot have covariance and it is not
equivalent to the Einstein-Cartan GR. The Hamiltonian of LQG can be written in
different variables; but they are originated from lapse and shift functions,
that is why, the \textquotedblleft spatial diffeomorphism\textquotedblright%
\ constraint is always there, as in the ADM gravity (\ref{eqnAM203}).
Moreover, and this was already demonstrated (e.g. \cite{HNS, Franke,
Schucker}), all new variables which are used in LQG can be converted into ADM
variables by canonical transformations, i.e. new formulations are equivalent
to the ADM formulation. The gauge transformation generated by constraints of
the ADM-inspired Hamiltonian of tetrad gravity is the \textquotedblleft
spatial diffeomorphism\textquotedblright;\ but the transformations generated
by the Hamiltonian of EC action are different \cite{Report, Trans}. So, in
reality LQG, using rigorous mathematics, studies quantization of the
ADM-inspired model and its gauge invariance and, because it is not covariant,
all well-known results follow: quantization of three-dimensional surfaces
\cite{RS}, Lorentz violation, etc. The suspicion of people not working in this
field about how a covariant theory of GR can lead to such results is correct.
The explanation of these effects originated from the classical Hamiltonian
mechanics and the theory of canonical transformations. These are no quantum
effects, but they just manifest the dependence of quantum effect on a
classical background, the ADM Hamiltonian. To show this, neither rigorous nor
elegant, but just a mathematical condition, the Poisson brackets
(\ref{eqnAM208}) and (\ref{eqnAM225}), that must be zero to keep two
Hamiltonians equivalent and the notion of a phase space are needed. This is
what any outsider to this field can and should read from the LQG
\textquotedblleft cards\textquotedblright. This situation can be perfectly
described using one of the famous paradoxical forms of Wheeler (e.g. see
\cite{Wheeler}): LQG is quantum \textquotedblleft gravitation without
gravitation\textquotedblright\ in Einstein's sense.

We discuss recent, current and coming soon (see next paragraph)
anniversaries/celebrations. But let us also mention a lesser known
anniversary: 200 years ago Sim\'{e}on-Denis Poisson published his work
\textquotedblleft Sur la Variation des Constantes arvitraires dans les
questions de M\'{e}canique\textquotedblright\ in the Journal de l'\'{E}cole
Polytechnique, Tome VIII, p. 266, D\'{e}cembre 1809, Paris. This is the first
appearance of what has become known as the Poisson brackets; and we have used
them to show that transition from EH or EC variables to ADM variables is not
canonical (see (\ref{eqnAM208}) and (\ref{eqnAM225})).

Let us return from the past to present days, and continue with one more
quotation from Thiemann \cite{Inside}: \textquotedblleft The `rules of the
game'\ have been written and are not tinkered with\textquotedblright. And that
is an important standard to which current research in GR must be held. The
rules of the game are those published by Poisson, in 1809, and by Einstein at
the beginning of the twentieth century. These are important rules which must
not be tinkered with, and must not be ignored.

However, seems to us that the main rule of the LQG \textquotedblleft
game\textquotedblright\ is to work with the ADM-inspired action which is not
covariant and, because of this, allows to obtain non-covariant results but, at
the same time, to present this as properties of the Einstein-Cartan action, as
scientific community at large still associates gravity with the name of
Einstein. Neither outside view on LQG (Nicolai, at al \cite{Outside}) nor
inside view on the same subject (Thiemann \cite{Inside}) disobey this rule,
and, as a result, LQG is also approaching its 25 anniversary.

Of course, there is a freedom of research (or variety of \textquotedblleft
games\textquotedblright\ in nowadays language) and one can choose to play with
the ADM-inspired metric or tetrad \textit{models} in a framework of
non-covariant theories; but one should not be surprised by the results which
are not covariant. Or one can study the covariant theories of GR (Einstein or
Einstein-Cartan) and in this case one ought be surprised if non-covariant
results follow.

We choose to work with the covariant Einstein formulation of GR and do not
convert his brilliant theory into \textquotedblleft common
currency\textquotedblright\footnote{The name given by Pullin \cite{Pulin2008}
to the ADM formulation in his Editorial note for republication of \cite{ADM}
in \cite{goldies}.}, i.e. we do not make any non-canonical change of variables
and we return to the total Hamiltonian (\ref{eqnAM31}), (\ref{eqnAM74}) of the
first order affine-metric Einstein GR, the main subject of this article. The
Hamiltonian formulation of any system leading to first class constraints
cannot be considered as complete without deriving the corresponding gauge
invariance. In addition, a restoration of gauge invariance is an important
consistency check, especially if we investigate such a complicated theory as GR.

In the course of our calculations it becomes clear that, despite different
starting points in our current consideration (affine-metric action
(\ref{eqnAM2})) and those based on different set of variables (see
(\ref{eqnAM43})) that were discussed in \cite{KK, KKM}, and especially due to
work of Ghalati and McKeon \cite{G/R} where the closure of Dirac procedure was
demonstrated for the first time, the results are very similar. One can say it
is remarkable or surprising, but nothing here is either \textquotedblleft
remarkable\textquotedblright\ or \textquotedblleft
surprising\textquotedblright\ because these results come from two equivalent
first order formulations of the same theory (see Appendix of \cite{KK}). Of
course, there are some differences; but secondary and tertiary constraints
(for the same particular choice) are exactly the same after the canonical
transformations (\ref{eqnAM132})-(\ref{eqnAM138}) were performed. Neither our
formulation, nor \cite{G/R} were converted into a \textquotedblleft
canonical\textquotedblright\ form (despite that some attempts to make a
formulation \textquotedblleft canonical\textquotedblright\ were made in the
novel approach of \cite{Novel}); and one should expect the complete
restoration of four-diffeomorphism as it was obtained for the second order
formulation of metric GR \cite{KKRV, Myths, FKK}.

The first steps of the restoration of gauge symmetry were made in
\cite{Novel}, with a truly surprising result for a covariant theory: the need
for a field dependent redefinition of gauge parameters to make the
transformations of $h^{00}$ (the only one that was calculated\footnote{To
avoid any confusion, we would like to emphasize that converting the results of
direct calculations into the ADM form, which is extensively discussed in
\cite{Novel}, was not used in the calculation of these transformations.})
\textquotedblleft correspond to diffeomorphism invariance\textquotedblright%
\ \cite{Novel}, exactly as in the conventional \textquotedblleft
canonical\textquotedblright\ formulations of metric and tetrad gravities
\cite{Saha, Pons, PonsSS}. Why does the Hamiltonian formulation of the first
order EH action, obtained without non-canonical changes of variables, give so
different result compared to the Hamiltonian formulation of the second order
of GR? In the restoration of gauge invariance, we used the Castellani
algorithm \cite{Castellani} but in \cite{Novel} the different method, the HTZ
ansatz \cite{HTZ}, was employed. This is the only difference, and a comparison
of the two different methods is the best point to start the search for
understanding of the apparently different gauge transformations for first and
second order formulations of GR. The field dependent redefinition of gauge
parameters of these two formulations could be just an artifact of a particular
method (if in one it is needed, but in another it is not). In the next
Section, we first apply the Castellani algorithm to the first order
formulation to obtain, as in \cite{Novel}, partial transformations (which is
enough to make a conclusion about the necessity of a field dependent
redefinition of gauge parameters). And then, we compare this with the results
obtained using the HTZ ansatz.

\section{Castellani algorithm}

In his paper \cite{Castellani} Castellani illustrated the application of his
method by considering Yang-Mills theory (the system with only primary and
secondary first class constraints). We also used this method to restore gauge
invariance in the Hamiltonian formulation of the second order metric GR
\cite{KKRV, Myths} and of the first order, tetrad-spin connection, formulation
of Einstein-Cartan action in the three dimensional case \cite{3D}. In the
Hamiltonian of the first-order affine-metric GR, we have tertiary constraints;
but the procedure of \cite{Castellani} is general and can be applied to
systems with any number of generations of constraints. We are not aware of an
application of the Castellani procedure to a realistic Hamiltonian with
tertiary constraints, and such an application is interesting by itself. Our
main goal in this, and in the following Section, is to analyze the appearance
of a field dependent redefinition of gauge parameters, not a complete
restoration of gauge invariance. So as in \cite{Novel}, we will calculate
transformations only partially but for all fields, to see whether this method
produces correct terms in the transformations of \textit{all fields}, contrary
to \cite{Novel}, where the transformation of only one field was found.

In the Castellani algorithm \cite{Castellani} the generator of the gauge
transformations for the Hamiltonian with first class constraints for the
system with tertiary constraints is given by%

\begin{equation}
G=\varepsilon^{\mu}G_{\left(  0\right)  \mu}+\dot{\varepsilon}^{\mu}G_{\left(
1\right)  \mu}+\ddot{\varepsilon}^{\mu}G_{\left(  2\right)  \mu}%
\label{eqnCAS1}%
\end{equation}
where $\varepsilon^{\mu}$ are the gauge parameters and $\dot{\varepsilon}%
^{\mu}$, $\ddot{\varepsilon}^{\mu}$ are their temporal derivatives. The number
of gauge parameters and their tensorial dimension are uniquely defined by
primary first class constraints, so for the formulation considered, the number
of parameters is equal to the dimension of spacetime, $D$. The functions
$G_{\left(  i\right)  \mu}$ are defined by the following iterative procedure
(see Eq. (16b) and for more details see also Section 5 of \cite{Castellani})%

\begin{equation}
G_{\left(  2\right)  \mu}=\Pi_{\mu}^{00}, \label{eqnCAS2}%
\end{equation}

\begin{equation}
G_{\left(  1\right)  \mu}+\left\{  G_{\left(  2\right)  \mu},H_{T}\right\}
=\int d\overrightarrow{y}\alpha_{\mu}^{\nu}\left(  \overrightarrow
{x},\overrightarrow{y}\right)  \Pi_{\nu}^{00}\left(  \overrightarrow
{y}\right)  , \label{eqnCAS3}%
\end{equation}

\begin{equation}
G_{\left(  0\right)  \mu}+\left\{  G_{\left(  1\right)  \mu},H_{T}\right\}
=\int d\overrightarrow{y}\beta_{\mu}^{\nu}\left(  \overrightarrow
{x},\overrightarrow{y}\right)  \Pi_{\nu}^{00}\left(  \overrightarrow
{y}\right)  , \label{eqnCAS4}%
\end{equation}

\begin{equation}
\left\{  G_{\left(  0\right)  \mu},H_{T}\right\}  =primary. \label{eqnCAS5}%
\end{equation}

Note that only primary constraints enter equations (\ref{eqnCAS2}%
)-(\ref{eqnCAS5}) explicitly. The functions $G_{\left(  2\right)  \mu}$ are
uniquely defined as primary constraints. The functions, $G_{\left(  1\right)
\mu}$ and $G_{\left(  0\right)  \mu}$, in general, are not just secondary or
tertiary constraints because, for example, different linear combinations of
tertiary constraints can be considered (see (\ref{eqnAM99}), (\ref{eqnAM101}),
and (\ref{eqnAM110})). This makes this method insensitive to our choice of
combinations of non-primary constraints and gives the same gauge invariance
regardless of what combinations we will call `tertiary constraints'. We will
illustrate this important point in more detail: for complicated theories the
possibility of working with different combinations of constraints, without
destroying its unique gauge symmetry, could give significant computational advantages.

Using PBs among the first class constraints and the total Hamiltonian, which
is given by (\ref{eqnAM31})-(\ref{eqnAM32}) with (\ref{eqnAM54a}) and
(\ref{eqnAM55})-(\ref{eqnAM57}), we can solve (\ref{eqnCAS2}) for $G_{\left(
1\right)  \mu}$%

\begin{equation}
G_{\left(  1\right)  \mu}=-\left\{  G_{\left(  2\right)  \mu},H_{T}\right\}
+\alpha_{\mu}^{\nu}\Pi_{\nu}^{00}=-\chi_{\mu}^{00}+\alpha_{\mu}^{\nu}\Pi_{\nu
}^{00}. \label{eqnCAS6}%
\end{equation}

In this equation the secondary constraints $\chi_{\mu}^{00}$ unambiguously
appear through $\left\{  \Pi_{\mu}^{00},H_{T}\right\}  $; and $G_{\left(
1\right)  \mu}$ becomes a secondary plus linear combination of primary
constraints with coefficient-functions $\alpha_{\mu}^{\nu}$ that have to be
found. To shorten the notation, we will not write integrals in equations that
involve coefficient functions, which in general might also depend on fields
and their derivatives (see, i.e. \cite{KKRV, Myths}). Only in the case of
contributions with derivatives, which is specific to field theories, a more
careful treatment is needed for finding the expressions of the corresponding
coefficient-functions. In this Section we restrict ourselves to some simple
steps of procedure mainly to discuss its important properties and to show that
it provides a strong indication for the correct restoration of the gauge
transformations for all fields. We will consider only a part of the
transformations that are produced by simple contributions from
coefficient-functions, which do not involve spatial derivatives, so our short
notation without integrals will not lead to any ambiguity. The complete
restoration of diffeomorphism invariance from the constraint structure of the
Hamiltonian of first order affine-metric GR using the Castellani method is in
progress and, of course, the complete calculations involve the spatial
derivatives of the fields and integral form of expressions with
coefficient-functions becomes important. Details of such calculations will be
reported elsewhere.

At the next step of the procedure, using (\ref{eqnCAS4}) and the above result
(\ref{eqnCAS6}) the functions $G_{\left(  0\right)  \mu}$ can be found. The
total Hamiltonian (in our condensed notation) is%

\begin{equation}
H_{T}=\dot{\Gamma}_{00}^{\mu}\Pi_{\mu}^{00}-\Gamma_{00}^{\mu}\chi_{\mu}%
^{00}+H_{c}^{\prime}~. \label{eqnCAS7}%
\end{equation}

In such a form, (\ref{eqnAM55})-(\ref{eqnAM57}), $H_{c}^{\prime}$ is
independent of \textquotedblleft primary fields\textquotedblright%
\ $\Gamma_{00}^{\mu}$ (fields for which conjugate momenta are primary
constraints) and this allows us to perform a few steps of calculation, which
are independent of a particular choice of tertiary constraints. Using this
form of the Hamiltonian, (\ref{eqnCAS7}) and after some rearrangement,
(\ref{eqnCAS4}) becomes%

\[
G_{\left(  0\right)  \mu}=-\left\{  \chi_{\mu}^{00},\Gamma_{00}^{\nu}\chi
_{\nu}^{00}\right\}  +\left\{  \chi_{\mu}^{00},H_{c}^{\prime}\right\}
-\Pi_{\nu}^{00}\left\{  \alpha_{\mu}^{\nu},\Pi_{\nu}^{00}\right\}  \dot
{\Gamma}_{00}^{\nu}-\alpha_{\mu}^{\nu}\chi_{\nu}^{00}+\Pi_{\nu}^{00}\left\{
\alpha_{\mu}^{\nu},\chi_{\gamma}^{00}\right\}  \Gamma_{00}^{\gamma}%
\]

\begin{equation}
-\Pi_{\nu}^{00}\left\{  \alpha_{\mu}^{\nu},H_{c}^{\prime}\right\}  +\beta
_{\mu}^{\gamma}\Pi_{\gamma}^{00}.\label{eqnCAS8}%
\end{equation}
At this stage of the calculation, both coefficient-functions $\alpha_{\mu
}^{\nu}$ and $\beta_{\mu}^{\gamma}$ enter (\ref{eqnCAS8}) while at the
previous step only $\alpha_{\mu}^{\nu}$ was present in (\ref{eqnCAS6}).

The last equation of Castellani algorithm, (\ref{eqnCAS5}), serves to find
unspecified coefficient-functions $\alpha_{\mu}^{\nu}$ and $\beta_{\mu}^{\nu}$
as all terms proportional to the secondary and tertiary constraints must be
identically zero%

\begin{equation}
\left\{  G_{\left(  0\right)  \mu},\dot{\Gamma}_{00}^{\nu}\Pi_{\nu}%
^{00}-\Gamma_{00}^{\nu}\chi_{\nu}^{00}+H_{c}^{\prime}\right\}  =primary.
\label{eqnCAS9}%
\end{equation}

After $\alpha_{\mu}^{\nu}$ and $\beta_{\mu}^{\nu}$ are found, we have the
following generator%

\begin{equation}
G=\varepsilon^{\mu}G_{\left(  0\right)  \mu}+\dot{\varepsilon}^{\mu}G_{\left(
1\right)  \mu}+\ddot{\varepsilon}^{\mu}G_{\left(  2\right)  \mu}=
\label{eqnCAS10}%
\end{equation}

\[
\varepsilon^{\mu}\left(  -\left\{  \chi_{\mu}^{00},\chi_{\nu}^{00}\right\}
\Gamma_{00}^{\nu}+\left\{  \chi_{\mu}^{00},H_{c}^{\prime}\right\}  -\left\{
\alpha_{\mu}^{\nu}\Pi_{\nu}^{00},\dot{\Gamma}_{00}^{\alpha}\Pi_{\alpha}%
^{00}\right\}  +\alpha_{\mu}^{\nu}\chi_{\nu}^{00}+\beta_{\mu}^{\gamma}%
\Pi_{\gamma}^{00}\right)  +\dot{\varepsilon}^{\mu}\left(  -\chi_{\mu}%
^{00}+\alpha_{\mu}^{\nu}\Pi_{\nu}^{00}\right)  +\ddot{\varepsilon}^{\mu}%
\Pi_{\mu}^{00}%
\]
that allows us to find the transformations of fields or combinations of
fields, $F$, using%

\begin{equation}
\delta F=\left\{  G,F\right\}  . \label{eqnCAS11}%
\end{equation}

Here, we again do not specify an $H_{c}^{\prime}$, which can be written in
different ways, as its form depends on a choice of tertiary constraints.
Obviously a particular choice of tertiary constraints can only affect the
transformations through the coefficient-functions. The transformations
generated by the part of the generator (\ref{eqnCAS10}), without
coefficient-functions is%

\begin{equation}
\delta F=\left\{  \varepsilon^{\mu}\left(  -\left\{  \chi_{\mu}^{00},\chi
_{\nu}^{00}\right\}  \Gamma_{00}^{\nu}+\left\{  \chi_{\mu}^{00},H_{c}^{\prime
}\right\}  \right)  -\dot{\varepsilon}^{\mu}\chi_{\mu}^{00}+\ddot{\varepsilon
}^{\mu}\Pi_{\mu}^{00},F\right\}  \label{eqnCAS12}%
\end{equation}
and it produces the same result, whatever combination of tertiary constraints
is considered. The complete restoration of the gauge generator, and especially
the gauge transformations of all fields, is a technically involved problem.
Solutions of second class constraints must be used or we have to go to the
reduced Lagrangian (\ref{eqnAM120}), which corresponds to the reduced
Hamiltonian, and find the momenta in terms of the coordinates from the
equations of motion, as it is described in the HTZ paper \cite{HTZ}. We
restrict our discussion to the first and relatively simple steps of
calculations: to single out a method of restoration that does not lead to contradictions.

Let us start from the derivation of the partial transformations of
$\Gamma_{00}^{\mu}$ to illustrate that the Castellani algorithm is independent
of a choice of tertiary constraints. We will show that it leads to the correct
contributions to the gauge transformations for all fields; and that these
transformations are equivalent to diffeomorphism without any field dependent
or even numerical redefinitions of gauge parameters.

First of all, we write $G_{\left(  0\right)  \mu}$ in components that allow us
to explicitly calculate some PBs%

\[
G_{\left(  0\right)  0}=\Gamma_{00}^{k}\chi_{k}^{00}+\left\{  \chi_{0}%
^{00},H_{c}^{\prime}\right\}  -\Pi_{\nu}^{00}\frac{\delta\alpha_{0}^{\nu}%
}{\delta\Gamma_{00}^{\gamma}}\dot{\Gamma}_{00}^{\gamma}%
\]

\begin{equation}
-\alpha_{0}^{\nu}\chi_{\nu}^{00}+\Pi_{\nu}^{00}\left\{  \alpha_{0}^{\nu}%
,\chi_{\gamma}^{00}\right\}  \Gamma_{00}^{\gamma}-\Pi_{\nu}^{00}\left\{
\alpha_{0}^{\nu},H_{c}^{\prime}\right\}  +\beta_{0}^{\gamma}\Pi_{\gamma}^{00},
\label{eqnCAS15}%
\end{equation}

\[
G_{\left(  0\right)  p}=-\chi_{p}^{00}\Gamma_{00}^{0}+\left\{  \chi_{p}%
^{00},H_{c}^{\prime}\right\}  -\Pi_{\nu}^{00}\frac{\delta\alpha_{p}^{\nu}%
}{\delta\Gamma_{00}^{\gamma}}\dot{\Gamma}_{00}^{\gamma}%
\]

\begin{equation}
-\alpha_{p}^{\nu}\chi_{\nu}^{00}+\Pi_{\nu}^{00}\left\{  \alpha_{p}^{\nu}%
,\chi_{\gamma}^{00}\right\}  \Gamma_{00}^{\gamma}-\Pi_{\nu}^{00}\left\{
\alpha_{p}^{\nu},H_{c}^{\prime}\right\}  +\beta_{p}^{\gamma}\Pi_{\gamma}^{00}.
\label{eqnCAS16}%
\end{equation}

The condition (\ref{eqnCAS9}) gives for (\ref{eqnCAS15})-(\ref{eqnCAS16}) the
following terms, which are not proportional to primary constraints,
\begin{equation}
\left\{  G_{\left(  0\right)  0},H_{T}\right\}  =\chi_{k}^{00}\dot{\Gamma
}_{00}^{k}-\Gamma_{00}^{k}\chi_{k}^{00}\Gamma_{00}^{0}+\Gamma_{00}^{k}\left\{
\chi_{k}^{00},H_{c}^{\prime}\right\}  \label{eqnCAS17}%
\end{equation}

\[
-\left\{  \left\{  \chi_{0}^{00},H_{c}^{\prime}\right\}  ,\chi_{\alpha}%
^{00}\right\}  \Gamma_{00}^{\alpha}+\left\{  \left\{  \chi_{0}^{00}%
,H_{c}^{\prime}\right\}  ,H_{c}^{\prime}\right\}  -\frac{\delta\alpha_{0}%
^{\nu}}{\delta\Gamma_{00}^{\alpha}}\dot{\Gamma}_{00}^{\alpha}\chi_{\nu}^{00}%
\]

\[
-\chi_{\nu}^{00}\frac{\delta\alpha_{0}^{\nu}}{\delta\Gamma_{00}^{\alpha}}%
\dot{\Gamma}_{00}^{\alpha}+\left\{  \alpha_{0}^{\nu}\chi_{\nu}^{00}%
,\chi_{\alpha}^{00}\right\}  \Gamma_{00}^{\alpha}-\alpha_{0}^{\nu}\left\{
\chi_{\nu}^{00},H_{c}^{\prime}\right\}  -\chi_{\nu}^{00}\left\{  \alpha
_{0}^{\nu},H_{c}^{\prime}\right\}
\]

\[
+\left\{  \alpha_{\mu}^{\nu},\chi_{\alpha}^{00}\right\}  \Gamma_{00}^{\alpha
}\chi_{\nu}^{00}-\left\{  \alpha_{\mu}^{\nu},H_{c}^{\prime}\right\}  \chi
_{\nu}^{00}+\beta_{0}^{\gamma}\chi_{\gamma}^{00}=0,
\]

\begin{equation}
\left\{  G_{\left(  0\right)  p},\dot{\Gamma}_{00}^{\alpha}\Pi_{\alpha}%
^{00}-\Gamma_{00}^{\alpha}\chi_{\alpha}^{00}+H_{c}^{\prime}\right\}  =
\label{eqnCAS18}%
\end{equation}

\[
-\chi_{p}^{00}\dot{\Gamma}_{00}^{0}+\left\{  -\chi_{p}^{00}\Gamma_{00}%
^{0},-\Gamma_{00}^{\nu}\chi_{\nu}^{00}\right\}  -\Gamma_{00}^{0}\left\{
\chi_{p}^{00},H_{c}^{\prime}\right\}  -\left\{  \left\{  \chi_{p}^{00}%
,H_{c}^{\prime}\right\}  ,\chi_{\nu}^{00}\right\}  \Gamma_{00}^{\nu}+\left\{
\left\{  \chi_{p}^{00},H_{c}^{\prime}\right\}  ,H_{c}^{\prime}\right\}
\]

\[
-\dot{\Gamma}_{00}^{\alpha}\frac{\delta\alpha_{p}^{\nu}}{\delta\Gamma
_{00}^{\alpha}}\chi_{\nu}^{00}-\chi_{\nu}^{00}\left\{  \alpha_{p}^{\nu}%
,\Pi_{\alpha}^{00}\right\}  \dot{\Gamma}_{00}^{\alpha}+\left\{  \alpha
_{p}^{\nu}\chi_{\nu}^{00},\chi_{\nu}^{00}\right\}  \Gamma_{00}^{\nu}%
-\alpha_{p}^{\nu}\left\{  \chi_{\nu}^{00},H_{c}^{\prime}\right\}  -\chi_{\nu
}^{00}\left\{  \alpha_{p}^{\nu},H_{c}^{\prime}\right\}
\]

\[
+\left\{  \alpha_{p}^{\nu},\chi_{\gamma}^{00}\right\}  \Gamma_{00}^{\gamma
}\chi_{\nu}^{00}-\left\{  \alpha_{p}^{\nu},H_{c}^{\prime}\right\}  \chi_{\nu
}^{00}+\beta_{p}^{\nu}\chi_{\nu}^{00}=0.
\]

The only terms in (\ref{eqnCAS17})-(\ref{eqnCAS18}) that can give
contributions proportional to tertiary constraints (hidden in $H_{c}^{\prime}%
$) are:%

\begin{equation}
+\Gamma_{00}^{k}\left\{  \chi_{k}^{00},H_{c}^{\prime}\right\}  -\left\{
\left\{  \chi_{0}^{00},H_{c}^{\prime}\right\}  ,\chi_{\alpha}^{00}\right\}
\Gamma_{00}^{\alpha}+\left\{  \left\{  \chi_{0}^{00},H_{c}^{\prime}\right\}
,H_{c}^{\prime}\right\}  -\alpha_{0}^{\nu}\left\{  \chi_{\nu}^{00}%
,H_{c}^{\prime}\right\}  =0, \label{eqnCAS19}%
\end{equation}

\begin{equation}
-\Gamma_{00}^{0}\left\{  \chi_{p}^{00},H_{c}^{\prime}\right\}  -\left\{
\left\{  \chi_{p}^{00},H_{c}^{\prime}\right\}  ,\chi_{\nu}^{00}\right\}
\Gamma_{00}^{\nu}+\left\{  \left\{  \chi_{p}^{00},H_{c}^{\prime}\right\}
,H_{c}^{\prime}\right\}  -\alpha_{p}^{\nu}\left\{  \chi_{\nu}^{00}%
,H_{c}^{\prime}\right\}  =0. \label{eqnCAS20}%
\end{equation}

Equations (\ref{eqnCAS19})-(\ref{eqnCAS20}) allow us to find the
coefficient-functions $\alpha_{\mu}^{\nu}$. Note that (\ref{eqnCAS19}%
)-(\ref{eqnCAS20}) are algebraic equations which are written in a general form
and are independent of a particular choice of tertiary constraints. But to
find the coefficient-functions we have to expand the expressions
(\ref{eqnCAS19})-(\ref{eqnCAS20}) and collect terms proportional to the
secondary and tertiary constraints. To do this we have to specify our choice.
A few choices were discussed in Section 4; but let us start from one
particular combination, (\ref{eqnAM74}) with the simplest PBs among
constraints, namely one that leads to zero PBs among secondary with tertiary
constraints: $\left\{  \chi_{\nu}^{00},\tau_{\mu}^{00}\right\}  =0$. The
simplicity of the PBs was also the reason in \cite{G/R} to use these
combinations to prove closure of the Dirac procedure. It was precisely this
choice that led to the conclusion in \cite{Novel} that diffeomorphism
invariance does not follow directly and a field dependent redefinition of
parameters is needed to find the \textquotedblleft
correspondence\textquotedblright\ with diffeomorphism using HTZ ansatz
\cite{HTZ}. So, we specify $H_{c}^{\prime}$ (note that until this moment all
equations were independent of a choice of tertiary constraints) and use the
following form%

\begin{equation}
H_{c}^{\prime}=\frac{1}{h^{00}}\tau_{0}^{00}+\frac{h^{0m}}{h^{00}}\tau
_{m}^{00}+A^{0}\chi_{0}^{00}+A^{m}\chi_{m}^{00}.\label{eqnCAS21}%
\end{equation}
With this $H_{c}^{\prime}$, and keeping only terms that lead to tertiary
constraints (needed to find $\alpha_{\mu}^{\nu}$) from (\ref{eqnCAS19}) and
(\ref{eqnCAS21}), we have%

\[
2\Gamma_{00}^{k}\tau_{k}^{00}+\left(  \frac{1}{h^{00}}\tau_{0}^{00}%
+\frac{h^{0m}}{h^{00}}\tau_{m}^{00}\right)  \Gamma_{00}^{0}-\alpha_{0}%
^{0}\left(  \frac{1}{h^{00}}\tau_{0}^{00}+\frac{h^{0m}}{h^{00}}\tau_{m}%
^{00}\right)  -\alpha_{0}^{k}\tau_{k}^{00}%
\]

\begin{equation}
+\left\{  \frac{1}{h^{00}}\tau_{0}^{00}+\frac{h^{0m}}{h^{00}}\tau_{m}%
^{00}+\left\{  \chi_{0}^{00},A^{0}\chi_{0}^{00}+A^{m}\chi_{m}^{00}\right\}
,H_{c}^{\prime}\right\}  =0. \label{eqnCAS22}%
\end{equation}

Let us, for simplicity, restrict our calculations even further and consider
only the dependence of the coefficient-functions $\alpha_{\mu}^{\nu}$ on
$\Gamma_{00}^{\mu}$. From the first line of (\ref{eqnCAS22}) we can uniquely
find how $\alpha_{0}^{0}$ and $\alpha_{0}^{k}$ depend on $\Gamma_{00}^{\mu}$
(there are no contributions proportional to $\Gamma_{00}^{\mu}$ in the second
line of (\ref{eqnCAS22})). Combining together the terms in the first line of
(\ref{eqnCAS22}) that are proportional to the tertiary constraints we obtain%

\begin{equation}
\left(  2\Gamma_{00}^{k}-\alpha_{0}^{k}+\frac{h^{0k}}{h^{00}}\Gamma_{00}%
^{0}-\frac{h^{0k}}{h^{00}}\alpha_{0}^{0}\right)  \tau_{k}^{00}+\left(
\frac{1}{h^{00}}\Gamma_{00}^{0}-\frac{1}{h^{00}}\alpha_{0}^{0}\right)
\tau_{0}^{00}=0.\label{eqnCAS23}%
\end{equation}
The second bracket of (\ref{eqnCAS23}) gives%

\begin{equation}
\alpha_{0}^{0}\left(  \Gamma\right)  =\Gamma_{00}^{0}.\label{eqnCAS24}%
\end{equation}
Using this result and the first bracket of (\ref{eqnCAS23}), we obtain%

\begin{equation}
\text{\ }\alpha_{0}^{k}\left(  \Gamma\right)  =2\Gamma_{00}^{k}%
.\label{eqnCAS25}%
\end{equation}
Similarly, from (\ref{eqnCAS20}) we find%

\begin{equation}
\alpha_{p}^{0}\left(  \Gamma\right)  =0\text{, \ \ \ }\alpha_{p}^{k}\left(
\Gamma\right)  =-\delta_{p}^{k}\Gamma_{00}^{0}. \label{eqnCAS26}%
\end{equation}

With these results we can also calculate the contributions to the
coefficient-function $\beta_{\mu}^{\nu}$ from the corresponding expressions
that are proportional to the secondary constraints. In (\ref{eqnCAS17}%
)-(\ref{eqnCAS18}), keeping only terms with $\beta_{\mu}^{\nu}$ and those with
temporal derivatives of $\dot{\Gamma}_{00}^{\mu}$, we have
\begin{equation}
\chi_{k}^{00}\dot{\Gamma}_{00}^{k}-\dot{\Gamma}_{00}^{\alpha}\frac
{\delta\alpha_{0}^{\nu}}{\delta\Gamma_{00}^{\alpha}}\chi_{\nu}^{00}-\chi_{\nu
}^{00}\frac{\delta\alpha_{0}^{\nu}}{\delta\Gamma_{00}^{\alpha}}\dot{\Gamma
}_{00}^{\alpha}+\beta_{0}^{0}\chi_{0}^{00}+\beta_{0}^{p}\chi_{p}%
^{00}=0,\label{eqnCAS27}%
\end{equation}

\begin{equation}
-\chi_{p}^{00}\dot{\Gamma}_{00}^{0}-\frac{\delta\alpha_{p}^{\nu}}{\delta
\Gamma_{00}^{\alpha}}\dot{\Gamma}_{00}^{\alpha}\chi_{\nu}^{00}-\chi_{\nu}%
^{00}\frac{\delta\alpha_{p}^{\nu}}{\delta\Gamma_{00}^{\alpha}}\dot{\Gamma
}_{00}^{\alpha}+\beta_{p}^{0}\chi_{0}^{00}+\beta_{p}^{k}\chi_{k}%
^{00}=0,\label{eqnCAS28}%
\end{equation}
which after substitution of (\ref{eqnCAS24})-(\ref{eqnCAS26}) gives%

\begin{equation}
\beta_{0}^{0}\left(  \dot{\Gamma}\right)  =2\dot{\Gamma}_{00}^{0}\text{,
\ \ }\beta_{0}^{k}\left(  \dot{\Gamma}\right)  =3\dot{\Gamma}_{00}^{k}~,
\label{eqnCAS29}%
\end{equation}

\begin{equation}
\beta_{p}^{0}\left(  \dot{\Gamma}\right)  =0\text{, \ \ \ }\beta_{p}%
^{k}\left(  \dot{\Gamma}\right)  =-\delta_{p}^{k}\dot{\Gamma}_{00}^{0}~.
\label{eqnCAS30}%
\end{equation}

Let us check the consistency of these simple partial results and consider
variation of $\delta\Gamma_{00}^{\mu}$%

\begin{equation}
\delta\Gamma_{00}^{\mu}=\left\{  G,\Gamma_{00}^{\mu}\right\}  .
\label{eqnCAS31}%
\end{equation}

For the relevant part of the generator (terms proportional to primary
constraints $\Pi_{\nu}^{00}$ and found in (\ref{eqnCAS24})-(\ref{eqnCAS26})
and (\ref{eqnCAS29})-(\ref{eqnCAS30}) $\Gamma$-dependent parts of $\alpha
_{\mu}^{\nu}$ and $\beta_{\mu}^{\nu}$), we have%

\begin{equation}
G=\varepsilon^{0}\left(  \dot{\Gamma}_{00}^{0}\Pi_{0}^{00}+\dot{\Gamma}%
_{00}^{k}\Pi_{k}^{00}\right)  +\dot{\varepsilon}^{0}\Gamma_{00}^{0}\Pi
_{0}^{00}+\dot{\varepsilon}^{0}2\Gamma_{00}^{k}\Pi_{k}^{00}-\dot{\varepsilon
}^{p}\Gamma_{00}^{0}\Pi_{p}^{00}+\ddot{\varepsilon}^{0}\Pi_{0}^{00}%
+\ddot{\varepsilon}^{p}\Pi_{p}^{00}.\label{eqnCAS32}%
\end{equation}
This part of the generator gives%

\begin{equation}
\delta\Gamma_{00}^{0}=-\varepsilon^{0}\dot{\Gamma}_{00}^{0}-\dot{\varepsilon
}^{0}\Gamma_{00}^{0}-\ddot{\varepsilon}^{0}, \label{eqnCAS33}%
\end{equation}

\begin{equation}
\delta\Gamma_{00}^{p}=-\varepsilon^{0}\dot{\Gamma}_{00}^{p}-2\dot{\varepsilon
}^{0}\Gamma_{00}^{p}+\dot{\varepsilon}^{p}\Gamma_{00}^{0}-\ddot{\varepsilon
}^{p}. \label{eqnCAS34}%
\end{equation}

We have to compare these partial results with the well-known transformations
of the components, $\Gamma_{00}^{\mu}$, under diffeomorphism invariance. Using
the general expression for the transformations of affine-connections under diffeomorphism,%

\begin{equation}
\delta_{diff}\Gamma_{\mu\nu}^{\lambda}=-\varepsilon_{,\mu\nu}^{\lambda}%
+\Gamma_{\mu\nu}^{\rho}\varepsilon_{,\rho}^{\lambda}-\varepsilon^{\rho}%
\Gamma_{\mu\nu,\rho}^{\lambda}-\Gamma_{\mu\rho}^{\lambda}\varepsilon_{,\nu
}^{\rho}-\Gamma_{\nu\rho}^{\lambda}\varepsilon_{,\mu}^{\rho}~,\label{eqnCAS35}%
\end{equation}
for two particular components of (\ref{eqnCAS33}) and (\ref{eqnCAS34}), we obtain%

\begin{equation}
\delta_{diff}\Gamma_{00}^{0}=\mathbf{-\varepsilon}_{,00}^{0}-\Gamma_{00}%
^{0}\partial_{0}\varepsilon^{0}-\varepsilon^{0}\partial_{0}\Gamma_{00}%
^{0}+\Gamma_{00}^{k}\partial_{k}\varepsilon^{0}-\varepsilon^{k}\partial
_{k}\Gamma_{00}^{0}-2\left(  \Sigma_{0k}+\Gamma_{km}^{m}\right)  \partial
_{0}\varepsilon^{k}, \label{eqnCAS36}%
\end{equation}

\[
\delta_{diff}\Gamma_{00}^{b}=-\varepsilon_{,00}^{b}+\Gamma_{00}^{0}%
\partial_{0}\varepsilon^{b}+\Gamma_{00}^{k}\partial_{k}\varepsilon^{b}%
\]

\begin{equation}
-\varepsilon^{0}\partial_{0}\Gamma_{00}^{b}-\varepsilon^{k}\partial_{k}%
\Gamma_{00}^{b}-2\Gamma_{00}^{b}\partial_{0}\varepsilon^{0}-2\Sigma_{0k}%
^{b}\partial_{0}\varepsilon^{k}+\frac{2}{D-1}\Sigma_{00}\partial
_{0}\varepsilon^{b}. \label{eqnCAS37}%
\end{equation}

Here we used the redefinitions (\ref{eqnAM6}), (\ref{eqnAM17}) and
(\ref{eqnAM19}) (from last two it follows that $\pi_{\mu\nu}=\Sigma_{\mu\nu}%
$). We do not perform such substitutions for $\Sigma_{0k}^{m}$ and
$\Gamma_{km}^{m}$ as they do not affect the parts of the transformations that
we shall compare with our partial result (\ref{eqnCAS33})-(\ref{eqnCAS34})
(they do not have terms that depend on primary variables $\Gamma_{00}^{\mu}$).

Comparing (\ref{eqnCAS33})-(\ref{eqnCAS34}) and (\ref{eqnCAS36}%
)-(\ref{eqnCAS37}), we can see that all terms with temporal derivatives of the
parameter $\varepsilon^{\mu}$ (up to the second order which requires tertiary
constraints, as we argued in \cite{KK, KKM}) and temporal derivatives of
primary fields, in the part of the generator (\ref{eqnCAS33})-(\ref{eqnCAS34})
we considered, coincide with diffeomorphism transformation (\ref{eqnCAS36}%
)-(\ref{eqnCAS37}) without any need for field dependent redefinition of the
gauge parameters.

If the transformations for some fields that were calculated using the same
generator require a field dependent redefinition of gauge parameters to
\textquotedblleft correspond\textquotedblright\ to diffeomorphism
transformations, then such a redefinition will obviously destroy
transformations (\ref{eqnCAS33})-(\ref{eqnCAS34}). If that happens then either
the Hamiltonian formulation is not correct (ordinary mistakes are possible in
such long calculations or non-canonical change of variables were performed) or
the method of restoration of the gauge invariance is incorrect or, perhaps,
sensitive to a choice of constraints; in that case such a method cannot be
called an algorithm.

There are a few more contributions to the transformations of different fields
that we can obtain using the part of the generator (\ref{eqnCAS10}), even
without knowledge of the coefficient-functions $\alpha_{\mu}^{\nu}$ and
$\beta_{\mu}^{\nu}$. This part of (\ref{eqnCAS10}), after substitution of
expressions for secondary constraints (\ref{eqnAM24}) and (\ref{eqnAM25}), is%

\begin{equation}
G^{\prime}=-\dot{\varepsilon}^{\mu}\chi_{\mu}^{00}=-\dot{\varepsilon}%
^{0}\left(  -h_{,k}^{0k}+h^{00}\pi_{00}-h^{km}\pi_{km}\right)  -\dot
{\varepsilon}^{k}\left(  h_{,k}^{00}+2h^{00}\pi_{0k}+2h^{m0}\pi_{km}\right)
.\label{eqnCAS50}%
\end{equation}
It allows us to find some contributions to the transformations of
$h^{\alpha\beta}$%

\begin{equation}
\delta h^{\alpha\beta}=\left\{  G^{\prime},h^{\alpha\beta}\right\}
\label{eqnCAS51}%
\end{equation}
that for different components leads to:%

\begin{equation}
\delta h^{pq}=\left\{  -\dot{\varepsilon}^{0}\left(  -h^{km}\pi_{km}\right)
-\dot{\varepsilon}^{k}\left(  2h^{m0}\pi_{km}\right)  ,h^{pq}\right\}
=-\dot{\varepsilon}^{0}h^{pq}+\dot{\varepsilon}^{p}h^{q0}+\dot{\varepsilon
}^{q}h^{p0}, \label{eqnCAS52}%
\end{equation}%

\begin{equation}
\delta h^{0p}=\left\{  -\dot{\varepsilon}^{k}\left(  2h^{00}\pi_{0k}\right)
,h^{0p}\right\}  =\dot{\varepsilon}^{p}h^{00}, \label{eqnCAS53}%
\end{equation}

\begin{equation}
\delta h^{00}=\left\{  -\dot{\varepsilon}^{0}\left(  h^{00}\pi_{00}\right)
,h^{00}\right\}  =\dot{\varepsilon}^{0}h^{00}. \label{eqnCAS54}%
\end{equation}

Using the known transformations for $g^{\alpha\beta}$ under diffeomorphism
invariance and the definition of $h^{\alpha\beta}$ (\ref{eqnAM19}), one obtains%

\begin{equation}
\delta_{diff}h^{\mu\nu}=h^{\mu\lambda}\varepsilon_{,\lambda}^{\nu}%
+h^{\nu\lambda}\varepsilon_{,\lambda}^{\mu}-\left(  h^{\mu\nu}\varepsilon
^{\lambda}\right)  _{,\lambda}\label{eqnCAS55}%
\end{equation}
which for different components of $h^{\alpha\beta}$ gives:%

\begin{equation}
\delta_{diff}h^{00}=h^{00}\varepsilon_{,0}^{0}+2h^{0p}\varepsilon_{,p}%
^{0}-\varepsilon^{0}h_{,0}^{00}-\varepsilon^{p}h_{,p}^{00}-h^{00}%
\varepsilon_{,p}^{p}~, \label{eqnCAS56}%
\end{equation}

\begin{equation}
\delta_{diff}h^{0p}=h^{00}\varepsilon_{,0}^{p}+h^{0m}\varepsilon_{,m}%
^{p}+h^{pm}\varepsilon_{,m}^{0}-\varepsilon^{0}h_{,0}^{0p}-h^{0p}%
\varepsilon_{,m}^{m}-\varepsilon^{m}h_{,m}^{0p}~, \label{eqnCAS57}%
\end{equation}

\begin{equation}
\delta_{diff}h^{pq}=h^{p0}\varepsilon_{,0}^{q}+h^{q0}\varepsilon_{,0}%
^{p}+h^{pm}\varepsilon_{,m}^{q}+h^{qm}\varepsilon_{,m}^{p}-h^{pq}%
\varepsilon_{,0}^{0}-h^{pq}\varepsilon_{,m}^{m}-\varepsilon^{0}h_{,0}%
^{pq}-\varepsilon^{m}h_{,m}^{pq}~. \label{eqnCAS58}%
\end{equation}

Similar to transformations of $\Gamma_{00}^{\mu}$ (\ref{eqnCAS33}%
)-(\ref{eqnCAS34}), all found in the (\ref{eqnCAS52})-(\ref{eqnCAS54})
contributions, are present in (\ref{eqnCAS56})-(\ref{eqnCAS58}) and cover all
terms with temporal derivatives of the gauge parameter $\varepsilon^{\mu}$.
So, field dependent redefinition of gauge parameters is also not needed.

The same part of the generator (\ref{eqnCAS50}) can be used to find partial
transformations of $\pi_{\alpha\beta}$%

\begin{equation}
\delta\pi_{\alpha\beta}=\left\{  G^{\prime},\pi_{\alpha\beta}\right\}
,\label{eqnCAS59}%
\end{equation}
which gives for components%

\begin{equation}
\delta\pi_{pq}=\left\{  \dot{\varepsilon}^{0}h^{km}\pi_{km},\pi_{pq}\right\}
=\dot{\varepsilon}^{0}\pi_{pq}, \label{eqnCAS60}%
\end{equation}

\begin{equation}
\delta\pi_{0p}=\left\{  \dot{\varepsilon}^{0}h_{,k}^{0k}-\dot{\varepsilon}%
^{k}2h^{m0}\pi_{km},\pi_{0p}\right\}  =-\frac{1}{2}\dot{\varepsilon}_{,p}%
^{0}-\dot{\varepsilon}^{k}\pi_{kp}, \label{eqnCAS61}%
\end{equation}

\begin{equation}
\delta\pi_{00}=\left\{  -\dot{\varepsilon}^{0}\left(  h^{00}\pi_{00}\right)
-\dot{\varepsilon}^{k}\left(  h_{,k}^{00}+2h^{00}\pi_{0k}\right)  ,\pi
_{00}\right\}  =-\dot{\varepsilon}^{0}\pi_{00}+\dot{\varepsilon}_{,k}%
^{k}-2\dot{\varepsilon}^{k}\pi_{0k}. \label{eqnCAS62}%
\end{equation}

Returning to our original redefinitions, (\ref{eqnAM17}) and (\ref{eqnAM19}),
we find that $\pi_{\alpha\beta}=\Sigma_{\alpha\beta}$. \ Now we can compare
(\ref{eqnCAS60})-(\ref{eqnCAS62}) with the transformations of $\pi
_{\alpha\beta}$ under diffeomorphism by using%

\begin{equation}
\delta_{diff}\pi_{\alpha\beta}=\delta_{diff}\Sigma_{\alpha\beta}%
\label{eqnCAS63}%
\end{equation}
and redefinition (\ref{eqnAM6}), which allows us to express $\Sigma
_{\alpha\beta}$ in terms of $\Gamma_{\alpha\beta}^{\mu}$ (transformation of
$\Gamma_{\alpha\beta}^{\mu}$ under diffeomorphism is known (\ref{eqnCAS35})).
From (\ref{eqnCAS63}), for the components of $\pi_{\alpha\beta}$, we obtain:%

\[
\delta_{diff}\pi_{pq}=-\varepsilon_{,pq}^{0}+\pi_{pq}\varepsilon_{,0}%
^{0}+\Gamma_{pq}^{m}\varepsilon_{,m}^{0}-\varepsilon^{0}\pi_{pq,0}%
\]

\begin{equation}
-\varepsilon^{m}\pi_{pq,m}-\left(  2\pi_{p0}+\Gamma_{pm}^{m}\right)
\varepsilon_{,q}^{0}-\pi_{pm}\varepsilon_{,q}^{m}-\left(  2\pi_{q0}%
+\Gamma_{qm}^{m}\right)  \varepsilon_{,p}^{0}-\pi_{qm}\varepsilon_{,p}^{m}~,
\label{eqnCAS64}%
\end{equation}

\[
\delta_{diff}\pi_{0p}=-\frac{1}{2}\varepsilon_{,p0}^{0}-\pi_{pm}%
\varepsilon_{,0}^{m}+\Sigma_{0p}^{m}\varepsilon_{,m}^{0}%
\]

\begin{equation}
-\frac{1}{2}\frac{D+1}{D-1}\pi_{00}\varepsilon_{,p}^{0}-\varepsilon^{0}%
\pi_{p0,0}-\varepsilon^{m}\pi_{p0,m}-\frac{1}{2}\Gamma_{00}^{0}\varepsilon
_{,p}^{0}-\pi_{m0}\varepsilon_{,p}^{m}+\frac{1}{2}\varepsilon_{,pm}^{m}~,
\label{eqnCAS65}%
\end{equation}

\begin{equation}
\delta_{diff}\pi_{00}=\varepsilon_{,0m}^{m}-2\pi_{m0}\varepsilon_{,0}%
^{m}-\varepsilon^{0}\pi_{00,0}-\varepsilon^{k}\pi_{00,k}+\Gamma_{00}%
^{m}\varepsilon_{,m}^{0}-\pi_{00}\varepsilon_{,0}^{0}~. \label{eqnCAS66}%
\end{equation}

Here, as in (\ref{eqnCAS36})-(\ref{eqnCAS37}), the solutions for $\Gamma
_{pq}^{m}$ and $\Sigma_{0p}^{m}$ must also be substituted (but it is not
needed at this stage). Compare (\ref{eqnCAS60})-(\ref{eqnCAS62}) with the
diffeomorphism transformation (\ref{eqnCAS64})-(\ref{eqnCAS66}); we see that
again all terms with temporal derivatives of the gauge parameter
$\varepsilon^{\mu}$ are exactly the same, including two contributions with
mixed spatio-temporal derivatives.

Note also that the last two transformations, (\ref{eqnCAS56})-(\ref{eqnCAS58})
and (\ref{eqnCAS60})-(\ref{eqnCAS62}), follow just from an explicit form of
the secondary constraints and obviously cannot be affected (in the Castellani
procedure) by a choice of tertiary constraints. The transformations
(\ref{eqnCAS33})-(\ref{eqnCAS34}) were obtained using the explicit form of
coefficient-functions for which a particular choice of tertiary constraints
(\ref{eqnCAS21}) was used; and there is possibility that different choices can
affect the result. Of course, from physical point of view, this is impossible
and if such happens, it would be an indication of problems with a method of
the restoration of gauge invariance. So, let us consider the effect of a
different choice of constraints; e.g. one which we discussed in Section 4. To
be explicit, let us consider (\ref{eqnAM110}) for which the Hamiltonian takes
the form%

\begin{equation}
H_{c}^{\prime\prime}=\tilde{\tau}_{0}^{00}+A^{0}\chi_{0}^{00}+A^{m}\chi
_{m}^{00}.\label{eqnCAS68}%
\end{equation}
For this choice, contribution (\ref{eqnCAS23}) is modified%

\begin{equation}
+2\Gamma_{00}^{k}\tau_{k}^{00}+\tilde{\tau}_{0}^{00}\Gamma_{00}^{0}-\alpha
_{0}^{0}\tilde{\tau}_{0}^{00}-\alpha_{0}^{k}\tau_{k}^{00};\label{eqnCAS69}%
\end{equation}
but solutions for coefficient-functions are the same as those given by
(\ref{eqnCAS24})-(\ref{eqnCAS25}). The remaining contributions calculated
earlier, for the original choice of constraints, are also the same. One can
easily check another choice that we considered in Section 4, (\ref{eqnAM101}),
and verify that it also preserves the parts of coefficient-functions
proportional to $\Gamma_{00}^{\mu}$. These particular examples illustrate the
general statement made in Section 4 that a choice of linear combinations of
constraints (or their PBs algebra) should not affect physical results (should
not, in particular, affect gauge invariance).

The above examples of partial restoration of gauge invariance using the
Castellani procedure, which were considered, lead to partial contributions to
the transformations of \textit{all fields} presented in the reduced
Hamiltonian of affine-metric formulation of GR. All contributions found are
exactly the same as the corresponding terms of diffeomorphism transformations;
and there is no need for a field dependent redefinition of gauge parameters.
Moreover, all these contributions are independent of a choice of tertiary
constraints, as it should be, if we respect the concept that the gauge
invariance is a unique characteristic of a gauge invariant theory. Of course,
to make the final conclusion, full calculations must be performed and all
contributions have to be found. These calculations are straightforward, but
quite laborious; the results will be reported elsewhere. Our main goal here is
to find the reason for the contradictory result of \cite{Novel} about
necessity to have \textquotedblleft correspondence\textquotedblright\ with
diffeomorphism invariance. Now, after we show that diffeomorphism invariance
follows directly from the Castellani procedure; but when used in \cite{Novel}
the HTZ ansatz does not give correct transformations. One can draw the
conclusion that the HTZ ansatz cannot be considered to be an algorithm for the
restoration of the gauge invariance and the need of field dependent
redefinitions found in \cite{Novel} is just an artifact of the method used. In
next Section we explicitly demonstrate the failure of HTZ and discuss the
possible reasons for this failure.

\section{Henneaux-Teitelboim-Zanelli ansatz}

In this Section we turn our attention to the HTZ approach \cite{HTZ}. This
approach was used in \cite{Novel} where the gauge invariance obtained for the
first order Einstein-Hilbert action differs from the results of the previous
Section where we follow the Castellani procedure. According to \cite{Novel},
the gauge invariance of first order formulation of GR is not a
four-diffeomorphism and a field dependent redefinition of gauge parameters is
needed to \textquotedblleft correspond to diffeomorphism
invariance\textquotedblright.

The appearance of such a \textquotedblleft correspondence\textquotedblright%
\ in \cite{Novel} cannot be explained by a non-canonical change of variables,
as it was in the case of ADM model \cite{Myths}. We have the Hamiltonian of
the same theory and one method, Castellani (see previous Section), leads to
four-diffeomorphism and another, HTZ (see \cite{Novel}), does not. This is
clearly a deficiency of HTZ approach and this is exactly what we want to
investigate and discuss in this Section.

There are two major differences between the HTZ approach and the Castellani
algorithm. The first one is the use of a so-called extended formalism: instead
of the total Hamiltonian (as in Castellani's) the extended formalism is the
starting point of the HTZ approach, where all primary constraints are included
in the Hamiltonian with Lagrange multipliers. The second one is so-called HTZ
ansatz - a generator of the gauge transformation is assumed to be a linear
combination of all first class constraints and all of them enter explicitly
\cite{HTZ}, contrary to Castellani's generator (\ref{eqnCAS2})-(\ref{eqnCAS5})
where only primary first class constraints explicitly enter a generator.
Moreover, the HTZ iterative procedure \cite{HTZ} to find field dependent
coefficients in front of the constraints (see (\ref{eqnHTZ1})), is started
from the end of the constraint chains (e.g. start from tertiary constraints
for the first order affine-metric GR).

The question of a \textquotedblleft total versus extended\textquotedblright%
\ Hamiltonian will not be discussed here. We only would like to mention that
in the particular application of the HTZ approach the \textquotedblleft gauge
fixing\textquotedblright\ of the Lagrange multipliers of non-primary first
class constraints is used. This effectively converts the extended Hamiltonian
into a total Hamiltonian (see Eq. (4.1) of \cite{HTZ}). Moreover, different
variations of the HTZ method were developed; and some of them are based
entirely on the total Hamiltonian, as in Castellani procedure (e.g.
\cite{BRR}) without even mentioning the extended Hamiltonian; but they lead to
the same results as HTZ approach. So, our interest is the HTZ ansatz, which is
also the essential part of all modifications of this approach (e.g. see Eq.
(5) of \cite{BRR}). In \cite{Novel} the gauge transformations were obtained
\textquotedblleft using a method very similar to the method of
HTZ\textquotedblright. We will consider the HTZ ansatz following the original
paper \cite{HTZ} (the reprint of this paper can also be found in Sections
3.2-3.3 of the book \cite{HTbook}).

According to the HTZ ansatz \cite{HTZ} for a system with first class
constraints, the generator is simply a linear combination of all first class
constraints. As in the Castellani algorithm, all second class constraints
should be eliminated in the preliminary step and the Dirac brackets should be
calculated \cite{HTZ}; this is exactly what we did in Section 2. In the case
of the Hamiltonian with tertiary constraints, the HTZ generator is given by
(see Eq. (4.2a) of \cite{HTZ})%

\begin{equation}
G=a^{\mu_{1}}\phi_{\mu_{1}}+a^{\mu_{2}}\phi_{\mu_{2}}+a^{\mu_{3}}\phi_{\mu
_{3}}~. \label{eqnHTZ1}%
\end{equation}

We slightly adjust the HTZ notation to make it more transparent for a
covariant theory. $\phi_{\mu_{i}}$ are the first class constraints of
different generations $i=1,2,3$ (primary, secondary, and tertiary) and
$a^{\mu_{i}}$ are functions of the canonical variables and inexpressible
velocities ($\dot{\Gamma}_{00}^{\nu}$ in the considered formulation) that are
iteratively defined from equation (see Eq. (4.2b) of \cite{HTZ} for $i\geq2$)%

\begin{equation}
\frac{Da^{\mu_{i}}}{Dt}+\left\{  a^{\mu_{i}},H_{c}\right\}  +\dot{\Gamma}%
^{\nu_{1}}\left\{  a^{\mu_{i}},\phi_{\nu_{1}}\right\}  -\sum_{j\geq i-1}%
a^{\nu_{j}}V_{\nu_{j}}^{~\mu_{i}}-\dot{\Gamma}^{\nu_{1}}\sum_{j\geq i}%
a^{\beta_{j}}C_{\nu_{1}\beta_{j}}^{\quad\mu_{i}}=0\label{eqnHTZ2}%
\end{equation}
where $C_{\nu_{1}\beta_{j}}^{\quad\mu_{i}}$ and $V_{\nu_{j}}^{~\mu_{i}}$ are
structure functions in the PBs of the primary constraints with the rest of
constraints ($C_{\nu_{1}\beta_{j}}^{\quad\mu_{i}}$) and in PB of any
constraints with the canonical Hamiltonian ($V_{\nu_{j}}^{~\mu_{i}}$)%

\begin{equation}
\left\{  \phi_{\mu_{1}},\phi_{\nu_{s}}\right\}  =\sum_{i\leq s}C_{\mu_{1}%
\nu_{s}}^{\quad\mu_{i}}\phi_{\mu_{i}}~, \label{eqnHTZ3}%
\end{equation}

\begin{equation}
\left\{  H_{c},\phi_{\mu_{s}}\right\}  =\sum_{i\leq s+1}V_{\mu_{s}}^{~\nu_{i}%
}\phi_{\nu_{i}}~,\label{eqnHTZ4}%
\end{equation}
$\frac{D}{Dt}$ is a short notation for (see Eq. (3.4c) of \cite{HTZ})%

\begin{equation}
\frac{D}{Dt}=\frac{\partial}{\partial t}+\dot{\Gamma}_{00}^{\nu}\frac
{\partial}{\partial\Gamma_{00}^{\nu}}+\ddot{\Gamma}_{00}^{\nu}\frac{\partial
}{\partial\dot{\Gamma}_{00}^{\nu}}+... \label{eqnHTZ5}%
\end{equation}

According to \cite{HTZ}, one has to start from the equation of the highest
order and \textquotedblleft without loss of generality\textquotedblright\ take
$a^{\mu_{3}}=\varepsilon^{\mu}$, i.e. functions (gauge parameters), which are
independent of the canonical variables that leads for $i=3$ to%

\begin{equation}
\frac{D\varepsilon^{\mu}}{Dt}+\left\{  \varepsilon^{\mu},H_{c}\right\}
+\dot{\Gamma}^{\nu_{1}}\left\{  \varepsilon^{\mu},\phi_{\nu_{1}}\right\}
-\sum_{j\geq2}a^{\nu_{j}}V_{\nu_{j}}^{~\mu_{3}}-\dot{\Gamma}^{\nu_{1}}%
\sum_{j\geq i}a^{\beta_{j}}C_{\nu_{1}\beta_{j}}^{\quad\mu_{i}}=0.
\label{eqnHTZ6}%
\end{equation}

We would like to emphasize that the independence of the gauge parameters of
the canonical variables is the starting point of this iterative procedure; and
without this assumption one will face the problem of solving variational
equations instead of algebraic ones. Despite the computational problems with
field dependent parameters, there are no \textit{a priori} criteria of what
possible dependence should be assumed. Let us, as the authors of \cite{HTZ}
suggested and as it was done in \cite{Novel} (see Eq. (145) of \cite{Novel}),
take $a^{\mu_{3}}\equiv\varepsilon^{\mu}$; i.e. as independent of the
phase-space variables function of spacetime coordinates.

This equation, (\ref{eqnHTZ6}), can be simplified as the result of
independence of $\varepsilon^{\mu}$ of the canonical variables: $\frac{D}%
{Dt}=\frac{\partial}{\partial t}$; and both PBs in (\ref{eqnHTZ6}) are zero.
Because the HTZ ansatz (\ref{eqnHTZ1}) and structure functions (\ref{eqnHTZ3}%
)-(\ref{eqnHTZ4}) explicitly depend on a choice of constraints, we have to
specify our choice from outset (note that in previous Section, using the
Castellani algorithm, we were able to perform some calculations without
referring to the explicit form of the tertiary constraints). Partial
restoration of gauge invariance using the HTZ approach for first order GR was
discussed in \cite{Novel} for a particular choice of tertiary constraints. So,
we also consider the same combinations, (\ref{eqnAM70}) and (\ref{eqnAM72})
(the same choice was used in previous Section which led directly to
four-diffeomorphism when the Castellani procedure was applied). For these
constraints the corresponding canonical Hamiltonian is%

\begin{equation}
H_{c}=-\Gamma_{00}^{\nu}\chi_{\nu}^{00}+\frac{1}{h^{00}}\tau_{0}^{00}%
+\frac{h^{0m}}{h^{00}}\tau_{m}^{00}+A^{0}\chi_{0}^{00}+A^{m}\chi_{m}^{00}.
\label{eqnHTZ7}%
\end{equation}

The PBs among the chosen constraints are the simplest, $\left\{  \chi_{\nu
}^{00},\tau_{\mu}^{00}\right\}  =0$, and primary constraints have zero PBs
with all secondary and tertiary constraints making all $C_{\nu_{1}\beta_{j}%
}^{\quad\mu_{i}}=0$ (see (\ref{eqnHTZ3})), which leads to a simple form of
equation (\ref{eqnHTZ6})%

\begin{equation}
\frac{\partial\varepsilon^{\mu}}{\partial t}-a^{\nu_{2}}V_{\nu_{2}}^{~\mu_{3}%
}-\varepsilon^{\nu}V_{\nu_{3}}^{~\mu_{3}}=0.\label{eqnHTZ8}%
\end{equation}
The explicit form of the structure functions $V_{\nu_{2}}^{~\mu_{3}}$
(\ref{eqnHTZ4}) for this choice of tertiary constraints is%

\begin{equation}
V_{0_{2}}^{~0_{3}}=-\frac{1}{h^{00}},\quad V_{0_{2}}^{~m_{3}}=-\frac{h^{0m}%
}{h^{00}},\quad V_{p_{2}}^{~m_{3}}=\delta_{p}^{m},\quad V_{p_{2}}^{~0_{3}%
}=0,\label{eqnHTZ9}%
\end{equation}
which allows to solve equation (\ref{eqnHTZ8}), which is algebraic with
respect to $a^{\nu_{2}}$%

\begin{equation}
a^{0_{2}}=-h^{00}\frac{\partial\varepsilon^{0}}{\partial t}+h^{00}%
\varepsilon^{\nu}V_{\nu_{3}}^{~0_{3}}, \label{eqnHTZ10}%
\end{equation}

\begin{equation}
a^{p_{2}}=\frac{\partial\varepsilon^{p}}{\partial t}-h^{0p}\frac
{\partial\varepsilon^{0}}{\partial t}+h^{0p}\varepsilon^{\nu}V_{\nu_{3}%
}^{~0_{3}}-\varepsilon^{\nu}V_{\nu_{3}}^{~p_{3}}. \label{eqnHTZ11}%
\end{equation}

Using (\ref{eqnAM91})-(\ref{eqnAM94}), the remaining structure functions
$V_{\nu_{3}}^{~\mu_{3}}$ can be easily found (note that they have only
dependence on canonical variables $h^{0\alpha}$), which makes $a^{\mu_{2}}$ to
be functions of the gauge parameters and variables $h^{0\alpha}$. This
immediately allows us to find the partial transformations of $h^{\alpha\beta}$
using part of the generator (\ref{eqnHTZ1}) with  $a^{\mu_{1}}$ and
$a^{\mu_{2}}$ (already found)%

\begin{equation}
\delta h^{\alpha\beta}=\left\{  h^{\alpha\beta},G\right\}  =\left\{
h^{\alpha\beta},...+a^{\mu_{2}}\chi_{\mu}^{00}+\varepsilon^{\mu}\tau_{\mu
}^{00}\right\}  =\frac{\delta}{\delta\pi^{\alpha\beta}}\left(  ...+a^{\mu_{2}%
}\chi_{\mu}^{00}+\varepsilon^{\mu}\tau_{\mu}^{00}\right)  . \label{eqnHTZ12}%
\end{equation}

This is especially simple for the component $h^{0\alpha}$, as the
corresponding momenta are present only in the secondary constraints, which
lead to%

\begin{equation}
\delta h^{00}=a^{\mu_{2}}\frac{\delta\chi_{\mu}^{00}}{\delta\pi^{00}}%
=a^{0_{2}}h^{00}, \label{eqnHTZ13}%
\end{equation}

\begin{equation}
\delta h^{0p}=a^{\mu_{2}}\frac{\delta\chi_{\mu}^{00}}{\delta\pi^{0p}}%
=a^{p_{2}}h^{00}.\label{eqnHTZ14}%
\end{equation}
After substitution of explicit form of $V_{\nu_{3}}^{~\mu_{3}}$ into
(\ref{eqnHTZ10})-(\ref{eqnHTZ11}), we obtain:%

\begin{equation}
a^{0_{2}}=-h^{00}\left[  \frac{\partial\varepsilon^{0}}{\partial
t}+\varepsilon_{,m}^{0}\frac{h^{0m}}{h^{00}}-\varepsilon^{0}\left(
\frac{h^{0m}}{h^{00}}\right)  _{,m}-\varepsilon^{m}\left(  \frac{1}{h^{00}%
}\right)  _{,m}+\varepsilon_{,m}^{m}\frac{1}{h^{00}}\right]  ,
\label{eqnHTZ15}%
\end{equation}

\begin{equation}
a^{p_{2}}=\dot{\varepsilon}^{p}-h^{0p}\dot{\varepsilon}^{0}+h^{00}%
e^{pm}\left(  \frac{1}{h^{00}}\right)  _{,m}\varepsilon^{0}-e^{pm}%
\varepsilon_{,m}^{0}- \label{eqnHTZ16}%
\end{equation}

\[
-\left(  \frac{h^{0p}}{h^{00}}\right)  _{,m}\varepsilon^{m}+\frac{h^{0m}%
}{h^{00}}\varepsilon_{,m}^{p}-\frac{h^{0p}}{h^{00}}\varepsilon_{,m}^{m}%
+h^{0p}\left(  \frac{1}{h^{00}}\right)  _{,m}\varepsilon^{m}+h^{0p}\left(
\frac{h^{0m}}{h^{00}}\right)  _{,m}\varepsilon^{0}-\frac{h^{0m}h^{0p}}{h^{00}%
}\varepsilon_{,m}^{0}~.
\]

Equation (\ref{eqnHTZ15}) is exactly the same as Eq. (146) of \cite{Novel};
but (\ref{eqnHTZ16}) has different signs in front of a few terms, as compared
to the similar Eq. (147) of \cite{Novel}. These transformations,
(\ref{eqnHTZ13})-(\ref{eqnHTZ14}), are not transformations under
four-diffeomorphism (see (\ref{eqnCAS56})-(\ref{eqnCAS57}) in previous
Section) and at most the so-called \textquotedblleft
correspondence\textquotedblright\ can be found by a field dependent
redefinition of the gauge parameters. Note that such a \textquotedblleft
redefinition\textquotedblright\ is not needed when the Castellani algorithm is
employed. The \textquotedblleft appropriate\textquotedblright\ redefinition of
gauge parameters was found in \cite{Novel}:%

\begin{equation}
\varepsilon^{0}=-\frac{1}{h^{00}}\xi^{0}, \label{eqnHTZ17}%
\end{equation}

\begin{equation}
\varepsilon^{i}=\xi^{i}-\frac{h^{0i}}{h^{00}}\xi^{0}.\label{eqnHTZ18}%
\end{equation}
\ Substitution of these expressions into (\ref{eqnHTZ15}) gives%

\begin{equation}
\delta h^{00}=h^{00}\xi_{,0}^{0}+2h^{0p}\xi_{,p}^{0}-\xi^{0}h_{,0}^{00}%
-\xi^{p}h_{,p}^{00}-h^{00}\xi_{,p}^{p}~. \label{eqnHTZ19}%
\end{equation}

As we mentioned above, our (\ref{eqnHTZ16}) has a few different signs compared
to the similar Eq. (147) of \cite{Novel}; this can probably explain why only a
transformation $\delta h^{00}$ was provided in \cite{Novel}. Substitution of
(\ref{eqnHTZ17}) and (\ref{eqnHTZ18}) into (\ref{eqnHTZ16}) gives%

\begin{equation}
\delta h^{0p}=h^{00}\xi_{,0}^{p}+h^{0m}\xi_{,m}^{p}+h^{pm}\xi_{,m}^{0}-\xi
^{0}h_{,0}^{0p}-h^{0p}\xi_{,m}^{m}-\xi^{m}h_{,m}^{0p}~. \label{eqnHTZ20}%
\end{equation}

According to the author of \cite{Novel} (\ref{eqnHTZ15}) \textquotedblleft is
IDENTICAL [Capital letters are ours] with diffeomorphism invariance ... IF we
substitute\textquotedblright\ (\ref{eqnHTZ17}) and (\ref{eqnHTZ18}), i.e.
perform field dependent redefinition of gauge parameters, which by the initial
assumption of the HTZ ansatz should be independent of the phase-space variables.

What is the significance of such a \textquotedblleft
correspondence\textquotedblright,\ especially if there is a different method
that leads directly to four-diffeomorphism? Firstly, by the transformations
produced by the HTZ ansatz are (\ref{eqnHTZ15})-(\ref{eqnHTZ16}) and their
derivation was based on the assumption (that was used in the course of
derivation) that the gauge parameters are field independent. So, a field
dependent redefinition of the gauge parameters is just a manipulation, not a
derivation; and this manipulation is in contradiction with what was used to
derive the gauge transformations. Secondly, why is this particular
redefinition of parameters chosen? If the gauge invariance of some theory is
not known \textit{a priori}, then it is meaningless to seek a correspondence
to this unknown gauge invariance precisely because it is unknown. In such a
case what should be called the gauge invariance if such manipulations are
allowed? Gauge invariance is a unique and very important characteristic of a
theory, and neither a Hamiltonian formulation nor methods of restoration that
lead to such ambiguities (or the need for such manipulations) can be accepted.
Returning to the discussion in Section 5, it is quite obvious that the HTZ
method cannot give a covariant result for this choice of tertiary constraints
because these constraints, (\ref{eqnAM70}) and (\ref{eqnAM72}), have different
dimensions, as do the corresponding gauge parameters. In the Castellani
procedure the choice of linear combinations of tertiary constraints is
irrelevant and any combination leads to the same unique gauge transformation,
whereas in the HTZ method, at least some of the combinations of tertiary
constraints definitely cannot lead to a covariant result. This gives a
limitation on the application of this method and imposes severe restrictions
on the possible operations with non-primary first class constraints.

Let us take a different combination of tertiary constraints which gives
$H_{c}^{\prime\prime}$ of (\ref{eqnCAS68}), one that we used in previous
Section where we showed that in this case the Castellani procedure produced
the same gauge transformations as using any other combination. Let us return
to (\ref{eqnHTZ8}), which is the first step of the iterative procedure, and
its form is the same for any choice of tertiary constraints (only structure
functions will be different). Now we consider the following linear
combinations of tertiary constraints%

\begin{equation}
\tilde{\tau}_{0}^{00}=\frac{1}{h^{00}}\tau_{0}^{00}+\frac{h^{0m}}{h^{00}}%
\tau_{m}^{00}\text{,\quad}\tilde{\tau}_{m}^{00}=\tau_{m}^{00}.
\label{eqnHTZ40}%
\end{equation}

In terms of these constraints the canonical Hamiltonian is%

\begin{equation}
H_{c}=-\Gamma_{00}^{\nu}\chi_{\nu}^{00}+\tilde{\tau}_{0}^{00}+A^{0}\chi
_{0}^{00}+A^{m}\chi_{m}^{00}.\label{eqnHTZ41}%
\end{equation}
In this case the structure functions $V_{\nu_{2}}^{~\mu_{3}}$ of the HTZ
ansatz become (as $\left\{  H_{c},\chi_{\mu}^{00}\right\}  =-\tilde{\tau}%
_{\mu}^{00}+$ terms proportional to $\chi_{\mu}^{00}$)%

\begin{equation}
V_{\nu_{2}}^{~\mu_{3}}=-\delta_{\nu}^{\mu}\label{eqnHTZ42}%
\end{equation}
and the solution to equation (\ref{eqnHTZ8}) is%

\begin{equation}
a^{\mu_{2}}=-\frac{\partial\varepsilon^{\mu}}{\partial t}+\varepsilon^{\nu
}V_{\nu_{3}}^{\text{~}\mu_{3}}.\label{eqnHTZ43}%
\end{equation}
Even without specifying the structure functions $V_{\nu_{3}}^{~\mu_{3}}$,
which are more complicated compared to the previous choice of tertiary
constraints, some contributions to the transformations can be calculated using
the general expression (\ref{eqnAM12})%

\begin{equation}
\delta h^{\alpha\beta}=\left\{  h^{\alpha\beta},G\right\}  =\left\{
h^{\alpha\beta},...+\frac{\partial\varepsilon^{\mu}}{\partial t}\chi_{\mu
}^{00}+...\right\}  =\frac{\partial\varepsilon^{\mu}}{\partial t}\frac
{\delta\chi_{\mu}^{00}}{\delta\pi^{\alpha\beta}}. \label{eqnHTZ44}%
\end{equation}

This partial contribution to the transformations leads exactly to the same
result as in (\ref{eqnCAS60})-(\ref{eqnCAS62}) (for this choice of
constraints, (\ref{eqnHTZ40}), equation (\ref{eqnHTZ44}) equals to
(\ref{eqnCAS59}) in previous Section), which are part of four-diffeomorphism
without any need for redefinition of gauge parameters, in full correspondence
with the original assumption of their independence of phase-space variables.

This is a clear demonstration of the sensitivity of the HTZ ansatz to the
choice of tertiary constraints. It cannot be considered an algorithm if such
ambiguities are possible. We have already discussed this in Section 4 (see
(\ref{eqnAM120})) using the arguments that invariance of the corresponding
Lagrangian cannot depend on a choice of tertiary constraints.

The simple examples, (\ref{eqnHTZ13})-(\ref{eqnHTZ16}) and (\ref{eqnHTZ44}),
are related to calculation of a small part of the full generator and
transformations that it produces, more discrepancies might appear if full
calculations are performed. If the HTZ ansatz for each choice of linear
combinations of tertiary first class constraints gives different
transformations, it cannot be considered a reliable method of finding gauge
invariance, which must be unique characteristics of a system. The only
possibility to reconcile the ambiguities of the HTZ ansatz with a unique gauge
invariance of a system is an existence of one particular choice of tertiary
constraints and this choice must be specified in the HTZ approach. How can
such a combination of tertiary constraints be found? It makes no sense to even
try. Especially, since it is not clear at all whether it is necessary to do
this if there exists the Castellani algorithm that allows us to work with any
combination of tertiary constraints without affecting the unique gauge
invariance of a system; and, at the same time, we can pick a combination of
constraints with which it is easier to perform calculations.

The possible existence of one special combination of constraints gives too
strong restriction on the HTZ approach and is opposite to the advantages of
the method that were stated in \cite{HTZ}. In particular, according to the
authors \cite{HTZ}, if the restriction that structure functions become
structure constants is imposed (see Eq. (4.5b) \cite{HTZ}) then the generators
can be written in the form given previously by Castellani. In reality, the
opposite is true: only with exactly such a restriction, (\ref{eqnHTZ42}), the
HTZ ansatz gives some meaningful results, contrary to the Castellani
algorithm, which is independent on a choice of linear combinations of tertiary
constraints, as it should be in a reliable algorithm. For completeness of
discussion of the HTZ ansatz we provide our arguments on what restrictions
might be imposed on a choice of constraints in the HTZ ansatz to recover the
same invariance with the Castellani algorithm (which is free from such
restrictions). These arguments are based on calculations that were performed
in \cite{Myths} where to restore the four-diffeomorphism in Dirac's
Hamiltonian formulation of second order GR we used both methods: Castellani's
and one based on the HTZ ansatz developed in \cite{BRR}. We obtained the same
result, four-diffeomorphism with no need for any field dependent redefinition
of parameters, by using both methods; and this, as we understood now, was an
\textquotedblleft accidental\textquotedblright\ result. The reason for this is
a very special choice of secondary constraints, as in (\ref{eqnAM201}), that
we used in \cite{Myths}; and they were defined as the result of the PBs of
primary constraints with the Hamiltonian. This is the only case where the HTZ
ansatz might work. One realistic field-theoretical example considered by the
authors of \cite{BRR} is Yang-Mills theory, which also leads to known gauge
transformations; it was also considered by Castellani \cite{Castellani} to
illustrate the use of his algorithm. But in case of Yang-Mills theory, the
secondary constraints were also defined as PBs of the corresponding primary
constraints with the Hamiltonian. It seems to us that only with such a
restriction on what we should call a secondary constraint the HTZ ansatz
reproduces the same results as the Castellani algorithm. In the first order
affine-metric formulation of GR, tertiary constraints also appear. It is not
clear how to find the \textquotedblleft right\textquotedblright\ combination
of constraints. The HTZ method might work if we define a \textquotedblleft
tertiary constraint\textquotedblright\ as everything that is produced as a
time development of the corresponding secondary first class constraint. In
contrast, according to the Dirac procedure at every step, when we consider a
time development of known constraints, we have, first of all, to single out
constraints which are already found, then the remaining part can be called a
new constraint. In the formulation considered here, first order metric-affine
GR, PBs of secondary first class constraints with the Hamiltonian give the
expressions in which we can isolate terms proportional to a linear combination
of secondary constraints. But this linear combination is not unique, and we do
not have any prescription to find which linear combination is preferable. In
addition, if all linear combinations of secondary constraints are equally
good, why do we have to use a particular and very special linear combination
of tertiary constraints? From our point of view, it is preferable to use the
method which is free from these problems.

Another important observation, based on our calculation in the second order EH
action using the Castellani or \cite{BRR} methods is that the amount of
calculation is the same in both cases; so there is no advantage in using the
HTZ ansatz even if we can figure out what restrictions should be imposed for
the general case with long chains of constraints. In contrast, our partial
calculations in the previous Section show that the independence of the final
result on which linear combinations of tertiary constraints are used, gives us
an advantage as some of them can drastically simplify the calculations.

The sensitivity of the HTZ ansatz and the problems related to it provide an
additional illustration of importance of primary first class constraints in
the Hamiltonian formulation that we discussed in Section 5. They are the true
Masters of the Hamiltonian formulation and because the Castellani approach
uses explicitly only primary constraints, it gives correct results and takes
care of the possible redefinition of all non-primary constraints. But methods
based on the HTZ ansatz, which starts from non-primary constraints, is an
attempt to interchange the roles of constraints and this is the reason for its
failure. The HTZ ansatz has some similarities with the non-canonical change of
variables in the ADM formulation: both treat primary constraints as
unimportant and both emphasize the role of non-primary constraints, either by
starting the iterative procedure from them or by completely fixing one
particular combination of secondary constraints that are reflected in giving
the special names: \textquotedblleft Hamiltonian\textquotedblright\ and
\textquotedblleft spatial diffeomorphism\textquotedblright\ constraints. Such
names are in contradiction with the covariance of General Relativity and with
the Dirac conjecture that all first class constraints are responsible for
gauge invariance. For example, among eight constraints of the second order
formulation of GR in four-dimensional spacetime, if we separate three of them
and call them a \textquotedblleft spatial diffeomorphism\textquotedblright%
\ constraint, for what invariance are the five remaining constraints responsible?

The origin of the failure of the HTZ ansatz is made especially clear from
another article \cite{GHP}, which was published almost simultaneously with
\cite{HTZ}. In \cite{GHP} the Castellani procedure for construction of a
generator of gauge transformations is considered, and the authors remark that
\textquotedblleft the problem is complicated by the fact that the chain
algorithm [the name used for the Castellani algorithm (\ref{eqnCAS2}%
-\ref{eqnCAS5})] strongly depends on the representation of the primary
first-class constraints surface that is adopted. More precisely, if the
primary first-class constraints are%

\begin{equation}
\phi_{a_{1}}=0,\text{ \ \ }a_{1}=1,...,m_{1}\label{eqnHTZ51}%
\end{equation}
and can each be taken as the head of a chain in [our (\ref{eqnCAS2})], it is
in general not true that the equivalent constraints%

\begin{equation}
\phi_{a_{1}}^{\prime}=M_{a_{1}}^{b_{1}}\left(  q,p\right)  \phi_{b_{1}%
}=0\label{eqnHTZ52}%
\end{equation}
also lead to consistent solution of [our (\ref{eqnCAS2}-\ref{eqnCAS5})]. So,
algorithm is not invariant under [our (\ref{eqnHTZ52})]. An algorithm ... that
is invariant under redefinition [our (\ref{eqnHTZ52})] has been recently
proposed in [our \cite{HTZ}].\textquotedblright\ 

So, according to \cite{GHP} the HTZ ansatz is built on the assumption of
invariance under (\ref{eqnHTZ52}) while the Castellani algorithm is not.

Of course, the Hamiltonian formulation \textquotedblleft strongly
depends\textquotedblright\ on the primary first class constraints; but this is
not a \textquotedblleft problem\textquotedblright. It is just the nature of
the Hamiltonian formulation that is reflected in the Castellani
algorithm\footnote{This kind of problem is similar to what is stated in the
first line of \textquotedblleft Golden Oldie\textquotedblright\ \cite{goldies}%
: \textquotedblleft The general coordinate invariance underlying the theory of
relativity creates basic problems in the analysis of the dynamics of the
gravitational field\textquotedblright. ADM variables is the solution of the
problem with covariance.}. As we discussed in Section 5, primary first class
constrains are special due to their dual nature: they are constraints and, at
the same time, they are phase-space variables. As constraints, they allow
redefinition (\ref{eqnHTZ52}) without any restriction. If primary constraints
are pure momenta (as it happens in most cases), then (\ref{eqnHTZ52}) would be
also the redefinition of phase-space variables and it must be canonical! So
this is not a surprise that an \textquotedblleft algorithm\textquotedblright%
\ which allows an arbitrary redefinition (\ref{eqnHTZ52}), leads to loss of a
unique gauge invariance because in general such a redefinition does not
correspond to a canonical transformation; and consequently, the new total
Hamiltonian will not be equivalent to the original one, as well as its
original gauge invariance will not be restored without some field dependent
redefinition of gauge parameters (at best). So any \textquotedblleft
algorithm\textquotedblright\ that is invariant under arbitrary (\ref{eqnHTZ52}%
) will unavoidably lead to the loss of gauge invariance for majority of
choices of $M_{a_{1}}^{b_{1}}\left(  q,p\right)  $.

Calculations with the Hamiltonian of GR are involved and to illustrate the
devastating effect of (\ref{eqnHTZ52}) the simple Hamiltonian formulation can
be considered, e.g. Maxwell Electodynamics for which the total Hamiltonian
\cite{Kurt, KK} is%

\[
H_{T}=\dot{A}^{0}\pi_{0}-A^{0}\partial_{k}\pi^{k}-\frac{1}{2}\pi_{k}\pi
^{k}+\frac{1}{4}\left(  \partial_{k}A_{m}-\partial_{m}A_{k}\right)  \left(
\partial^{k}A^{m}-\partial^{m}A^{k}\right)  .
\]
One can take, for example, the following redefinition%

\[
\pi_{0}^{\prime}=A^{k}\pi_{k}\pi_{0}%
\]
and try to consider the time development of this new primary constraint,
closure of the Dirac procedure, algebra of constraints and the restoration of
gauge invariance in this case.

\section{Conclusion}

In this paper we presented a detailed derivation of the Hamiltonian for the
first order affine-metric formulation of GR, including restoration of its
unique gauge invariance, four-dimensional diffeomorphism, using the Castellani
algorithm, and we demonstrated that four-diffeomorphism can be lost as the
result of a non-canonical change of variables or by using methods of
restoration which are sensitive to a choice of linear combinations of
non-primary first class constraints. These results are based on mathematical
derivations and as such do not need any interpretation or discussion. They can
be disproved by indication of mistake(s) or must be accepted. One additional,
\textquotedblleft conventional\textquotedblright, option is just to ignore
them saying that it is the well-known fact that \textquotedblleft the
canonical treatment breaks the symmetry between space and time in general
relativity and the resulting algebra of constraints is not the algebra of four
diffeomorphism\textquotedblright\ \cite{Pullin} and, because of this, only by
some unjustified manipulations the \textquotedblleft correspondence to
diffeomorphism invariance\textquotedblright\ can be accomplished.

We address this article to the readers who make their judgement based on the
results, not on the correspondence of results to \textquotedblleft
conventional wisdom\textquotedblright. Such readers, as well as
\textquotedblleft conventional\textquotedblright\ ones, do not need a long
conclusion to make their minds and, because of this, we stop our discussion here.

\vspace{5mm}

\textbf{ACKNOWLEDGMENTS}

\vspace{5mm}

The authors would like to thank D.G.C. McKeon for attracting their attention
to the necessity of the field dependent redefinition of the gauge parameters
in the Hamiltonian formulation of affine-metric Einstein's GR when applying
the Henneaux-Teitelboim-Zanelli approach.

The authors are grateful to A.M. Frolov, A.V. Zvelindovsky and especially to
P.G. Komorowski, for numerous discussions and suggestions. The partial support
of The Huron University College Faculty of Arts and Social Science Research
Grant Fund is greatly acknowledged.

\appendix

\section{Solving the second class constraints}

We outline the last step of the Hamiltonian reduction, the elimination of two
additional pairs of the phase-space variables $\left(  \Pi_{k}^{0m}%
,\Sigma_{0m}^{k}\right)  $ and $\left(  \Pi_{kp}^{m},\Gamma_{kp}^{m}\right)
$. The part of the canonical Hamiltonian, $H_{c}$, with terms proportional to
$\Sigma_{0m}^{k}$ and $\Gamma_{km}^{p}$ (two last lines of (\ref{eqnAM23})),
the only source of contributions into secondary second class constraints
(\ref{eqnAM28}) and (\ref{eqnAM29}), is%

\begin{equation}
H_{c}\left(  \Sigma_{0m}^{k},\Gamma_{km}^{p}\right)  =-h^{00}\Sigma_{0k}%
^{m}\Sigma_{0m}^{k}+2h^{k0}\left(  \Gamma_{mp}^{p}\Sigma_{0k}^{m}-\Gamma
_{km}^{p}\Sigma_{0p}^{m}\right)  -h^{km}2\pi_{kp}\Sigma_{0m}^{p}-2h_{,k}%
^{p0}\Sigma_{0p}^{k} \label{eqnA1}%
\end{equation}

\[
+h^{km}\left(  2\Gamma_{pq}^{q}\Gamma_{km}^{p}-\Gamma_{kp}^{p}\Gamma_{mq}%
^{q}-\Gamma_{kq}^{p}\Gamma_{mp}^{q}\right)  -2h^{k0}\pi_{00}\Gamma_{km}%
^{m}+2h^{km}\left(  \pi_{0p}\Gamma_{km}^{p}-2\pi_{0k}\Gamma_{mq}^{q}\right)
-h_{,k}^{pq}\Gamma_{pq}^{k}+2h_{,k}^{pk}\Gamma_{pm}^{m}.
\]
The first line of (\ref{eqnA1}) can be presented in the following form%

\begin{equation}
H_{c}\left(  \Sigma_{0m}^{k}\right)  =-h^{00}\Sigma_{0k}^{m}\Sigma_{0m}%
^{k}-2\Sigma_{0k}^{m}\tilde{D}_{m}^{0k}\label{eqnA2}%
\end{equation}
where%

\begin{equation}
\tilde{D}_{m}^{0k}=h_{,m}^{k0}+h^{qk}\pi_{qm}+h^{q0}\Gamma_{qm}^{k}%
-h^{k0}\Gamma_{mp}^{p}.\label{eqnA3}%
\end{equation}
Note that $\tilde{D}_{m}^{0k}$ is not a traceless combination.

The secondary constraint $\chi_{k}^{0m}$, (\ref{eqnAM28}), can be obtained
from the variation of (\ref{eqnA1}) (using the fundamental PB $\left\{
\Sigma_{0m}^{k},\Pi_{q}^{0p}\right\}  $ from (\ref{eqnAM11}))%

\begin{equation}
\dot{\Pi}_{k}^{0m}=\left\{  \Pi_{k}^{0m},H_{c}\right\}  =-\frac{\delta H_{c}%
}{\delta\Sigma_{0m}^{k}}=\chi_{k}^{0m}=h^{00}\Sigma_{0k}^{m}+\tilde{D}%
_{k}^{0m}-\frac{\delta_{k}^{m}}{D-1}\tilde{D}_{n}^{0n}. \label{eqnA4}%
\end{equation}

This constraint obviously has a non-zero PB with the primary constraint
$\Pi_{q}^{0p}$; and this pair of constraints $\left(  \Pi_{k}^{0m},\chi
_{0m}^{k}\right)  $ is of second class and the corresponding pair of
variables, $\left(  \Pi_{k}^{0m},\Sigma_{0m}^{k}\right)  ,$ can be eliminated.
Solving $\chi_{k}^{0m}=0$ from (\ref{eqnA4}) for $\Sigma_{0k}^{m}$ and
substituting the pair%

\begin{equation}
\Pi_{q}^{0p}=0,\text{ \ \ \ }\Sigma_{0k}^{m}=-\frac{1}{h^{00}}\left(
\tilde{D}_{k}^{0m}-\frac{1}{D-1}\tilde{D}_{n}^{0n}\delta_{k}^{m}\right)
\label{eqnA5}%
\end{equation}
into (\ref{eqnAM23}) gives the next reduction with the following change in the
canonical Hamiltonian $H_{c}$,%

\begin{equation}
H_{c}\left(  \Sigma_{0m}^{k}\right)  =H_{c}\left(  \Sigma_{0m}^{k}\text{ from
}Eq.(\ref{eqnA5})\right)  =-\frac{1}{h^{00}}\tilde{D}_{b}^{0a}\tilde{D}%
_{a}^{0b}+\frac{1}{D-1}\frac{1}{h^{00}}\tilde{D}_{n}^{0n}\tilde{D}_{a}%
^{0a}.\label{eqnA6}%
\end{equation}
Separating in $\tilde{D}_{b}^{0a}$ contributions proportional to $\Gamma
_{qm}^{k}$ we write%

\begin{equation}
\tilde{D}_{m}^{0k}=D_{m}^{0k}+D_{m}^{0k}\left(  \Gamma_{qm}^{k}\right)
\label{eqnA7}%
\end{equation}
where%

\begin{equation}
D_{m}^{0k}=h_{,m}^{k0}+h^{qk}\pi_{qm},\label{eqnA8}%
\end{equation}
and%

\begin{equation}
D_{m}^{0k}\left(  \Gamma_{qm}^{k}\right)  =h^{q0}\Gamma_{qm}^{k}-h^{k0}%
\Gamma_{mp}^{p}.\label{eqnA9}%
\end{equation}
Note that the trace of $D_{m}^{0k}\left(  \Gamma_{qm}^{k}\right)  $ is zero
and for (\ref{eqnA6}); we obtain%

\begin{equation}
H_{c}\left(  \Sigma_{0m}^{k}\right)  =-\frac{1}{h^{00}}D_{b}^{0a}D_{a}%
^{0b}+\frac{1}{D-1}\frac{1}{h^{00}}D_{n}^{0n}D_{a}^{0a}-\frac{1}{h^{00}}%
D_{b}^{0a}D_{a}^{0b}\left(  \Gamma_{qm}^{k}\right)  -\frac{1}{h^{00}}%
D_{b}^{0a}\left(  \Gamma_{qm}^{k}\right)  D_{a}^{0b}\left(  \Gamma_{qm}%
^{k}\right)  . \label{eqnA10}%
\end{equation}

The first two terms in (\ref{eqnA10}) correspond to the second line of
(\ref{eqnAM32a}) and the last two terms of (\ref{eqnA10}) must be combined
with the second line of (\ref{eqnA1}). Now the part of the canonical
Hamiltonian that depends on $\Gamma_{qm}^{k}$ can be written as%

\begin{equation}
H_{c}\left(  \Gamma_{qm}^{k}\right)  =e^{km}\left(  2\Gamma_{pq}^{q}%
\Gamma_{km}^{p}-\Gamma_{kp}^{p}\Gamma_{mq}^{q}-\Gamma_{kq}^{p}\Gamma_{mp}%
^{q}\right)  -\tilde{D}_{m}^{qk}\Gamma_{qk}^{m}. \label{eqnA11}%
\end{equation}

In the quadratic part of (\ref{eqnA11}), the combination $e^{km}$ naturally
arises and $\tilde{D}_{m}^{qk}$, in the part linear in $\Gamma_{qk}^{m}$, can
be written in manifestly symmetric form%

\begin{equation}
\tilde{D}_{m}^{qk}=-2h^{kq}\pi_{0m}+h_{,m}^{kq}-\frac{h^{q0}}{h^{00}}%
D_{m}^{0k}-\frac{h^{k0}}{h^{00}}D_{m}^{0q} \label{eqnA12}%
\end{equation}

\[
+\left(  h^{k0}\pi_{00}+2h^{pk}\pi_{0p}-h_{,p}^{kp}+\frac{h^{c0}}{h^{00}}%
D_{c}^{0k}\right)  \delta_{m}^{q}+\left(  h^{q0}\pi_{00}+2h^{pq}\pi
_{0p}-h_{,p}^{qp}+\frac{h^{c0}}{h^{00}}D_{c}^{0q}\right)  \delta_{m}^{k}%
\]
where $D_{m}^{0k}$ is defined as (\ref{eqnA8}) (or (\ref{eqnAM34}) in the main
text). This part of $H_{c}$, (\ref{eqnA11}), leads to the secondary second
class constraints (\ref{eqnAM29})%

\begin{equation}
\dot{\Pi}_{x}^{yz}=\left\{  \Pi_{x}^{yz},H_{c}\right\}  =-\frac{\delta H_{c}%
}{\delta\Gamma_{yz}^{x}}=\chi_{x}^{yz}=0.\label{eqnA14}%
\end{equation}
and the last pair of phase-space variables can be eliminated: $\Pi_{m}^{kp}=0$
and the solution of (\ref{eqnA14}) for $\Gamma_{kp}^{m}$. Performing the
variation $\frac{\delta H_{c}}{\delta\Gamma_{yz}^{x}}$ in (\ref{eqnA14}) we obtain%

\begin{equation}
2e^{yz}\Gamma_{xq}^{q}-e^{ky}\Gamma_{kx}^{z}-e^{kz}\Gamma_{kx}^{y}+\delta
_{x}^{z}\left(  e^{km}\Gamma_{km}^{y}-e^{ky}\Gamma_{kp}^{p}\right)
+\delta_{x}^{y}\left(  e^{km}\Gamma_{km}^{z}-e^{kz}\Gamma_{kp}^{p}\right)
=\tilde{D}_{x}^{yz}. \label{eqnA15}%
\end{equation}

The way to solve (\ref{eqnA15}) for $\Gamma_{kx}^{z}$ is similar to the
Einstein proof of equivalence of the first and second order formulations of
metric GR \cite{Einstein} (see also \cite{KK}, Appendix A). The solution of
(\ref{eqnA15}) is based on the subsequent elimination of \textquotedblleft
traces\textquotedblright. Contracting (\ref{eqnA15}) with $\delta_{y}^{x}$ we obtain%

\begin{equation}
e^{km}\Gamma_{km}^{z}-e^{kz}\Gamma_{kp}^{p}=\frac{1}{D-1}\tilde{D}_{x}^{xz}.
\label{eqnA16}%
\end{equation}

The left-hand side of (\ref{eqnA16}) is exactly the combination which appears
in brackets with a Kronecker delta in equation (\ref{eqnA15}). This allows us
to write (\ref{eqnA15}) as%

\begin{equation}
2e^{yz}\Gamma_{xq}^{q}-e^{ky}\Gamma_{kx}^{z}-e^{kz}\Gamma_{kx}^{y}=\tilde
{D}_{x}^{yz}-\frac{\delta_{x}^{z}}{D-1}\tilde{D}_{p}^{py}-\frac{\delta_{x}%
^{y}}{D-1}\tilde{D}_{p}^{pz}\equiv D_{x}^{yz}.\label{eqnA17}%
\end{equation}
This combination, $D_{x}^{yz}$, is also used in the reduced Hamiltonian
(\ref{eqnAM32a}); and its explicit form is written in (\ref{eqnAM35}).

We still have to eliminate the trace $\Gamma_{xq}^{q}$ in (\ref{eqnA17}).
Contracting (\ref{eqnA17}) with $h_{yz}$ we find%

\begin{equation}
\Gamma_{xq}^{q}=\frac{1}{2\left(  D-2\right)  }h_{yz}D_{x}^{yz}\label{eqnA18}%
\end{equation}
and equation (\ref{eqnA17}) takes the final form%

\begin{equation}
-e^{ky}\Gamma_{kx}^{z}-e^{kz}\Gamma_{kx}^{y}=D_{x}^{yz}-e^{yz}\frac{1}{\left(
D-2\right)  }h_{pq}D_{x}^{pq}\equiv\hat{D}_{x}^{yz}. \label{eqnA19}%
\end{equation}

The solution of (\ref{eqnA19}) can be found using the Einstein permutation, as
it was done in \cite{Einstein} and \cite{KK} (Appendix A). We need to have
three free indices in the same position. To obtain this, we contract
(\ref{eqnA19}) with $e^{nx}$ to obtain%

\begin{equation}
-e^{nx}e^{ky}\Gamma_{kx}^{z}-e^{nx}e^{kz}\Gamma_{kx}^{y}=e^{nx}\hat{D}%
_{x}^{yz}. \label{eqnA20}%
\end{equation}

Now we perform a permutation in indices $n,y$ and $z$ and add the resulting
equations in the following order $\left(  nyz\right)  +\left(  zny\right)
-\left(  yzn\right)  $. That gives us%

\begin{equation}
e^{nx}e^{kz}\Gamma_{kx}^{y}=-\frac{1}{2}\left(  e^{nx}\hat{D}_{x}^{yz}%
+e^{zx}\hat{D}_{x}^{ny}-e^{yx}\hat{D}_{x}^{zn}\right)  .\label{eqnA21}%
\end{equation}
Contracting (\ref{eqnA21}) with $h_{an}h_{bz}$ we obtain the solution for
$\Gamma_{ba}^{y}$,%

\begin{equation}
\Gamma_{ba}^{y}=-\frac{1}{2}\left(  h_{bz}\hat{D}_{a}^{yz}+h_{an}\hat{D}%
_{b}^{ny}-h_{an}h_{bz}e^{yx}\hat{D}_{x}^{zn}\right)  .\label{eqnA22}%
\end{equation}
The last pair of phase-space variables $\left(  \Pi_{x}^{yz},\Gamma_{yz}%
^{x}\right)  $ can be eliminated (the last Hamiltonian reduction).

Substitution of (\ref{eqnA22}) into the part of the canonical Hamiltonian
(\ref{eqnA11}) gives the third line of (\ref{eqnAM32a}) that is expressed in
terms of the combination $D_{x}^{yz}$. Note that in (\ref{eqnA11}) we have
$\tilde{D}_{x}^{yz}$ and by using our redefinition (\ref{eqnA17}) can be also
written in terms of $D_{x}^{yz}$%

\begin{equation}
\tilde{D}_{x}^{yz}=D_{x}^{yz}-D_{p}^{py}\delta_{x}^{z}-D_{p}^{pz}\delta
_{x}^{y}. \label{eqnA23}%
\end{equation}

This completes the Hamiltonian reduction of the affine-metric formulation of GR.

\end{document}